\pdfoutput=1

\documentclass[reprint,aps,prb,twocolumn,floatfix,superscriptaddress]{revtex4-2}
\usepackage{graphicx}
\usepackage{amssymb}
\usepackage{bm}
\usepackage{amsmath}
\usepackage{dsfont}
\usepackage{lineno}
\usepackage[english]{babel}
\usepackage[utf8]{inputenc}
\usepackage{xcolor}
\RequirePackage[T1]{fontenc}
\RequirePackage{lmodern}
\usepackage{appendix}
\usepackage{float}
\usepackage{algpseudocode}
\usepackage{algorithm}
\usepackage{mathtools}
\usepackage{filecontents}
\usepackage{physics}
\usepackage{comment}
\usepackage{bbm}
\usepackage{siunitx}
\usepackage{mathtools}
\usepackage{hyperref}
\usepackage[nameinlink,capitalise]{cleveref}
\usepackage{orcidlink}
\usepackage{color} % for the command \textcolor
\usepackage{soul} % for the command \hl

\begin{document}

\preprint{APS/123-QED}

\def\E{\mathbb E}
\def\H{\mathcal H}
\def\N{\mathcal N}
\def\P{\mathbb P}
\def\Q{\mathbb Q}
\def\R{\mathbb R}
\def\bfp{\mathbf p}
\def\bfq{\mathbf q}
\def\bfx{\mathbf x}
\def\V{\mathbb V}
\def\argmax{\mathrm{argmax}}
\def\argmin{\mathrm{argmin}}
\def\sgn{\mathrm{sgn}}
\newcommand\given[1][]{\:#1\vert\:}

\title{The effects of alloy disorder on strongly-driven flopping mode qubits in Si/SiGe}

\author{Merritt P. R. \surname{Losert} } 
\altaffiliation[Current address: ]{mlosert@umd.edu}
\affiliation{University of Wisconsin-Madison, Madison, WI 53706 USA}

\author{Utkan G{\"u}ng{\"o}rd{\"u}}
\affiliation{Laboratory for Physical Sciences, College Park, Maryland 20740, USA}
\affiliation{Department of Physics, University of Maryland, College Park, Maryland 20742, USA}

\author{S. N. \surname{Coppersmith}}
\affiliation{School of Physics, University of New South Wales, Sydney, New South Wales 2052, Australia}

\author{Mark Friesen}
\affiliation{University of Wisconsin-Madison, Madison, WI 53706 USA}

\author{Charles Tahan}
\altaffiliation[Current address: ]{Department of Physics, University of Maryland, College Park, Maryland 20742, USA}
\affiliation{Laboratory for Physical Sciences, College Park, Maryland 20740, USA}

\date{\today}

\date{\today}% It is always \today, today,

\begin{abstract}
In Si quantum dot systems, large magnetic field gradients are needed to implement spin rotations via electric dipole spin resonance (EDSR).
By increasing the effective electron dipole, flopping mode qubits can provide faster gates with smaller field gradients.
Moreover, operating in the strong-driving limit can reduce their sensitivity to charge noise.
However, alloy disorder in Si/SiGe heterostructures randomizes the valley energy splitting and the valley phase difference between dots, enhancing the probably of valley excitations while tunneling between the dots, and opening a leakage channel.
In this work, we analyze the performance of flopping mode spin qubits in the presence of charge noise and alloy disorder, and we optimize these qubits for a variety of valley configurations, in both weak and strong charge-noise regimes.
When the charge noise is weak, high fidelity qubits can be implemented across a wide range of valley parameters, provided the electronic pulse is fine-tuned for a given valley configuration.
When the charge noise is strong, high-fidelity pulses can still be engineered, provided the valley splittings in each dot are relatively large and the valley phase difference is relatively small.
We analyze how charge noise-induced fluctuations of the inter-dot detuning, as well as small shifts in other qubit parameters, impact qubit fidelities.
We find that strongly driven pulses are less sensitive to detuning fluctuations but more sensitive to small shifts in the valley parameters, which can actually dominate the qubit infidelities in some regimes.
Finally, we discuss schemes to tune devices away from poor-performing configurations, enhancing the scalability of flopping-mode-based qubit architectures. 
\end{abstract}

\maketitle

\section{Introduction}

Over the past decade, quantum dot spin qubits in Si/SiGe heterostructures have emerged as a leading quantum computing platform \cite{Loss:1998p120, Zwanenburg:2013p961, Burkard:2023p025003}.
These qubits are small, have long coherence times, and can be fabricated using standard semiconductor processing technology, making them naturally scalable.
Furthermore, single- and two-qubit fidelities above 99\% have been demonstrated in these systems, demonstrating their promise as scalable qubit platforms \cite{Yoneda:2018p102, Xue:2022p343, Noiri:2022p338, Mills:2022p5130}.

One of the primary sources of infidelity in Si spin qubits is charge noise, or fluctuating electric fields at the location of the spin qubit \cite{Burkard:2023p025003}.
These electric field fluctuations typically have a $1/f$ spectral density, leading to low-frequency (or quasistatic) fluctuations of the qubit parameters, resulting in dephasing \cite{Dutta:1981p497, Culcer:2009p073102, Bermeister:2014p192102, Shehata:2023p045305}.
These fluctuating fields are thought to arise from the motion of charges, primarily in the oxide layers of the heterostructure \cite{Saraiva:2022p2105488}, and they are the dominant source of infidelity in many state-of-the-art spin qubits \cite{Yoneda:2018p102, Struck:2020p40, Connors:2022p940, Xue:2022p343, Noiri:2022p338, Mills:2022p5130}.

It is possible to mitigate the impact of charge noise by designing qubits and gate operations that are inherently insensitive to these electric field fluctuations.
One such qubit is the flopping-mode qubit \cite{Benito:2019p125430, Croot:2020p012006, Hu:2023p134002, Teske:2023p035302, Young:2025p064042}.
In this system, a magnetic field gradient across the double dot induces synthetic spin-orbit coupling in the system.
By driving the inter-dot detuning, causing the electron to oscillate between the two dots, we can induce magnetic field oscillations that generate single-qubit rotations in a process known as electric dipole spin resonance (EDSR).
Compared to conventional single quantum dot qubits, flopping mode qubits maximize the dipole moment of the electron wavefunction, resulting in faster Rabi oscillations \cite{Benito:2019p125430}.
It has also been proposed to operate flopping mode qubits in the so-called ``strong driving'' regime, where the electron is strongly driven back and forth between the two dots \cite{Teske:2023p035302}.
This style of gate operation has two advantages: (1) it takes advantage of the full magnetic field gradient between the dots, resulting in fast gate operations, and (2) the electron spends less time at the symmetric point, where the spin quantization axis is most sensitive to changes in detuning.
In particular, it has been shown that by carefully shaping the detuning pulses, fast and high-fidelity single-qubit gates can be implemented, even in the presence of significant charge noise \cite{Teske:2023p035302}.

Another challenge for spin qubits in Si/SiGe is the presence of two low-lying conduction-band valley states \cite{Zwanenburg:2013p961, Burkard:2023p025003}.
Reported values of the valley-energy splitting are as large as \SI{300}{\micro\electronvolt} or as low as \SI{30}{\micro\electronvolt} \cite{Borselli:2011p123118, Shi:2011p233108, Zajac:2015p223507, Scarlino:2017p165429, Mi:2017p176803, Mi:2018p161404, Ferdous:2018p26, Neyens:2018p243107, Borjans:2019p044063, Hollmann:2020p034068, Oh:2021p125122, Chen:2021p044033, Esposti:2024p32, Volmer:2024p61}.
Even valley splitting measurements taken on the same chip are quite variable \cite{Chen:2021p044033} and sensitive to small displacements in the quantum dot position \cite{Hollmann:2020p034068,Dodson:2022p146802,Volmer:2024p61, Volmer:2025arXiv}.
Recent theoretical advances have demonstrated that random-alloy disorder can explain this significant variability of the valley splitting, even between neighboring dots on the same device \cite{Wuetz:2022p7730, McJunkin:2022p7777, Losert:2023p125405, Lima:2023p025004, Lima:2024p036202}.

Such valley splitting variability can create problems for flopping mode qubits \cite{Young:2025p064042}.
Since tunneling between dots need not preserve the valley quantum number \cite{Shiau:2007p195345}, valley state variability can induce leakage out of the computational subspace.
Weaker driving can be used to avoid this valley leakage, at the expense of increased qubit sensitivity to charge noise. 
Thus, there is a tradeoff between valley leakage and charge noise in strongly-driven flopping mode qubits.
In this work, we closely examine this tradeoff.
We develop an algorithm that optimizes pulses for strongly-driven flopping-mode qubits in the presence of charge noise, while accounting for differences in the valley configurations in each dot.
We show that high-fidelity single-qubit gates can be realized in these systems for a wide variety of valley parameters, as long as charge noise remains relatively weak (e.g., for detuning fluctuations on the order of 1~\SI{}{\micro\electronvolt}).
In cases where charge noise is stronger (e.g., for detuning fluctuations of order 15~\SI{}{\micro\electronvolt}) we can still implement high-fidelity single-qubit gates, provided that the valley splittings in each dot are large, and the valley phase difference is relatively small.
All of these proposed pulses require some degree of fine-tuning for the particular valley configuration of a double-dot.

In addition to noise-induced detuning fluctuations, we also examine how fluctuations in the local electric fields can lead to small quasistatic shifts in the other qubit parameters, including the valley splitting in each dot, the local orbital ground-state energy in each dot, and the inter-dot tunnel coupling.
While we find that noise-induced fluctuations in the local ground-state energy and tunnel coupling are relatively unimportant, fluctuations in the valley parameters can dominate the qubit infidelity, especially when the valley splitting is small and the charge noise fluctuations are large.
However, we also find that slightly weaker qubit driving can significantly mitigate these errors, and we propose a pulse shape that is more robust to these valley fluctuations.
Thus, we consider several sources of infidelity, including spatially varying valley parameters, charge noise fluctuations of the qubit detuning, and fluctuations of other key qubit parameters.
Table~\ref{table:intro} summarizes where in the paper each of these effects is analyzed.

We note that the effects of valley splitting variability on flopping mode qubits has also been analyzed in Ref.~[\citenum{Young:2025p064042}], where the authors employ realistic atomistic simulations of a double-dot device. 
Our work differs in two respects: first, we focus on lower-dimensional effective models, allowing us to perform many simulations across a wide variety of qubit parameters, and second, we focus mainly on the strong-driving regime.

The paper is organized as follows.
In Sec.~\ref{sec:flopping_mode_summary}, we review our model of the flopping mode qubit, and we discuss the impact of alloy disorder on the valley parameters of the qubit.
In Sec.~\ref{sec:gate_op}, we discuss our algorithm for optimizing flopping-mode qubits, taking into account charge noise and differences in the valley configuration.
In Sec.~\ref{sec:valley_dependence}, we discuss in detail the performance of flopping mode qubits under varying valley configurations.
In Sec.~\ref{sec:fine_tuning}, we address the generalizability of flopping mode qubits, and we come to the conclusion that these qubits should be fine-tuned for any given valley configuration.
In Sec.~\ref{sec:valley_fluctuations}, we analyze charge-noise-induced fluctuations to the qubit parameters besides the detuning, including inter-dot differences of the orbital ground state energies in each dot, the tunnel coupling between dots, and the valley parameters in each dot.
In Sec.~\ref{sec:scalability}, we evaluate schemes to avoid low-fidelity configurations, and we discuss the implications for scalable quantum processors based on the flopping mode qubit.

\begin{table}[]
\def\arraystretch{1.3}
\begin{ruledtabular}
\begin{tabular}{ l l }% syntax for siunitx v2; for v1 use "tabformat"
 Valley splitting and valley phase & Sec.~\ref{sec:valley_dependence} \\ 
 Detuning fluctuations & Sec.~\ref{sec:valley_dependence}   \\ 
 Valley fluctuations & Sec.~\ref{sec:valley_fluctuations} \\
 Inter-dot ground state energy flucts. & Sec.~\ref{sec:valley_fluctuations} and  App.~\ref{app:orbital} \\
 Tunnel coupling fluctuations & Sec.~\ref{sec:valley_fluctuations} and App.~\ref{app:tunnel_coupling} \\
\end{tabular}
\end{ruledtabular}
\caption{\label{table:intro} Summary of the potential sources of infidelity in flopping-mode qubits, and where we analyze them in this work.}
\end{table}

\section{Modelling strongly-driven flopping mode qubits} \label{sec:flopping_mode_summary}

\begin{figure}
    \centering
    \includegraphics[width=8cm]{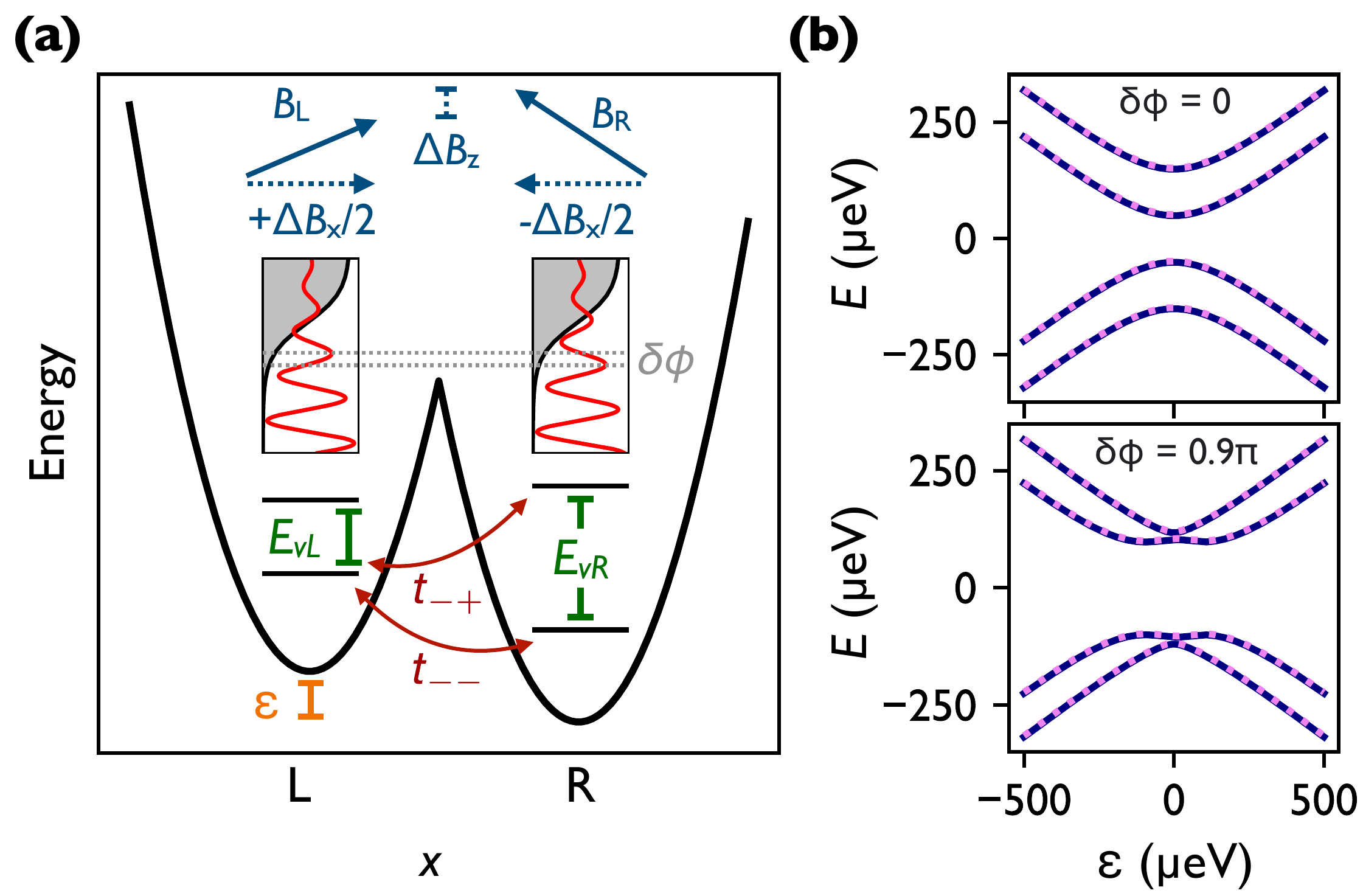}
    \caption{Schematic illustration of the flopping mode qubit. (a) Structure of the double quantum dot system, where the left and right dots are labeled L and R.
    Here, the detuning $\varepsilon$ controls the relative energy levels of the two dots. The valley splittings of the left and right dots are defined as $E_{vL}$ and $E_{vR}$. If the valley phase difference between the two dots satisfies $\delta \phi \neq 0$, then tunneling between the dots does not preserve the valley index, and both valley-preserving and valley-flipping tunneling processes are allowed. Two such processes are indicated in red, where $t_{--}$ and $t_{-+}$ are defined in Eq.~(\ref{eq:tunneling_matrix_elements}). Differences in the local magnetic fields of the two dots, labeled $B_L$ and $B_R$, enable spin rotations via EDSR as the electron is driven back and forth between the dots. (b) Example energy spectra of the double dot system as a function of detuning $\varepsilon$, with representative valley splittings given by $E_{vL} = E_{vR} = 100$~\SI{}{\micro\electronvolt}. The top panel shows the spectrum when the difference in the valley phase is given by $\delta \phi = 0$, while for the bottom panel, $\delta \phi = 0.9 \pi$. 
    Here, we use solid blue and dashed red to indicate the different spins states, which appear degeneratre for the parameters considered here. 
    All other parameters used in these calculates are specified in Sec.~\ref{sec:flopping_mode_summary}.}
    \label{fig:qubit_schematic}
\end{figure}

\subsection{Model Hamiltonian}

A flopping-mode qubit consists of a single-electron double quantum dot with a magnetic field difference between the dots.
By modulating the detuning $\varepsilon$, we can drive the dot back and forth between the left ($L$) and right ($R$) dot.
Since the field difference induces a synthetic spin-orbit coupling, resonant modulation induces rotations between the spin states.
Each dot also has two low-lying valley states, characterized by inter-valley couplings for each dot $\Delta_{L(R)} = e^{i \phi_{L(R)}} | \Delta_{L(R)}|$, where the valley splitting $E_{vL(R)} = 2|\Delta_{L(R)}|$ and $\phi_{L(R)}$ is the valley phase \cite{Friesen:2007p115318}.
These valley parameters are, in general, different for each dot, complicating the qubit operation, as described below.
A schematic of this qubit is shown in Fig.~\ref{fig:qubit_schematic}(a).

To model strongly-driven flopping mode qubits, we start with an 8-level Hamiltonian spanned by the charge, spin, and valley degrees of freedom.
The full Hamiltonian is given by
\begin{align} \label{eq:ham}
H(\varepsilon) = & \frac{\varepsilon(t)}{2} \tau_z + t_c \tau_x + \frac{g \mu_B B_z}{2}\sigma_z \nonumber\\
&+ \frac{g \mu_B}{2}(\Delta B_x \sigma_x + \Delta B_z \sigma_z) \tau_z \nonumber\\
&+ \frac{\tau_0 + \tau_z}{2} |\Delta_L| (\gamma_x \cos\phi_L - \gamma_y \sin\phi_L) \nonumber\\
&+ \frac{\tau_0 - \tau_z}{2} |\Delta_R| (\gamma_x \cos\phi_R - \gamma_y \sin\phi_R) ,
\end{align}
where $\tau_j$ are the Pauli operators operating in the basis of the left and right dots (i.e., the ``charge'' basis), defined as $\tau_z = |L\rangle \langle L| - |R\rangle \langle R|$ and $\tau_x = |L\rangle \langle R| + |R \rangle \langle L|$, $\sigma_j$ are Pauli operators in spin space, and $\gamma_j$ are the Pauli operators in valley space.
The detuning between the left and right dots is given by $\varepsilon(t)$, and $t_c$ is the tunnel coupling between the dots. 
We find that a relatively large tunnel coupling is critical for charge noise resilience in strongly-driven flopping-mode qubits (see Appendix~\ref{app:weak_tc}), consistent with simulations in Ref.~[\citenum{Teske:2023p035302}].
For simplicity, we fix our tunnel coupling to be $t_c = 100$~\SI{}{\micro\electronvolt} throughout this work, a large but experimentally feasible value. 
We assume magnetic field gradients of $\Delta B_x = 2$~\SI{}{\milli\tesla} and $\Delta B_z = 0.4$~\SI{}{\milli\tesla}, and a constant background magnetic field of $B_z = 20$~\SI{}{\milli\tesla}, where we have also adopted these parameters from Ref.~[\citenum{Teske:2023p035302}].
We can equivalently write this Hamiltonian in a basis that simultaneously diagonalizes the left and right-dot valleys \cite{Zhao:2022p064043}:
\begin{align} \label{eq:ham_diag_valley}
H(\varepsilon) \; = \; & \frac{\varepsilon(t)}{2} \tau_z + t_c U_v \tau_x U_v^\dag \nonumber \\
& + \frac{g \mu_B B_z}{2}\sigma_z + \frac{g \mu_B}{2}(\Delta B_x \sigma_x + \Delta B_z \sigma_z) \tau_z  \nonumber \\
& + \frac{1}{2}(|\Delta_L| + |\Delta_R|) \gamma_z + \frac{1}{2}(|\Delta_L| - |\Delta_R|) \tau_z \gamma_z ,
\end{align} 
where
\begin{align} \label{eq:u_rot_valley}
U_v & = P_L U_L + P_R U_R,  \\
U_{L(R)} & = \frac{1}{\sqrt{2}} \left[ \gamma_0 + i \cos(\phi_{L(R)}) \gamma_y + i \sin(\phi_{L(R)}) \gamma_x \right], \nonumber
\end{align}
and $P_{L(R)} = |L(R) \rangle \langle L(R) |$ projects the system into the left (right) subspace.
Equation~(\ref{eq:ham_diag_valley}) forms the basis of our gate optimization procedure, described in Sec.~\ref{sec:gate_op}.

To gain intuition for how pulses to the detuning parameter, $\varepsilon(t)$, induce spin rotations, we make the simplistic assumption that the valley and charge splittings are sufficiently large that the evolution is adiabatic with respect to excited valley and charge states.
(We derive a precise condition on the time derivative $\dot \varepsilon$ to avoid orbital excitations in Appendix~\ref{app:analysis}; we consider valley excitations in Sec.~\ref{sec:gate_op}.)
We also assume the spin terms in the Hamiltonian are small enough to be treated as perturbations (i.e.~$B_z$, $\Delta B_z$, and $\Delta B_x \ll t_c$, $|\Delta_L|$, and $|\Delta_R|$).
In this regime, we can replace $\tau_z$ by its expectation value in the ground-state spin subspace, $\langle \tau_z \rangle$, given by
\begin{equation} \label{eq:tau_z_ep}
    \langle \tau_z \rangle_\varepsilon = \frac{1}{2} \left( {_\varepsilon\langle 0 | \tau_z | 0 \rangle_\varepsilon} + {_\varepsilon\langle 1 | \tau_z | 1 \rangle_\varepsilon} \right).
\end{equation}
Here, $ \langle \tau_z \rangle_\varepsilon$ describes the relative occupation of the double dot and  $|0\rangle_\varepsilon$ and $|1\rangle_\varepsilon$ are the ground and first-excited (spin) instantaneous eigenstates of the system for a given detuning $\varepsilon$.
We take the average of both the ground and first-excited spin states to avoid slightly biasing our computation in favor of either computational basis state.
This approximation results in a simple two-level spin Hamiltonian
\begin{equation} \label{eq:ham_2level}
    H_\text{eff}(\varepsilon) = \frac{g \mu_B B_z}{2} \sigma_z + \frac{g \mu_B }{2}\langle \tau_z \rangle_\varepsilon \left( \Delta B_x  \sigma_x + \Delta B_z \sigma_z \right),
\end{equation}
where we have discarded valley and charge terms.
If we also discard the relatively small $\Delta B_z$ term, we have the simple Hamiltonian of a driven two-level system, where the driving amplitude is given by $g \mu_B \Delta B_x \langle \tau_z \rangle / 2$.
By pulsing the detuning $\varepsilon(t)$, we modulate the ground state charge expectation value $\langle \tau_z \rangle_\varepsilon$ from -1 to 1.
And, by driving at the resonant frequency $E_z/\hbar$, where $E_z = g\mu_B B_z$ is the Zeeman splitting, we can perform a rotation about the $x$-axis in spin space by angle $\theta$:
\begin{equation} \label{eq:ideal_pulse}
    U_\theta = \cos(\theta/2) - i \sin(\theta/2) \sigma_x.
\end{equation}
Including virtual $z$ gates, these $U_\theta$ are enough to implement all single-qubit rotations.
In this work, we choose the representative value $\theta = \pi$.

\subsection{Alloy disorder and valley splitting} \label{eq:valley_theory}

One potential difficulty for spin qubits in general, and for strongly-driven flopping mode qubits in particular, is the different valley parameters of the two dots.
Recent theoretical work suggests that the valley splitting in most realistic devices is primarily due to the random-alloy disorder in the SiGe buffer layers~\cite{Wuetz:2022p7730, Losert:2023p125405}.
In this regime, termed the \textit{disordered} valley splitting regime, both the valley splitting and the valley phase of the two dots are variable.
It is possible to create heterostructures with deterministically large valley splittings, for example by engineering very sharp quantum well interfaces $\leq 3$ atomic monolayers (ML) in width~\cite{Losert:2023p125405}, or by a combining shear strain and Ge concentration modulations~\cite{Woods:2024p54}.
However, these deterministically enhanced devices are relatively difficult to fabricate.
In this paper, we restrict ourselves to devices in the disordered valley splitting regime.

In the disordered valley splitting regime, the statistics of both the magnitude and phase of the (complex) inter-valley coupling can be described rather simply.
Refs.~[\citenum{Wuetz:2022p7730}] and~[\citenum{Losert:2023p125405}] show, using effective mass theory, that the variance of the inter-valley coupling for a single quantum dot, averaged over configurations of random alloy disorder, is given by
\begin{multline} \label{eq:sigma_delta}
    \sigma_\Delta^2 \coloneqq \V [\Delta] \\
    = \frac{1}{\pi} \left[ \frac{a_0^2 \Delta E_c }{8 a_\text{dot} (X_w - X_s)} \right]^2 \sum_l |\psi_\text{env}(z_l)|^4 \bar X_l (1 - \bar X_l) ,
\end{multline}
where $a_0 = 0.543$~\SI{}{\nano\meter} is the lattice constant of the Si cubic unit cell, $\Delta E_c$ is the conduction band offset between the Si quantum well and the SiGe barriers ($\approx$ \SI{150}{\milli\electronvolt} for a typical quantum well), $X_w$ is the nominal Si concentration in the quantum well, $X_s$ is the Si concentration in the SiGe barrier (substrate) region, and $a_\text{dot} = \sqrt{\hbar / m_t \omega_\text{orb}}$ is the dot radius, assuming the dot is confined in an isotropic harmonic confinement potential with characteristic level spacing of $\hbar \omega_\text{orb}$, and $m_t = 0.19 m_e$ is the transverse effective mass of an electron in Si.
% here
The sum in Eq.~(\ref{eq:sigma_delta}) is taken over the atomic layers in the heterostructure, and the quantities $\bar X_l$ are the expected Si concentrations at layer $l$, when averaged over the whole device.
Finally, $\psi_\text{env}$ is a 1D envelope function for the quantum dot wavefunction, ignoring effects of valley-orbit coupling.
In the disordered valley-splitting regime, the average valley splitting is given by \cite{Losert:2023p125405}
\begin{equation} \label{eq:avg_Ev}
E_v = \sqrt{\pi} \sigma_\Delta.
\end{equation}
From Eqs.~(\ref{eq:sigma_delta}) and (\ref{eq:avg_Ev}), we see that the valley splitting depends strongly on the Ge concentration in the quantum well, $1 - \bar X_l$.
Devices with nonzero minimum Ge concentrations ($G_\text{min}$) have been proposed to boost average $E_v$. 
However, this also increases $\sigma_\Delta$, which leads to larger spatial variations in the valley splitting. 
We examine the consequences of this spatial variability in
Secs.~\ref{sec:valley_fluctuations} and \ref{sec:scalability}.

We note that $\Delta$ is a complex quantity, so $\V[\Re[\Delta]] = \V[\Im[\Delta]] = \sigma_\Delta^2 / 2$.
Thus, between two dots, the inter-dot valley phase difference $\delta \phi = \phi_L - \phi_R$ is randomized uniformly between $-\pi$ and $\pi$, assuming the dots are spatially well-separated and there is no deterministic inter-valley coupling.
We now show that this creates complications for strongly driven flopping-mode qubits.
To see this, we expand the tunneling term in Eq.~(\ref{eq:ham_diag_valley}) in the basis $\{ |L,+\rangle, |L,-\rangle, |R,+\rangle, |R,-\rangle \}$ where $\pm$ labels the ground/excited valleys:
\begin{align}
   t_c U_v \tau_x U_v^\dag & = 
   \begin{pmatrix}
    0 & 0 & t_{++} & t_{+-} \\
    0 & 0 & t_{-+} & t_{--} \\
    t_{++}^* & t_{-+}^* & 0 & 0 \\
    t_{+-}^* & t_{--}^* & 0 & 0
   \end{pmatrix} ,
\end{align}
where the matrix elements are given by
\begin{equation}  \label{eq:tunneling_matrix_elements}
\begin{gathered}
     t_{++} = t_{--}^* = \frac{t_c}{2} \left(1 + e^{i (\phi_{L} - \phi_{R})} \right) , \\
t_{+-} = -t_{-+}^* = \frac{t_c}{2} \left( e^{i \phi_L} - e^{i \phi_R} \right).
\end{gathered}
\end{equation}
Now, if $\phi_L = \phi_R$, then $t_{+-} = t_{-+} = 0$, and the tunneling between the left and right dot is valley-preserving.
In Fig.~\ref{fig:qubit_schematic}(b), we plot the qubit spectrum as a function of $\varepsilon$ for this case, observing no valley anticrossing.
However, if $\phi_L \neq \phi_R$, valley-flipping tunneling between the left and right dots is allowed. 
As a result, a valley anticrossing emerges in the qubit spectrum, as seen in Fig.~\ref{fig:qubit_schematic}(b), for $\delta \phi = 0.9 \pi$.
We highlight two of these tunneling elements by red arrows in Fig.~\ref{fig:qubit_schematic}(a).
Thus, if $\delta \phi \neq 0$ and the detuning is strongly driven, the dot can tunnel into the excited valley, resulting in leakage outside the logical subspace.
In this paper, we examine the consequences of the valley leakage pathway for pulse optimization (Sec.~\ref{sec:gate_op}), pulse fidelity (Secs.~\ref{sec:valley_dependence}-\ref{sec:valley_fluctuations}), and scalability (Sec.~\ref{sec:scalability}).

\section{Gate optimization} \label{sec:gate_op}

\begin{figure*}
    \centering
    \includegraphics[width=14cm]{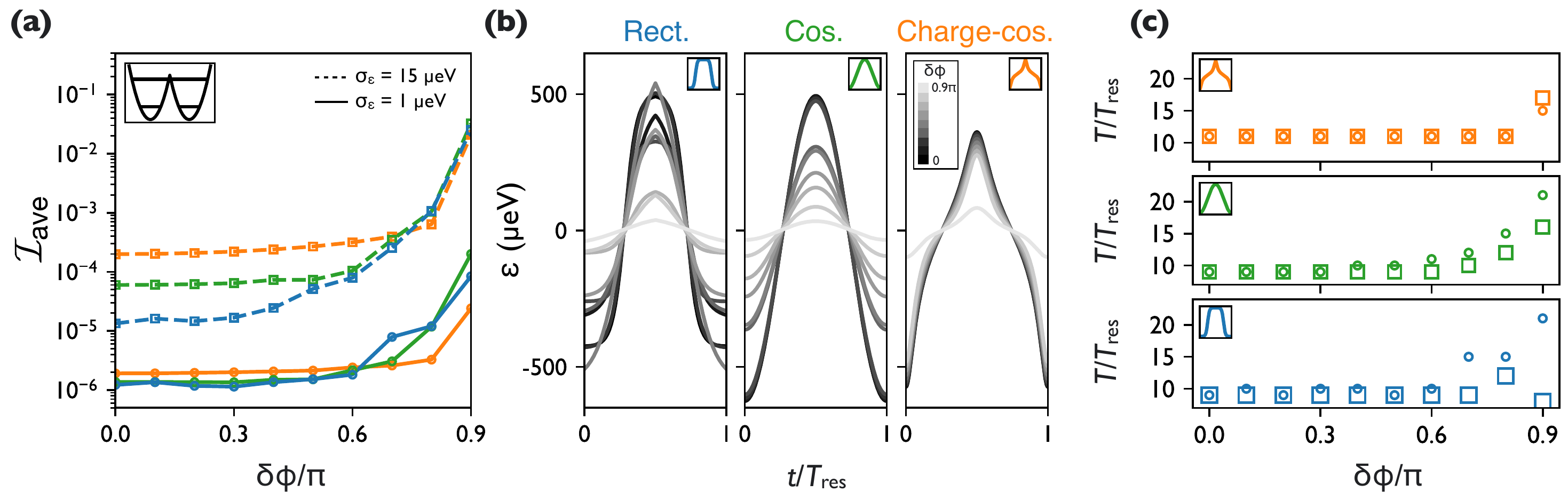}
    \caption{Performance comparison of the three pulse families described in Sec.~\ref{sec:gate_op}, for the valley splitting values $E_{vL} = E_{vR} = 100$~\SI{}{\micro\electronvolt}, as indicated in the inset of (a). 
    (a) Infidelities of the optimized pulses as a function of the valley phase difference $\delta \phi$, including the effects of charge noise, for the three pulse families: rectangular (blue), cosine (green), and charge-cosine (orange). We consider two charge noise regimes: an optimistic regime given by $\sigma_{\varepsilon} = 1$~\SI{}{\micro\electronvolt} (solid lines, circle markers), and a pessimistic regime given by $\sigma_\varepsilon = 15$~\SI{}{\micro\electronvolt} (dashed lines, square markers), where $\sigma_\varepsilon$ is the standard deviation of the noise-induced fluctuations to the detuning $\varepsilon$, which we relate to local electric field fluctuations in Eq.~(\ref{eq:sigma_Ex}). (b) One period of the optimized pulse shapes, for the three pulse families indicated in the insets. 
    Results here are shown for the case $\sigma_\varepsilon = 1$~\SI{}{\micro\electronvolt} and for the $\delta \phi$ values indicated by the grayscale. 
    (c) The total pulse time, in units of $T_\text{res}$, for the same optimized pulse results shown in (a), using the same color-coding scheme. The rectangular pulse is found to yield better results for most $\delta \phi$ values, especially for large noise amplitudes, as consistent with Ref.~[\citenum{Teske:2023p035302}].}
    \label{fig:phase_dependence}
\end{figure*}

In this section, we outline a straightforward procedure to optimize gates in strongly driven flopping mode qubits, given a fixed set of qubit parameters.
We analyze these pulses and their resulting fidelities, focusing on the role of the valley splitting in each dot and the valley phase difference between dots.
We also introduce a pulse shape designed to reduce valley excitations, which we show in Sec.~\ref{sec:valley_fluctuations} is more resilient to small noise-induced fluctuations in the valley parameters.

\subsection{Pulse shapes}

We consider three pulse types in this work, whose characteristic shapes are illustrated in the insets of Fig.~\ref{fig:phase_dependence}(b).
First, we consider cosine and smoothed-rectangular pulses similar to those considered in Ref.~[\citenum{Teske:2023p035302}]:
\begin{equation} \label{eq:pulse_cos_rect}
\begin{split}
    \varepsilon_\text{cos}(t) &= C + A \cos( 2\pi t/T_\text{res} ) \\
    \varepsilon_\text{rect}(t) &= C + A 
    \begin{cases}
        \tanh(2R ( \tilde t - (1 + c_\text{dc})/4) ), & \tilde t \leq 1/2 \\
        -\tanh(2R ( \tilde t - (3-c_\text{dc})/4) ), & \tilde t > 1/2
    \end{cases}
\end{split}
\end{equation}
where $T_\text{res}$ is the mean resonant period $2 \pi \hbar / E_z$ (corresponding to the resonant frequency $f_\text{res} = E_z / 2 \pi \hbar$), $E_z = g \mu_B B_z$ is the Zeeman splitting, and $\tilde t = t / T_\text{res} \text{ mod } 1$. 
These are both AC pulse shapes, which drive the electron back and forth across the double-dot, and they are both periodic in $T_\text{res}$.
(Since we restrict ourselves to resonant pulse frequencies, these pulses will be most optimal for field gradients $\Delta B_x$, $\Delta B_z \ll B_z$.)
In both pulses, $A$ represents the pulse amplitude, and $C$ is a constant offset.
In all simulations, we limit the maximum detuning $|\varepsilon| \leq \varepsilon_\text{max}$, where $\varepsilon_\text{max} = 1$~\SI{}{\milli\electronvolt}.
In the rectangular pulse, $R$ represents the steepness of the transition, limited to $R \leq 20$, and $c_\text{dc}$ is a duty-cycle coefficient, determining how far through the tanh function the pulse moves, which we restrict between 0 and 1.
The smooth rectangular pulse parameterization enables pulses that drive quickly through the anticrossing.
This can limit the qubit's sensitivity to charge noise in some regimes, as analyzed in Ref.~\cite{Teske:2023p035302}. 
However, strong driving can also cause leakage outside the qubit subspace, particularly in unfavorable valley configurations; we analyze this tradeoff in the following sections.

We also consider a third pulse form, which we term the ``charge-cosine'' or ``cc'' pulse, defined as a cosine function within the charge subspace:
\begin{align} \label{eq:pulse_spin_cos}
    \langle \tau_z \rangle_\text{cc}(t) &= C + A \cos (2 \pi t/T_\text{res}).
\end{align}
Again, $A$ is the pulse amplitude, in this space, and $C$ is the pulse offset.
As we see in Fig.~\ref{fig:phase_dependence}(b), the charge-cosine pulse drives more softly across the charge anticrossing, compared to the strong driving of the rectangular and cosine pulses.
This reduces valley excitations, thereby reducing the qubit's sensitivity to noise-induced valley fluctuations, as we demonstrate in Sec.~\ref{sec:valley_fluctuations}.
Since this pulse is defined in terms of the charge expectation value $\langle \tau_z \rangle$, we need a map from $\langle \tau_z \rangle \rightarrow \varepsilon$ to determine the detuning function $\varepsilon(t)$.
We obtain this map by diagonalizing $H(\varepsilon)$ across the full range of $\varepsilon$, computing $\langle \tau_z \rangle_\varepsilon$ according to Eq.~(\ref{eq:tau_z_ep}).
Then, by inverting the relationship in Eq.~(\ref{eq:tau_z_ep}), we obtain $\varepsilon(t)$.
Examples of these optimized pulses are shown in Fig.~\ref{fig:phase_dependence}(b), where we illustrate just one period of each pulse.

\subsection{Pulse optimization algorithm}

We now outline the algorithm used to optimize our pulses. 
Several optimization parameters were introduced in Eqs.~(\ref{eq:pulse_cos_rect}) and~(\ref{eq:pulse_spin_cos}).
Additionally, we need to optimize the total pulse length, $T = n T_\text{res}$, which we restrict to an integer multiple of the resonant period.
In general, there is an inverse relationship between pulse length and driving strength: longer pulses require less driving strength to implement the same qubit rotation.
Because making pulses longer and weaker tends to increase the charge noise sensitivity but decrease the leakage and infidelity errors, we expect to observe an interplay between these effects, with optimal fidelities occurring at a particular value of $n=n^*$, representing the optimal multiple of $T_\text{res}$.
To optimize over both $n$ and the pulse parameters, we adopt a three-stage approach.
In the first stage of our algorithm, we try to narrow down the potential range of $n^*$, employing a heuristic cost function to optimize the pulse shape for each possible $n$.
In the second stage, we take the five best-performing pulses from stage 1, and we fine-tune their parameters.
In the final stage, we perform a more thorough and computationally expensive estimate of the pulse infidelity for these five fine-tuned pulses, allowing us to select the best one.
We provide more details on each stage below, and we provide a graphical summary of this procedure in Fig.~\ref{fig:flowchart}.

In the first stage, we seek to narrow the range of possible values of $n^*$.
For each possible pulse of length $n T_\text{res}$, we identify the pulse parameters that minimize the following heuristic cost function:
\begin{equation} \label{eq:cost_func}
    \mathcal{C} = \mathcal{I}_0 + \mathcal{L}_\text{charge} + \mathcal{L}_\text{leak},
\end{equation}
where $\mathcal{I}_0$ is the trace infidelity of the pulse in the absence of any detuning fluctuations (i.e.,~in a zero-noise environment), $\mathcal{L}_\text{charge}$ is an estimate of the infidelity due to charge noise \textit{within} the computational subspace, and $\mathcal{L}_\text{leak}$ is an estimate of the infidelity due to noise-induced leakage \textit{outside} of the computational subspace.
The form of this cost function is chosen to balance computational efficiency and accuracy; minimizing $\mathcal{C}$ should produce pulses that achieve the desired rotation (by minimizing $\mathcal{I}_0$) while simultaneously minimizing noise-induced errors (by minimizing $\mathcal{L}_\text{charge}$ and  $\mathcal{L}_\text{leak}$).
More details on the components of $\mathcal{C}$ are given in Appendix~\ref{app:algorithm}.
We define the trace infidelity of a general (noisy) pulse $\mathcal{I}_{\delta \epsilon}$ as follows \cite{Gungordu:2022p023155}:
\begin{equation} \label{eq:tr_inf}
    \mathcal{I}_{\delta \varepsilon} = 1 - \left[\frac{ \vert \text{tr} \left( \mathcal{P}U_{\delta \varepsilon}(T) \mathcal{P}^\dagger U_\theta^\dagger \right)\vert }{2} \right]^2.
\end{equation}
Here, $U_{\delta\varepsilon}(T)$ is the propagator of the pulse evaluated at time $T$, in the presence of a quasistatic detuning fluctuation $\delta \varepsilon$ (see below) \footnote{The longest pulses reported in Figs.~2, 3, and 13 are all less than 50 multiples of the resonant period, $T_{res} = 2 \pi \hbar / E_z \approx 1.8$~\SI{}{\nano\second}. For the magnetic parameters considered in this work, this corresponds to a maximum pulse time $T_{max} \leq 90$~\SI{}{\nano\second}. The spectral density of $1/f$ noise on this short timescale is very small, compared to the spectral density at larger timescales. Compared to the fast gate time, noise acting on these larger timescales will be effectively quasistatic.}.
The propagator is determined by solving the Schrodinger equation
\begin{equation} \label{eq:schrodinger_eq_propagator}
    i \hbar \dot U_{\delta \varepsilon} = H(\varepsilon(t) + \delta \varepsilon) U_{\delta \varepsilon}
\end{equation}
where $\varepsilon(t)$ is the pulse shape we seek to optimize.
Note that $\mathcal{I}_0$ is the trace infidelity obtained when $\delta \varepsilon = 0$ in Eq.~(\ref{eq:schrodinger_eq_propagator}).
The projector $\mathcal{P} = |0(T)\rangle \langle 0(T) | + |1(T) \rangle \langle 1(T)|$ projects the system into the subspace spanned by the instantaneous ground and first excited states at time $T$, and $U_\theta$ is the ideal pulse defined in Eq.~(\ref{eq:ideal_pulse}).

At the second stage of optimization, we take the five pulse lengths that give the smallest $\mathcal{C}$ values from stage one, $\{n_{1..5} \}$, and fine-tune the optimization at these fixed values.
By keeping five possible solutions in this way, we increase the likelihood of finding the true $n^*$, while avoiding the computational overhead of further analyzing every possible $n$.
To perform the fine-tuning, we randomize and re-optimize the pulse parameters for each of the fixed $n_{1..5}$ values.
Since the quantum evolution of our flopping mode system is highly nontrivial, especially in regimes with significant valley excitations, this randomization and re-optimization increases the likelihood of finding the true global minimum of the cost function.
We find this procedure is especially important for the rectangular pulse family, which has more free parameters than the other pulse shapes.
We then repeat this procedure ten times for each $n_{1..5}$ value, finally selecting the pulse parameters that yield the lowest overall $\mathcal{C}$ value.

In the final stage of the optimization algorithm, we still keep all five of the now-fine-tuned solutions for pulse lengths $n_{1..5}$.
For each of these solutions, we now perform a more thorough estimate of the pulse infidelity.
We account for charge noise as a quasistatic fluctuation to the detuning, defined as $\tilde \varepsilon(t) = \varepsilon_\text{pulse}(t) + \delta \varepsilon$. 
Here, we assume $\delta \varepsilon$ is sampled from a normal distribution $p$ with zero mean and standard deviation $\sigma_\varepsilon$. 
We consider two charge noise regimes in this work, defined by an optimistic value of $\sigma_\varepsilon^\text{opt} = 1$~\SI{}{\micro\electronvolt}, or a more pessimistic value of $\sigma_\varepsilon^\text{pes} = 15$~\SI{}{\micro\electronvolt}.
For comparison, we note that $1/f$ noise with power spectral density $A_{1 \mathrm{Hz}} = 1$~\SI{}{\micro\electronvolt\per\sqrt{\hertz}}, similar to the charge noise spectral densities at 1~Hz reported in Refs.~\citenum{Connors:2019p165305, Kranz:2020p2003361, Connors:2022p940, Esposti:2024p32}, will have a standard deviation $\sigma_\varepsilon^\mathrm{exp} = \sqrt{\int_{f_\mathrm{min}}^{f_\mathrm{max}}df \; A_{1 \mathrm{Hz}}^2/f} = A_{1 \mathrm{Hz}} \sqrt{\log(f_\mathrm{max} / f_\mathrm{min})}$, where $f_\mathrm{min}$ and $f_\mathrm{max}$ set the minimum and maximum bounds on the noise frequency.
Assuming $f_\mathrm{min}$ set by a 10 minute recalibration time, and $f_\mathrm{max}$ set by the minimum pulse length $\sim 10$~\SI{}{\nano\second}, we have $\sigma_\varepsilon^\mathrm{exp} \approx 5$~\SI{}{\micro\electronvolt}--between our optimistic and pessimistic noise strengths.
To estimate the gate infidelity in the presence of charge noise, we sample 51 values of $\delta \varepsilon$ from distribution $p$, ranging from $-4\sigma_\varepsilon$ to $4\sigma_\varepsilon$, and then compute the weighted average of the resulting infidelities, as
\begin{equation} \label{eq:tot_inf}
    \mathcal{I}_{\text{avg}} = \sum_{-4\sigma_\varepsilon}^{4\sigma_\varepsilon}
    p(\delta \varepsilon)\mathcal{I}_{\delta \varepsilon}  ,
\end{equation}
where $\mathcal{I}_{\delta \varepsilon}$ is defined in Eq.~(\ref{eq:tr_inf}).
Finally, we select the pulse with the lowest $\mathcal{I}_\text{avg}$ value from the five pulses under consideration.

\section{Dependence of gate fidelity on the inter-valley coupling} \label{sec:valley_dependence}

\begin{figure}
    \centering
    \includegraphics[width=8cm]{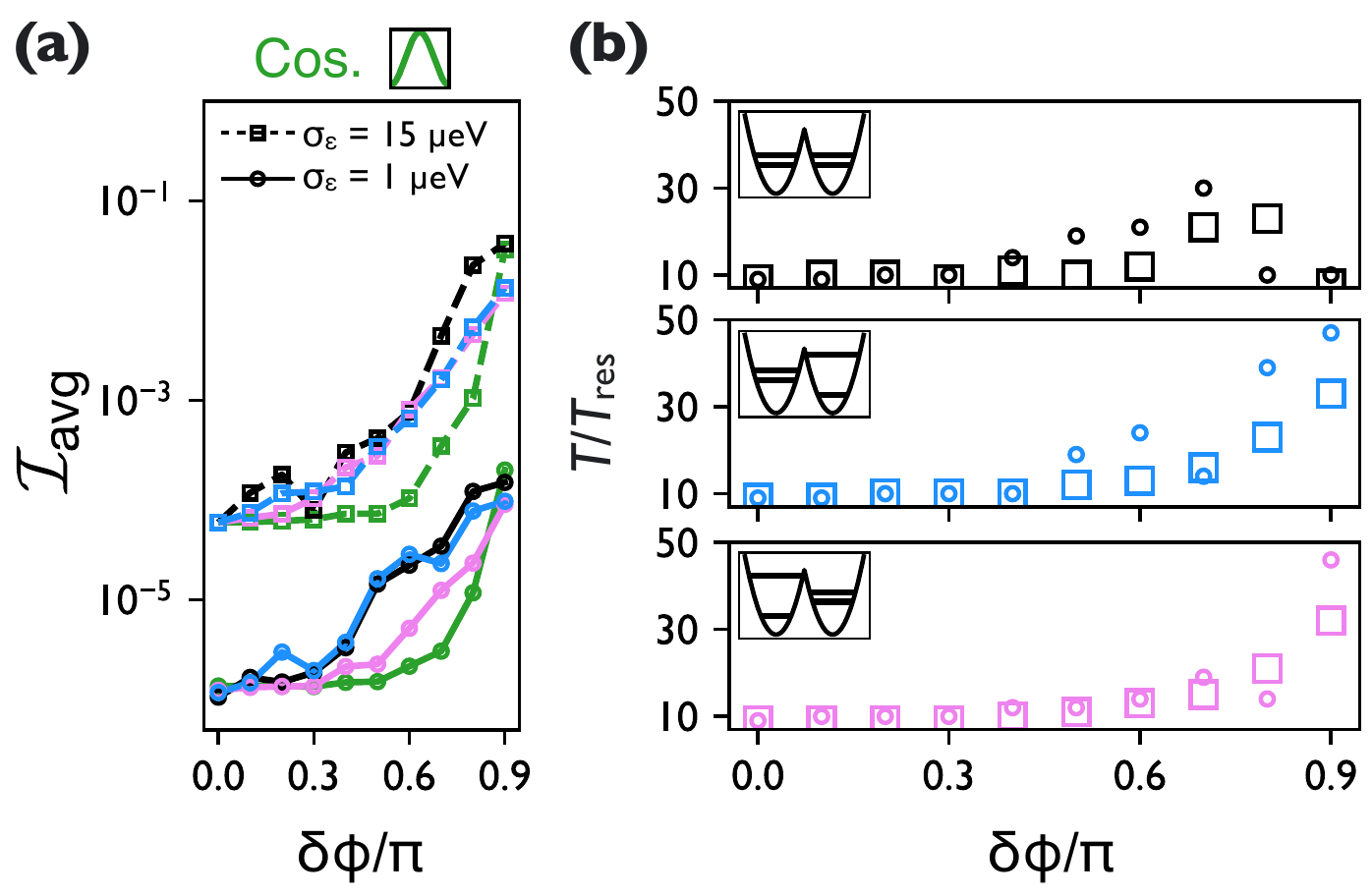}
    \caption{Evaluation of the cosine pulse family for a range of valley configurations. 
    (a) The average infidelity $\mathcal{I}_\text{avg}$, and (b) the corresponding pulse lengths, for the cases $\sigma_\varepsilon = 1$~\SI{}{\micro\electronvolt} (circles and solid lines) and $\sigma_\varepsilon = 15$~\SI{}{\micro\electronvolt} (squares and dashed lines). 
    Here, we take $E_{vL} = 100$ and $E_{vR} = 20$~\SI{}{\micro\electronvolt} (pink), $E_{vL} = 20$ and $E_{vR} = 100$~\SI{}{\micro\electronvolt} (blue), and $E_{vL} = E_{vR} = 20$~\SI{}{\micro\electronvolt} (black). 
    We also show the results for $E_{vL} = E_{vR} = 100$~\SI{}{\micro\electronvolt} (green), previously reported in Fig.~\ref{fig:phase_dependence}(a). 
    Note that the insets in (b) indicate the valley splitting configurations. 
    Although the case of high valley splittings (green data) generally gives better results, the other cases also achieve good fidelities, except when valley phase differences and charge noise fluctuations are large.  }
    \label{fig:valley_difference}
\end{figure}

Using the pulse optimization procedure outlined above, we now evaluate the performance of the three pulse families described in Eqs.~(\ref{eq:pulse_cos_rect}) and~(\ref{eq:pulse_spin_cos}) in the presence of variations in the inter-valley coupling between the two dots. 
We start with a favorable valley splitting configuration, $E_{vL} = E_{vR} = 100$~\SI{}{\micro\electronvolt} [indicated by the inset in Fig.~\ref{fig:phase_dependence}(a)], where we vary the valley phase difference $\delta \phi$ from $0$ to $0.9\pi$. 
The resulting average infidelities, for both optimistic and pessimistic charge noise configurations, are plotted in Fig.~\ref{fig:phase_dependence}(a).
The corresponding optimized pulse shapes are plotted in Fig.~\ref{fig:phase_dependence}(b) [for $\sigma_\varepsilon = 1$~\SI{}{\micro\electronvolt}] and the pulse lengths are plotted in Fig.~\ref{fig:phase_dependence}(c).
We note that some optimal rectangular pulses in Fig.~\ref{fig:phase_dependence}(b) contain cusps at their center.
This is an artifact of our pulse definition in Eq.~(\ref{eq:pulse_cos_rect}); in reality, all pulses should be smooth.
A different parameterization of the pulse shape could avoid this feature; however, it does not noticeably impact any of our results.

In the optimistic charge noise regime, pulse infidelities are fairly uniform around $~10^{-6}$, for each of the pulse families, across a wide range of $\delta \phi$.
In the pessimistic charge noise regime, the smoothed rectangular pulses yield the best results for the case of smaller phase differences, as consistent with~\citenum{Teske:2023p035302}, while the charge-cosine pulse yields the worst results.
This behavior is easy to understand, since the rectangular pulse allows for more rapid driving across the double-dot, so the system spends less time near $\varepsilon = 0$, where it is most sensitive to charge noise fluctuations.
On the other hand, the charge-cosine pulse is driven more gently through $\varepsilon = 0$, so it is more sensitive to charge noise.
As the valley phase difference increases above $\sim$$0.6\pi$, the fidelities of all pulse families and both charge noise regimes start to deteriorate.
This is due to the emergence of a valley anticrossing, visible in the energy spectra shown in Fig.~\ref{fig:qubit_schematic}(b), which causes strongly-driven pulses to leak into the excited valley state.
As a result, the optimal pulses become weaker as shown in Fig.~\ref{fig:phase_dependence}(b), and longer as shown in Fig.~\ref{fig:phase_dependence}(c).
These pulses spend more time around $\varepsilon = 0$, and are hence more sensitive to charge noise.
In contrast, for the lowest-fidelity configurations (e.g., the rectangular pulse shape with $\delta \phi = 0.9\pi$ and pessimistic charge noise), much shorter pulses are sometimes preferred, as seen in Fig.~\ref{fig:phase_dependence}(c).
These pulses have very poor fidelities, however.
(See Appendix~\ref{app:pulse_behavior} for further discussion.)

Next, we examine pulses in less favorable valley configurations.
In Fig.~\ref{fig:valley_difference}(a), we plot optimized infidelities for the cosine pulse as a function of $\delta \phi$ for three valley splitting regimes: $E_{vL} = 100$~\SI{}{\micro\electronvolt} and $E_{vR} = 20$~\SI{}{\micro\electronvolt} (pink), $E_{vL} = 20$~\SI{}{\micro\electronvolt} and $E_{vR} = 100$~\SI{}{\micro\electronvolt} (blue), and $E_{vL} = E_{vR} = 20$~\SI{}{\micro\electronvolt} (black).
For comparison, we also include infidelities for $E_{vL} = E_{vR} = 100$~\SI{}{\micro\electronvolt} that were reported in Fig.~\ref{fig:phase_dependence} (green).
As expected, as we reduce the valley splitting in one or both dots, gate infidelity increases for most $\delta \phi$.
This effect is small for $\delta \phi \lesssim 0.3\pi$, but worsens for larger phase differences.
Additionally, as demonstrated in Fig.~\ref{fig:valley_difference}(b), longer pulses are generally favored for larger $\delta \phi$, as the emergence of the valley anticrossing forces weaker detuning pulses to avoid valley excitations.
(Again, for the lowest $E_v$ and largest $\delta \phi$ in Fig.~\ref{fig:valley_difference}(b), we do see a preference for very short pulses; see Appendix~\ref{app:pulse_behavior} for a discussion.)
Nonetheless, as shown in Fig.~\ref{fig:phase_dependence} for the optimistic charge noise regime (circle markers), gate infidelities $< 10^{-4}$ can be achieved for $\delta \phi$ up to $0.8 \pi$.
This picture is qualitatively similar for the other two pulse families as well (see Appendix~\ref{app:pulse_behavior}).
Thus, as long as charge noise is limited, we expect that strongly driven flopping mode qubits can be achieved with good fidelity across a wide range of valley parameters.

\begin{figure*}
    \centering
    \includegraphics[width=12cm]{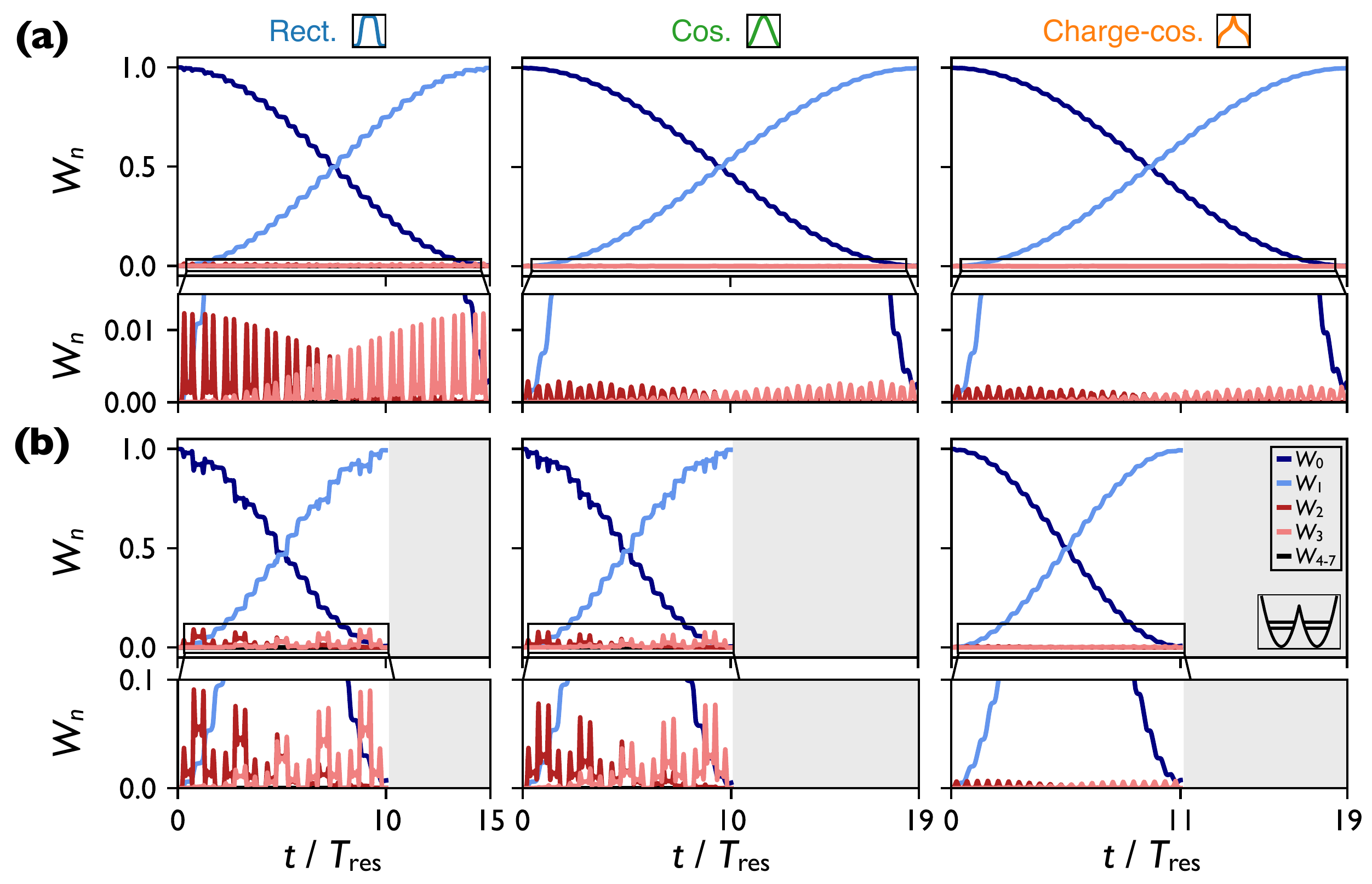}
    \caption{Optimized wavefunction solutions, including the excited states, for each of the three pulse families (indicated at the top), assuming the parameters $E_{vL} = E_{vR} = 20$~\SI{}{\micro\electronvolt} and $\delta \phi = 0.5\pi$. 
    Here, we plot the occupations (i.e., the weights, $W$, defined in the main text) of the instantaneous eigenstates, including the ground (dark blue) and first excited (light blue) spin states, which constitute the logical states, and the third (dark red) and fourth (light red) states, which are excited valley states and constitute leakage. 
    Results are shown for the noise levels (a) $\sigma_\varepsilon = 1$~\SI{}{\micro\electronvolt} and (b) $\sigma_\varepsilon = 15$~\SI{}{\micro\electronvolt}. 
    The lower panels in (a) and (b) show zoomed-in views of the excited valley states. To highlight the relative length of the various pulses, we plot them all on the same $x$-axis scale. For the parameters considered here, the charge-cosine pulse shape consistently minimizes excited valley leakage, since it drives more slowly through the valley anticrossing.}
    \label{fig:eigval_evol}
\end{figure*}

There are some notable differences between fidelities obtained for the different pulse families.
To illustrate this, we consider a representative case defined by the parameters $E_{vL} = E_{vR} = 20$~\SI{}{\micro\electronvolt}, $\delta \phi = 0.5\pi$, and $\sigma_\varepsilon = 1$~\SI{}{\micro\electronvolt}.
We begin the simulation in the ground state for each of the pulse shapes, and then compute the time evolution of the wavefunction $|\psi(t) \rangle$.
We can express $|\psi\rangle$ in the basis of instantaneous eigenstates as $|\psi(t)\rangle = \sum_n c_n(t) |n(t)\rangle$, where $c_n(t)$ is the component of $|\psi\rangle$ in eigenstate $|n\rangle$ at time $t$.
In Fig.~\ref{fig:eigval_evol}(a), we plot the state occupations $W_n = |c_n(t)|^2$, where we color the two lowest (spin) eigenstates shades of blue, and the next two excited (valley) eigenstates shades of red.
In the lower portion of each panel, we plot a zoomed-in picture of the evolution of the lowest two valley eigenstates.
While none of the pulse schemes achieves the goal of zero leakage, the leakage is visibly much larger for the rectangular pulse case.
This behavior can be explained by faster passage through the anticrossing for those pulses, yielding a larger Landau-Zener tunneling between the ground and excited valley states.
In addition to leakage, the larger excited valley occupation causes the evolution of the rectangular pulses to become more sensitive to charge-noise-induced fluctuations of other parameters, such as the inter-valley coupling.
We analyze these effects in greater detail in Sec.~\ref{sec:valley_fluctuations}.

As we increase the strength of the charge noise, the effects of Landau-Zener excitations become even more apparent, as shown in Fig.~\ref{fig:eigval_evol}(b).
Here, the simulations are performed with the same parameters as Fig.~\ref{fig:eigval_evol}(a), but with a larger charge noise parameter $\sigma_\varepsilon = 15$~\SI{}{\micro\electronvolt}.
We first note that the optimal pulses are all shorter than the those obtained in Fig.~\ref{fig:eigval_evol}(a).
This can be understood because charge noise effects are compounded in longer pulses, so the optimization procedure favors shorter pulses in this case.
Additionally, we note that the excited valley occupation is much larger for the rectangular and cosine pulses than for the charge-cosine pulse, because these two pulses have much stronger driving through the valley anticrossing.
As a result, these two pulse shapes are much more sensitive to small fluctuations of the valley parameters due to charge noise.
In fact, for the rectangular and cosine pulses, we expect small fluctuations in the valley parameters to be the dominant source of infidelity in the parameter regime studied in Fig.~\ref{fig:eigval_evol}, as we discuss in Sec.~\ref{sec:valley_fluctuations}.
On the other hand, by avoiding valley excitations, the charge-cosine pulse is relatively resilient to these valley fluctuations.

\section{Fine-tuning is required to achieve high gate fidelities} \label{sec:fine_tuning}

\begin{figure}
    \centering
    \includegraphics[width=8cm]{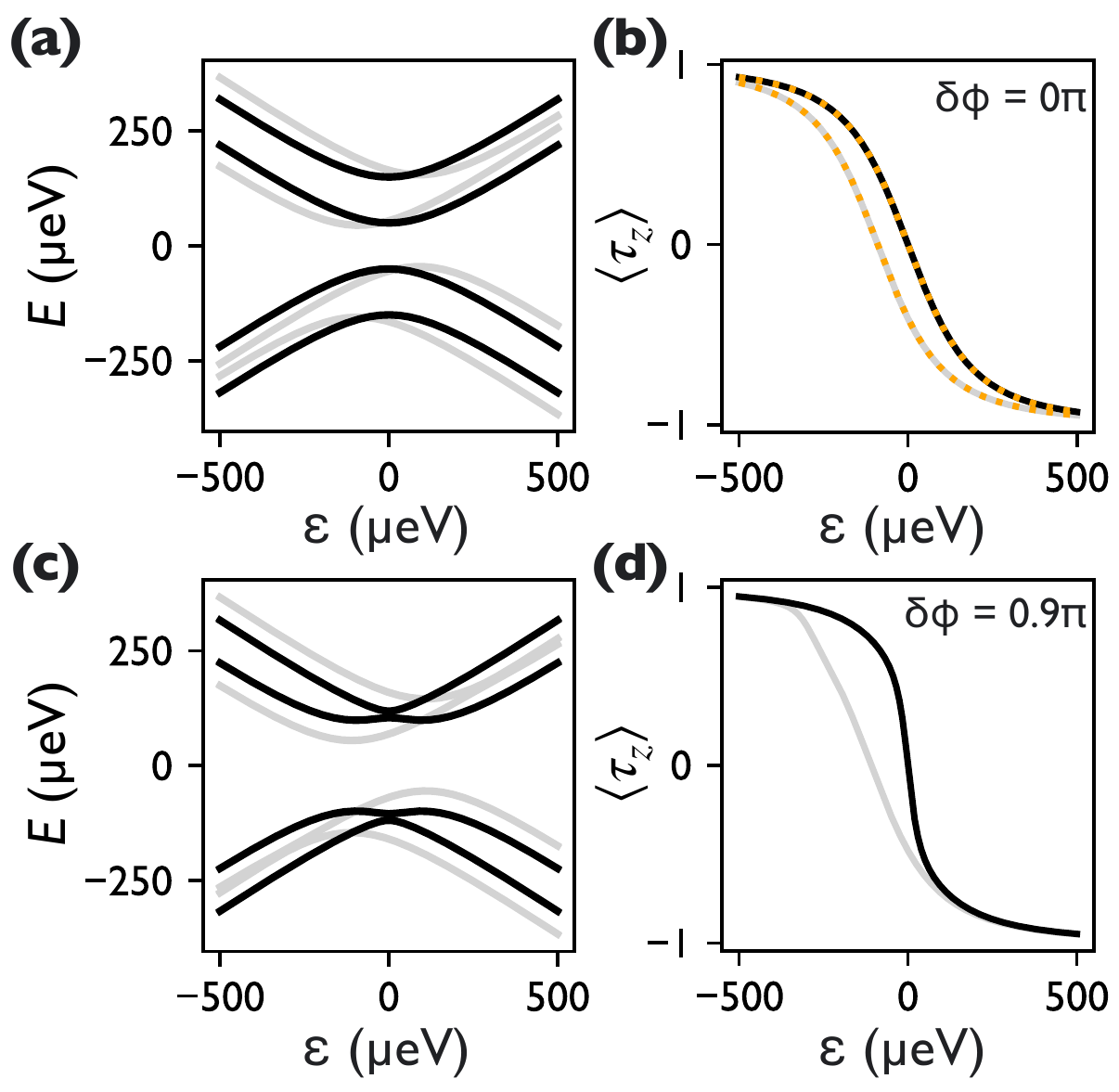}
    \caption{Locally varying valley splittings and valley phases require individualized pulse optimizations. 
    (a) Energy dispersion of a double dot as a function of the detuning $\varepsilon$, for valley splittings $E_{vL} = 20$ and $E_{vR} = 200$~\SI{}{\micro\electronvolt} (light gray), or $E_{vL} = E_{vR} = 100$~\SI{}{\micro\electronvolt} (black), where the valley phase difference is $\delta \phi = 0$ in both cases. 
    (Note that the spin splitting is too small to be resolved here.) 
    (b) The relative dot occupation, $\langle \tau_z \rangle$, computed numerically with Eq.~(\ref{eq:tau_z_ep}), for the same two cases and color codings shown in (a).
    (Note that the gray and black lines lie beneath the orange dotted lines.) 
    Here, the orange dotted lines represent the analytical results for the $\delta \phi = 0$ case, derived in Appendix~\ref{app:analysis} and computed using Eq.~(\ref{eq:tau_z_ep_theory_supp}). 
    (c), (d) The same quantities as (a) and (b), where we now set $\delta \phi = 0.9\pi$. 
    The differences in the dot occupations in (b) and (d) lead to different spin evolutions via Eq.~(\ref{eq:ham_2level}).}
    \label{fig:valley_dependent_tuning}
\end{figure}

In our optimization procedure, we fine-tune each pulse for a given set of valley parameters.
%The pulses described above were all fine-tuned to a particular valley configuration.
While it would be convenient to use the same pulse shape for all valley configurations, we now show that individual fine-tuning is necessary to achieve very high-fidelity qubit operations.
The main reason is that, in the disordered valley splitting regime, the valley parameters can vary significantly from dot to dot.
As shown here, gates optimized for one set of valley parameters typically perform poorly when the parameters change.

We previously showed that the gate fidelity, and the excited valley leakage in particular, depends on the pulse shape at the valley anticrossing.
Here, we show that the optimal pulse shape, and the nature of the anticrossing itself, depends strongly on the valley parameters of the double dot.
In Fig.~\ref{fig:valley_dependent_tuning}, we plot the energy dispersion as a function of detuning $\varepsilon$, for several different valley configurations.
Figure~\ref{fig:valley_dependent_tuning}(a) shows the energy dispersion for two such configurations, each with $\delta \phi = 0$; here, the light gray lines correspond to $E_{vL} = 20$~\SI{}{\micro\electronvolt} and $E_{vR} = 200$~\SI{}{\micro\electronvolt}, while the black lines correspond to $E_{vL} = E_{vR} = 100$~\SI{}{\micro\electronvolt}.
The energy dispersion changes significantly between these two configurations because the valley splitting is of the same order of magnitude as the tunnel coupling.
These distinct energy dispersions also result in distinct spin behaviors, as reflected in the effective spin Hamiltonian of Eq.~(\ref{eq:ham_2level}).
For example, in Fig.~\ref{fig:valley_dependent_tuning}(b), we plot the charge expectation value $\langle \tau_z \rangle_\varepsilon$, defined in Eq.~(\ref{eq:tau_z_ep}), for the same valley configurations as Fig.~\ref{fig:valley_dependent_tuning}(a).
For this scenario, corresponding to $\delta \phi = 0$, we can also derive an analytical formula for $\langle \tau_z \rangle_\varepsilon$ [see Eq.~(\ref{eq:tau_z_ep_theory_supp}) in  Appendix~\ref{app:analysis}], which we plot as orange dotted lines in Fig.~\ref{fig:valley_dependent_tuning}(b). 
We observe even greater differences in the energy dispersion and in $\langle \tau_z \rangle_\varepsilon$ when the valley phase difference $\delta \phi\neq 0$, as seen in Figs.~\ref{fig:valley_dependent_tuning}(c) and \ref{fig:valley_dependent_tuning}(d), for the case $\delta \phi = 0.9 \pi$.

\begin{figure}
    \centering
    \includegraphics[width=8cm]{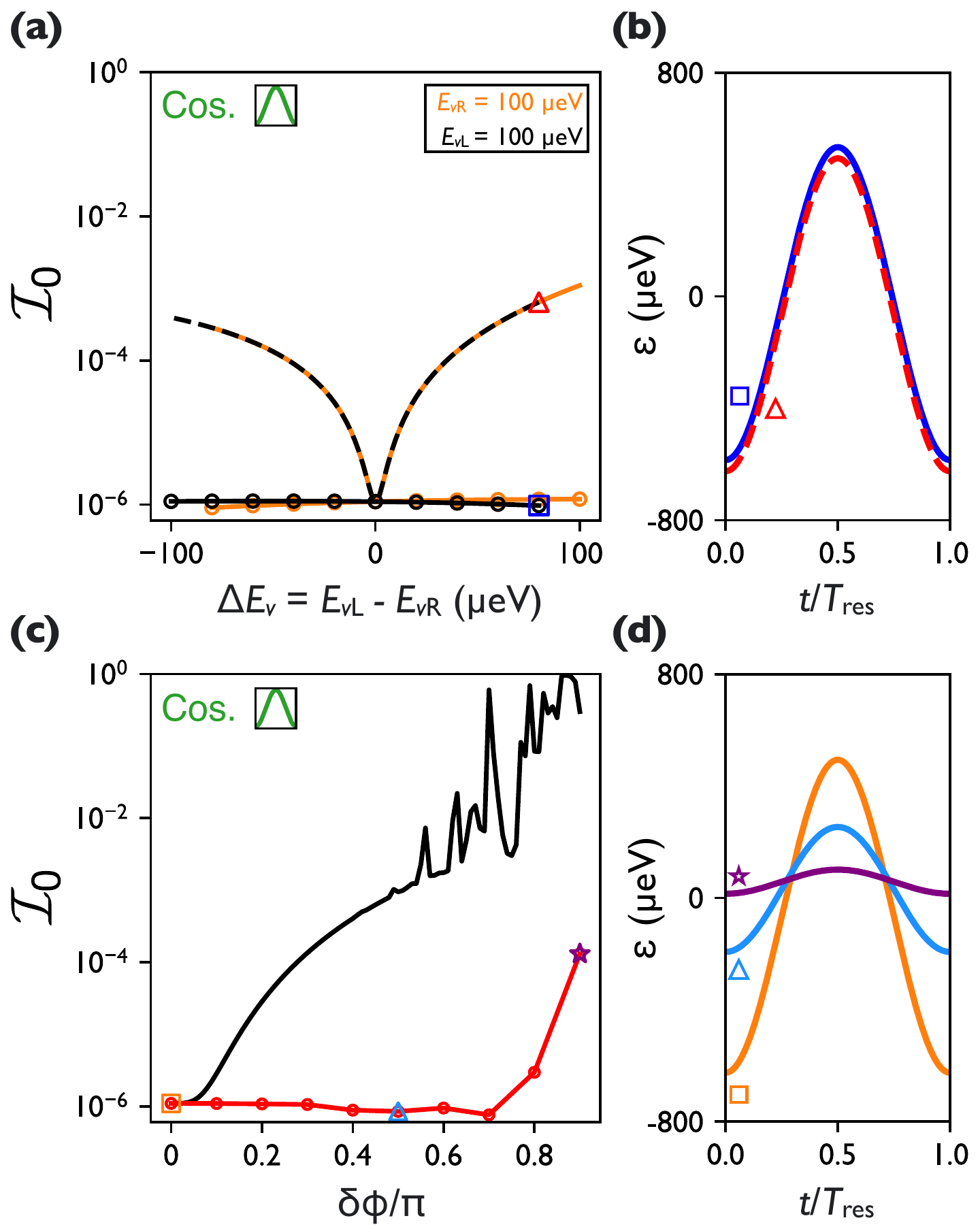}
    \caption{Failing to account for local valley parameters leads to poor gate fidelities. 
    (a) Gate infidelities $\mathcal{I}_0$ for the cosine pulse shape, in the absence of charge noise. 
    The open circles show infidelities for pulses optimized at $\delta\phi=0$, for two cases: (i) $E_{vL}$ is varied while holding $E_{vR} = 100$~\SI{}{\micro\electronvolt} fixed (orange circles), or (ii) $E_{vR}$ is varied while holding $E_{vL} = 100$~\SI{}{\micro\electronvolt} fixed (black circles).
    The dashed orange and black lines show infidelities obtained when the pulse shape is optimized just once, for the case $E_{vL} = E_{vR} = 100$~\SI{}{\micro\electronvolt}.
    This single pulse is then applied to a range of $\Delta E_v$ for which it was not intended, at the same valley parameters as the orange and black circles.
    (b) Cosine pulse shapes, optimized at $\delta\phi=0$, for two cases: (i) $E_{vL} = E_{vR} = 100$~\SI{}{\micro\electronvolt} (red triangle), or (ii) $E_{vL} = 100$~\SI{}{\micro\electronvolt} and $E_{vR} = 20$~\SI{}{\micro\electronvolt} (blue square).
    Although the pulse shapes are very similar, the corresponding infidelities can differ by orders of magnitude [e.g., the red triangle and blue square in (a)].
    (c) Cosine pulse shapes are optimized for $E_{vL} = E_{vR} = 100$~\SI{}{\micro\electronvolt} and a range of $\delta \phi$ values, in the absence of charge noise, yielding the infidelities plotted as red circles.
    The black line shows the infidelities obtained for a single pulse shape, optimized for the fixed value $\delta\phi=0$.
    (d) Three optimized pulse shapes are shown for the settings indicated by the square, triangle, and star markers in (c).
    Unlike (b), these pulses show clear differences in their amplitude and vertical offset. }
    \label{fig:valley_free_pulse}
\end{figure}

We now show how failing to account for the local variations of the valley splitting and the valley phase can result in poor gate fidelities.
We consider the cosine pulse shape, optimized for the parameters $E_{vL} = E_{vR} = 100$~\SI{}{\micro\electronvolt} and $\delta \phi = 0$. 
In this case, a fully optimized pulse can achieve an infidelity of $\mathcal{I}_0 \approx 10^{-6}$ in the absence of charge noise, as shown in Fig.~\ref{fig:valley_free_pulse}(a).
From this baseline, we then vary $E_{vL}$ (solid orange line) or $E_{vR}$ (dashed black line) without re-optimizing, plotting the resulting pulse infidelity as a function of $\Delta E_v = E_{vL} - E_{vR}$ in Fig.~\ref{fig:valley_free_pulse}(a). 
Away from the optimized point ($\Delta E_v = 0$), in either direction, we find that $\mathcal{I}_0$ quickly grows.
For comparison, we also plot results for the same $\Delta E_v$ parameters (orange and black circles) using optimized pulse shapes, for which the infidelities remain near $10^{-6}$.
Thus, failing to account for valley splitting variations can cause orders of magnitude increases in the infidelity.

In Fig.~\ref{fig:valley_free_pulse}(b), we also highlight how the differences between correctly and incorrectly optimized pulse shapes can be quite subtle. 
Here, we plot one period of pulses optimized for the cases $E_{vL} = E_{vR} = 100$~\SI{}{\micro\electronvolt} (red dashed) and $E_{vL} = 100$~\SI{}{\micro\electronvolt} and $E_{vR} = 20$~\SI{}{\micro\electronvolt} (blue solid), both for $\delta \phi = 0$.
The detuning values of these two pulses differ by only $\sim$\SI{40}{\micro\electronvolt}; however, the infidelities they induce can differ by about three orders of magnitude, as indicated by the red triangle and blue square in Fig.~\ref{fig:valley_free_pulse}(a), for the case $E_{vL} = 100$~\SI{}{\micro\electronvolt} and $E_{vR} = 20$~\SI{}{\micro\electronvolt}.

The outcomes from optimal and sub-optimal pulses also depends on variations of $\delta \phi$. 
In Fig.~\ref{fig:valley_free_pulse}(c), we plot the infidelity $\mathcal{I}_0$ of pulses optimized for $E_{vL} = E_{vR} = 100$~\SI{}{\micro\electronvolt}, for the range of phase differences $\delta \phi\in (0,0.9 \pi)$ (red circles), in the absence of charge noise.
Three typical, optimized pulse shapes are shown in Fig.~\ref{fig:valley_free_pulse}(d), corresponding to settings indicated by the orange square, the blue triangle, and the purple star in Fig.~\ref{fig:valley_free_pulse}(c).
We then consider the performance of a single, fixed pulse, optimized for $\delta \phi = 0$, for different values of $\delta \phi$ (black).
We see that the infidelity of the incorrectly optimized pulse rises quickly by about three orders of magnitude.

In summary, we have shown that high-fidelity gates in silicon-based flopping-mode qubits require careful, valley-dependent fine tuning.

\section{Other sources of infidelity} \label{sec:valley_fluctuations}

\begin{figure}
    \centering
    \includegraphics[width=8cm]{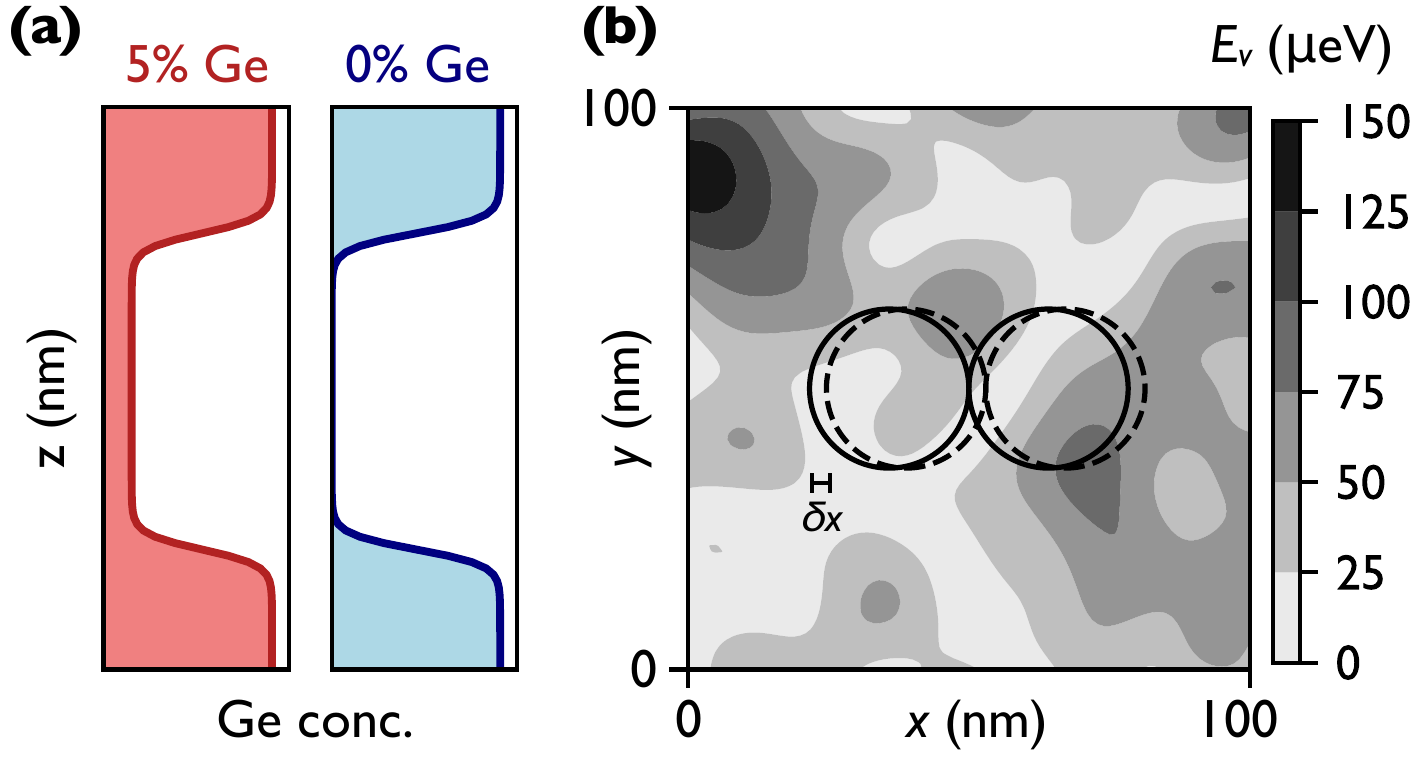}
    \caption{Schematic illustration of how charge noise causes valley splitting fluctuations. (a) Illustrations of the two heterostructure types we consider in this work: a quantum well (QW) with 5\% Ge at its center (red), and a conventional \SI{10}{\nano\meter} QW with no Ge at its center (blue). (b) Simulation of $E_v$ across a $100 \times 100$~\SI{}{\nano\meter\squared} region of heterostructure for a conventional \SI{10}{\nano\meter} QW. The two circles of radius $a_\text{dot}$ represent a double quantum dot, which is displaced by an amount $\delta x$ in the presence of a lateral electric field fluctuation, thus experiencing a fluctuation of $E_v$ on each dot. }
    \label{fig:Ev_shift_schematic}
\end{figure}

In the optimization procedure described in Sec.~\ref{sec:gate_op}, the pulses were designed to account for detuning fluctuations in a double quantum dot, which typically arise from charge impurities in the heterostructure~[\citenum{Ercan:2022p247701}].
For example, a fluctuating charge impurity gives rise to lateral electric field fluctuations (i.e., charge noise), $\delta \mathbf{E}_\text{lat} = \delta E_{\text{lat},x} \hat x + \delta E_{\text{lat},y} \hat y$.
If the double-dot axis is aligned with $\hat x$, then the fluctuator produces a detuning fluctuation of magnitude
\begin{equation} \label{eq:delta_detun}
    \delta \varepsilon = e d \delta E_{\text{lat},x},
\end{equation}
where $d$ is the inter-dot separation.
Following previous work~[\citenum{Culcer:2009p073102, Bermeister:2014p192102, Krzywda:2021p075439, Benito:2019p125430, Teske:2023p035302}], we use Eq.~(\ref{eq:delta_detun}) to make our fidelity estimates for spin qubits in Si/SiGe.
We also adopt the fixed values $d = 60$~\SI{}{\nano\meter} and $t_c = 100$~\SI{}{\micro\electronvolt}, although we note these parameters are variable in real devices.

The ``direct'' detuning fluctuations of Eq.~(\ref{eq:delta_detun}) are not the only possible sources of qubit infidelity.
For example, the quasistatic lateral electric field fluctuations $\delta E_{\text{lat},x}$ also cause the center of the dot confinement potential to shift by 
\begin{equation} \label{eq:delta_x_E_field}
    \delta x =  \frac{e \delta E_{\text{lat},x}}{m_t \omega_\text{orb}^2},
\end{equation}
resulting in a small shift of the dot center, which causes the dot to sample a slightly different disorder landscape caused by the inevitable random-alloy disorder and interface steps in Si/SiGe heterostructures.
(For shifts along $x$, we obtain standard deviations $\sigma_{\delta x} \approx 0.0017$~\SI{}{\nano\meter} for $\sigma_\varepsilon = 1$~\SI{}{\micro\electronvolt}, and $\sigma_{\delta x} \approx 0.025$~\SI{}{\nano\meter} for $\sigma_\varepsilon = 15$~\SI{}{\micro\electronvolt}.)
In turn, such disorder causes small shifts of the qubit parameters in Eq.~(\ref{eq:ham}), including the orbital levels, the tunnel coupling, and the inter-valley coupling.
This effect is illustrated schematically in Fig.~\ref{fig:Ev_shift_schematic}(b), where we have simulated a map of $E_v$ across a 100~$\times$~\SI{100}{\nano\meter^2} region of heterostructure using the methods of Ref.~[\citenum{Losert:2024p040322}]. 
The circles represent a double quantum dot which has been shifted by an amount $\delta x$ in response to the field fluctuation.
(The shift has been greatly exaggerated here for visual effect.)
By shifting the dot position in this fluctuating $E_v$ landscape, charge noise then induces a small modulation of the valley coupling parameters $\Delta_L$ and $\Delta_R$.
The same mechanism causes fluctuations of the local ground-state energy in each dot, as well as the tunnel couplings between the dots.
We analyze the impact of these fluctuations below.

To simulate the disorder landscape, we must first specify the details of heterostructure.
As described in Sec.~\ref{eq:valley_theory}, the average valley splitting depends on the disorder parameter $\sigma_\Delta$, which in turn depends on the wavefunction overlap with high-Ge layers in the heterostructure.
In heterostructures with greater Ge overlap, we expect larger fluctuations of the qubit parameters, as discussed above.
To make concrete estimates, we analyze two representative heterostructures, illustrated in Fig.~\ref{fig:Ev_shift_schematic}(a): a conventional \SI{10}{\nano\meter} quantum well (QW) with no additional Ge (labeled 0\% Ge) and a heterostructure with uniform 5\% Ge in the QW.
Such high-Ge heterostructures have been proposed to boost the average valley splitting in quantum dots [\citenum{Wuetz:2022p7730, Losert:2023p125405}], as would benefit flopping mode qubits.
For example, using Eqs.~(\ref{eq:sigma_delta}) and (\ref{eq:avg_Ev}), we estimate $\bar E_v \approx 360$~\SI{}{\micro\electronvolt} in the 5\% Ge QW, while $\bar E_v \approx 49$~\SI{}{\micro\electronvolt} for the conventional 0\% Ge QW.
This large $\bar E_v$ comes at a cost, however, with increased sensitivity to charge noise in some regimes, as discussed below.
More details on the design of these heterostructures and the estimation of their average valley splittings are presented in Appendix~\ref{app:heterostructure_design}.

\subsection{Fluctuations of the local ground-state orbital energy and tunnel coupling} \label{sec:Egsandtcflucts}

We first consider the effect of small, indirect shifts in the dots' local ground-state orbital energies $E_\text{gs}$, due to alloy disorder and interface steps, in combination with charge noise.
If the fluctuating ground-state energy $\delta E_\text{gs}$ differs between the dots, it causes effective detuning fluctuations $\delta \varepsilon_\text{eff} = \delta E_\text{gs,L} - \delta E_\text{gs,R}$.
We study these fluctuations in Appendix~\ref{app:orbital}, finding their effect to be somewhat smaller than the more-dominant direct detuning fluctuations $\delta \varepsilon$, defined in Eq.~(\ref{eq:delta_detun}).
For simplicity, we therefore do not include $\delta \varepsilon_\text{eff}$ fluctuations in our analysis.

We also consider fluctuations $\delta t_c$ of the tunnel coupling $t_c$, due to alloy disorder and interface steps, in combination with charge noise, as discussed in Appendix~\ref{app:tunnel_coupling}. 
For the realistic charge noise regimes considered here, these fluctuations are also found to be weak, of order $\lesssim 1$~\SI{}{\micro\electronvolt}.
For simplicity, we also exclude these effects from our analysis.

\subsection{Valley coupling fluctuations}
\label{sec:valleycoupling}

\begin{figure}
    \centering
    \includegraphics[width=8cm]{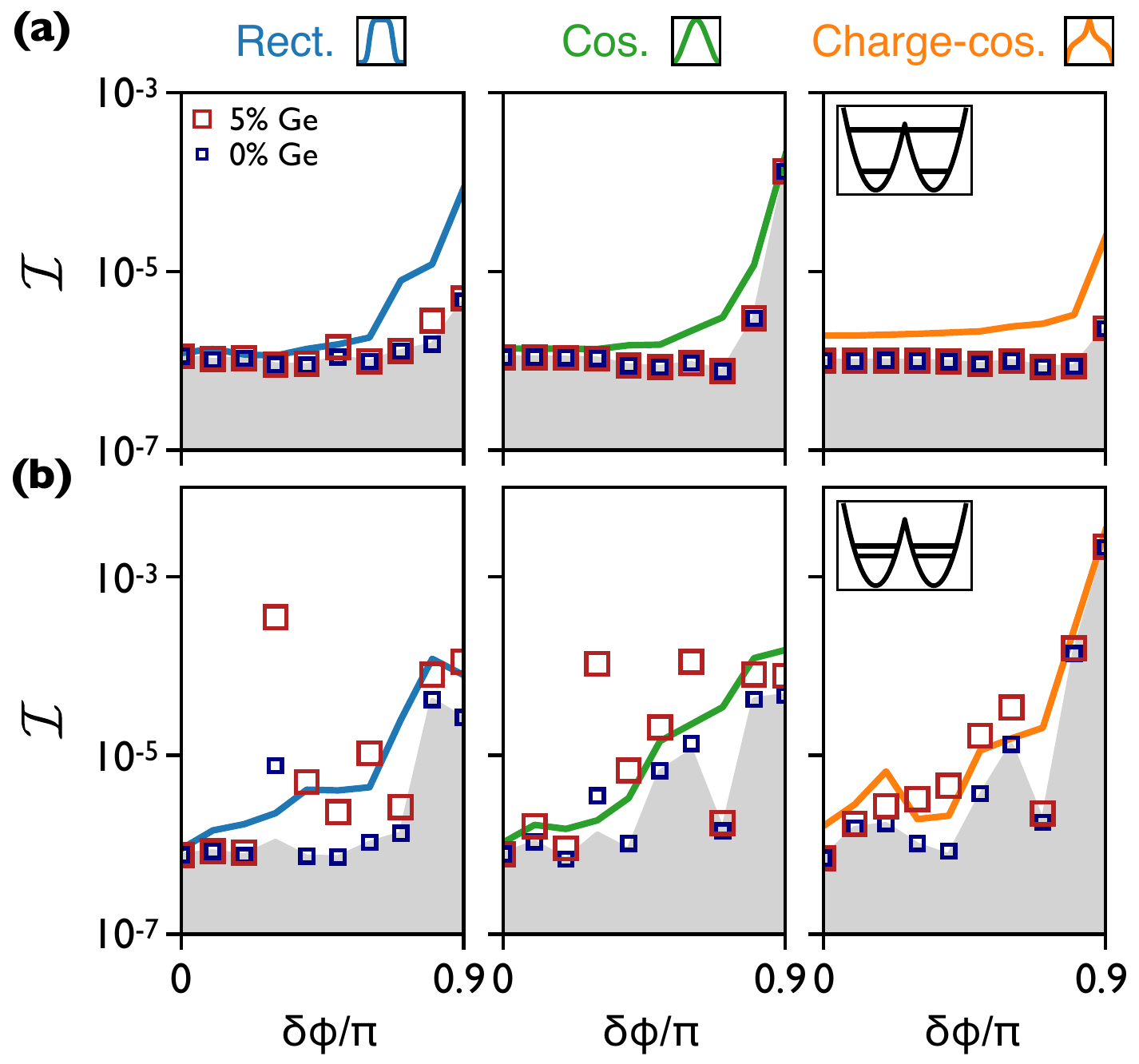}
    \caption{Infidelities due to $\delta\varepsilon$ and $\delta\Delta$ fluctuations in the \textit{optimistic} charge-noise regime, $\sigma_\varepsilon = 1$~\SI{}{\micro\electronvolt}. 
    For each pulse family (rectangular, cosine, and charge-cosine, indicated at the top), we plot the expected infidelities due to $\delta\varepsilon$ fluctuations $\mathcal{I}_\text{ave}$ (colored lines), as defined in Eq.~(\ref{eq:tot_inf}), for (a) the favorable valley configuration $E_{vL} = E_{vR} = 100$~\SI{}{\micro\electronvolt}, and (b) the unfavorable configuration $E_{vL} = E_{vR} = 20$~\SI{}{\micro\electronvolt}, as a function of the valley phase difference $\delta \phi\in[0,0.9\pi]$. 
    (The insets in the rightmost panels indicate the valley splitting configuration.) 
    The $\mathcal{I}_\text{ave}$ curves plotted in (a) are the same as in Fig.~\ref{fig:phase_dependence}(a). 
    The expected infidelities due to $\delta\Delta$ fluctuations, as described in Sec.~\ref{sec:valleycoupling}, are also plotted for a conventional \SI{10}{\nano\meter} QW (blue squares) and a 5\% Ge QW (red squares). 
    The gray region indicates the baseline pulse infidelity with no charge noise and serves as an effective lower bound.}
    \label{fig:vary_Ev_fidelity_opt}
\end{figure}

Next, we analyze the impact of small shifts of the inter-valley coupling parameters $\Delta_L$ and $\Delta_R$ due to lateral electric field fluctuations.
Unlike fluctuations considered in Sec.~\ref{sec:Egsandtcflucts}, these effects can dominate qubit infidelities in some regimes.
To explore this behavior, we begin by estimating the expected size of the fluctuations.
As derived in Ref.~[\citenum{Losert:2023p125405}], the spatial covariance between the (complex) inter-valley coupling in a single dot, measured at two locations separated by a distance $\delta r$, is given by 
\begin{multline} \label{eq:delta_cov}
    \text{Cov}[\mathrm{Re} \:\Delta_1, \mathrm{Re}\: \Delta_2] = \text{Cov}[\mathrm{Im} \:\Delta_1, \mathrm{Im}\: \Delta_2] \\
    = \frac{1}{2} e^{-\delta r^2 / 2 a_\text{dot}^2} \sigma_\Delta^2 ,
\end{multline}
where $\Delta_1$ and $\Delta_2$ are the inter-valley couplings at positions separated by a distance $\delta r$, and $\delta r$ is meant to represent the the shift caused by charge noise.
The difference between these inter-valley couplings $\delta \Delta = \Delta_2 - \Delta_1$ is then described by a complex-normal distribution with variance given by
\begin{equation} \label{eq:var_delta_diff}
    \V [ \mathrm{Re} \:\delta \Delta ] = \V [ \mathrm{Im} \: \delta \Delta ] = \frac{1}{2} \left( 1 - e^{-\delta r^2  /2 a_\text{dot}^2}\right) \sigma_\Delta^2.
\end{equation}
We note that this result was obtained for the special case of an isolated, isotropic, harmonically confined dot, which is not fully accurate, for example in the case where the dot is part of a double dot with a large tunnel coupling.
However, we expect it to provide a good estimate in many cases, particularly when $a_\text{dot}$ accurately describes the dot radius.

To determine the dot-position shifts, $\delta r$, we need to know the lateral electric field fluctuations $\delta E_\text{lat}$ that are caused by charge noise.
Statistically, we expect $\delta E_{\text{lat},x}$ to be normally distributed with a mean of zero.
We also expect the field fluctuations to be randomly oriented in the $x$-$y$ plane, such that $\sigma_{E} \coloneqq \sigma_{E_x} = \sigma_{E_y}$, where $\sigma_{E_x}$ is the variance of $\delta E_{\text{lat},x}$.
We can numerically estimate these quantities by relating them to $\sigma_\varepsilon$.
From Eq.~(\ref{eq:delta_detun}), we thus obtain
\begin{equation} \label{eq:sigma_Ex}
    \sigma_\varepsilon = e d \sigma_{E_{x}} .
\end{equation}

We now wish to compute the infidelities caused by the charge-noise-induced inter-valley-coupling fluctuations, $\delta \Delta$.
To simultaneously account for fluctuations occurring in both dots, we use the following Monte Carlo simulation procedure.
First, we specify the charge noise $\sigma_\varepsilon$ and the valley parameters $E_{vL}$, $E_{vR}$, and $\delta \phi$.
We also specify a heterostructure -- either a conventional (0\% Ge) QW or a 5\% Ge QW -- which determines the disorder parameter $\sigma_\Delta$.
We then optimize a pulse shape to match these parameters, following the steps described in Sec.~\ref{sec:gate_op}.
Finally, we repeat the following steps 500 times:
(1) Randomly generate a lateral field fluctuation in the $\hat x$ and $\hat y$ directions according to Eq.~(\ref{eq:sigma_Ex}); 
(2) For these $\delta E_{\text{lat},x}$ and $\delta E_{\text{lat},y}$, compute the dot displacement $\delta r = \sqrt{\delta x^2 + \delta y^2}$, where $\delta x$ and $\delta y$ are given by Eq.~(\ref{eq:delta_x_E_field}); 
(3) For this $\delta r$, randomly generate the inter-valley-coupling fluctuations for the two dots, $\delta \Delta_L$ and $\delta \Delta_R$, according to Eq.~(\ref{eq:var_delta_diff}); (4) Simulate the evolution of the qubit with these modified Hamiltonian valley parameters.
By averaging the infidelities over these 500 iterations, we obtain an estimate for the average infidelity caused by the $\Delta$ fluctuations, for a given set of valley parameters.
We note that this estimate includes only charge-noise-induced inter-valley-coupling fluctuations ($\delta\Delta$), not direct detuning fluctuations ($\delta \varepsilon$).

With these tools, we first consider the optimistic charge-noise regime, $\sigma_\varepsilon = 1$~\SI{}{\micro\electronvolt}, and the favorable valley configuration $E_{vL} = E_{vR} = 100$~\SI{}{\micro\electronvolt}, as shown in Fig.~\ref{fig:vary_Ev_fidelity_opt}(a) for all three pulse families. 
Here, the infidelity caused by $\delta\Delta$ fluctuations is shown for a conventional \SI{10}{\nano\meter} QW (blue squares) and a 5\% Ge QW (red squares). 
We also plot the infidelity caused by $\delta\varepsilon$ fluctuations, as defined in Eq.~(\ref{eq:tot_inf}) [colored lines, showing the same data as Fig.~\ref{fig:phase_dependence}(a)].
The gray boundaries highlight the baseline infidelities $\mathcal{I}_0$, which include neither  $\delta\varepsilon$ fluctuations nor $\delta\Delta$ fluctuations, serving as an effective lower bound on the fidelity for each pulse shape.
The results shown in Fig.~\ref{fig:vary_Ev_fidelity_opt}(a) suggest that infidelities with $\delta\Delta$ fluctuations are not much worse than the baseline infidelities, while infidelities with detuning fluctuations can be much larger.
Importantly, we see that the total infidelity (estimated by summing the $\delta\varepsilon$ and $\delta\Delta$ contributions) can be lower than $10^{-4}$ for a wide range of valley phase values with $\delta \phi < 0.9 \pi$, for all three pulse families.

In Fig.~\ref{fig:vary_Ev_fidelity_opt}(b), we also consider the less favorable valley configuration $E_{vL} = E_{vR} = 20$~\SI{}{\micro\electronvolt}, using the same plotting and color-coding scheme. 
In this case, $\delta\Delta$ fluctuations are no longer negligible, producing infidelities that often exceed those caused by $\delta\varepsilon$ fluctuations.
This effect is more pronounced for the 5\% Ge QW, because the same charge noise often induces larger $\delta \Delta$.
Indeed, for the 5\% Ge QW, there can be infidelities exceeding $10^{-4}$, even for small $\delta \phi$ values, although such spurious behavior is less common for the charge-cosine pulse shape, due to its weaker driving through the anticrossing.
In this latter case, the total infidelity, including both $\delta\varepsilon$ and $\delta\Delta$ fluctuations,  remains below $10^{-4}$ for $\delta \phi \leq 0.7$.

\begin{figure}
    \centering
    \includegraphics[width=8cm]{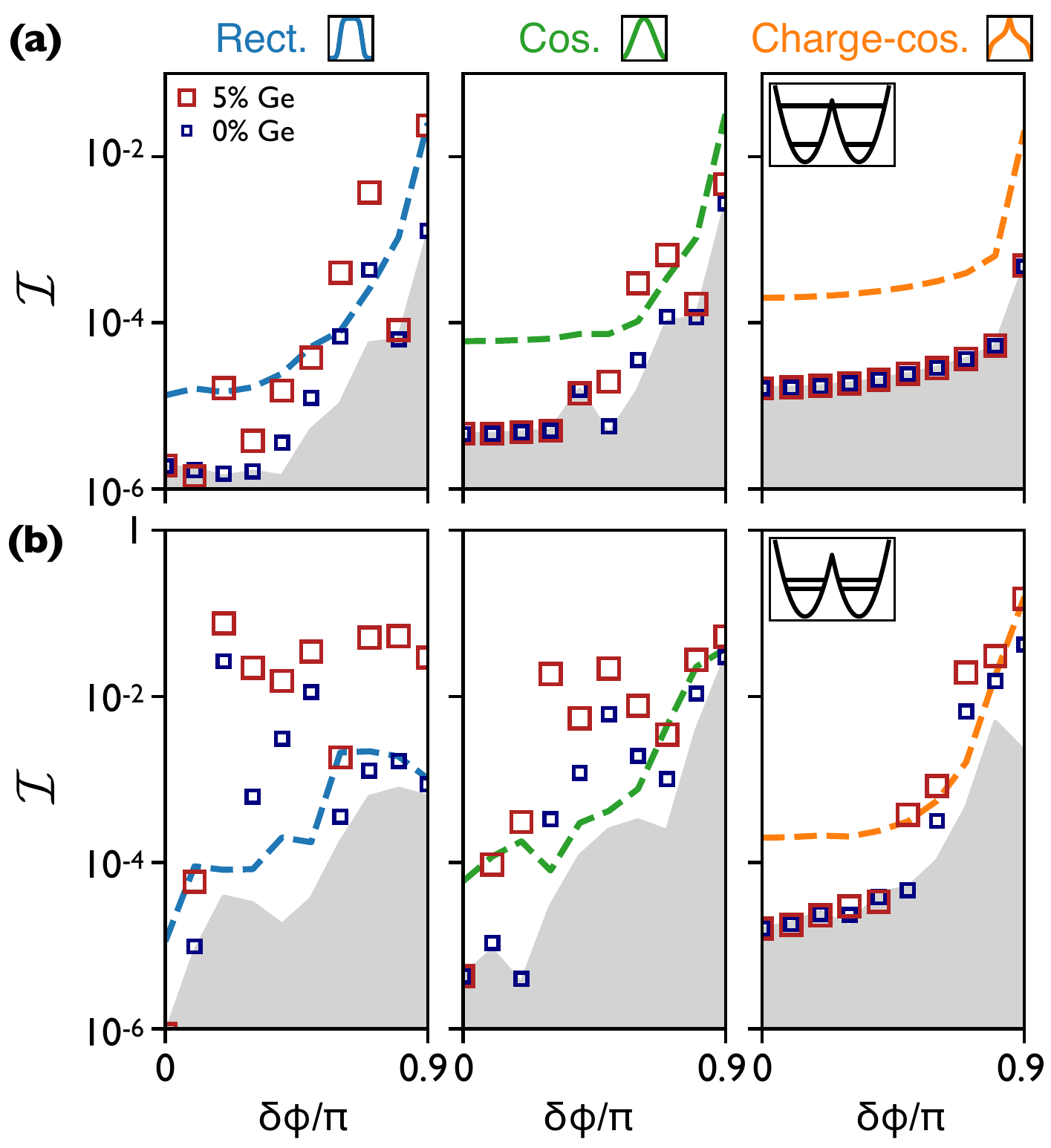}
    \caption{Infidelities due to $\delta\varepsilon$ and $\delta\Delta$ fluctuations in the \textit{pessimistic} charge-noise regime, $\sigma_{\varepsilon} = 15$~\SI{}{\micro\electronvolt}. 
    Except for this change in $\sigma_\varepsilon$, all calculation procedures and plotting styles are the same as Fig.~\ref{fig:vary_Ev_fidelity_opt}. 
    We again consider (a) the favorable valley configuration $E_{vL} = E_{vR} = 100$~\SI{}{\micro\electronvolt}, and (b) the unfavorable configuration $E_{vL} = E_{vR} = 20$~\SI{}{\micro\electronvolt}, where we plot the averaged infidelities as a function of the valley phase difference $\delta \phi\in[0,0.9\pi]$. 
    The $\mathcal{I}_\text{ave}$ curves (colored dashed lines) plotted in (a) are the same as in Fig.~\ref{fig:phase_dependence}(a).
    Compared to Fig.~\ref{fig:vary_Ev_fidelity_opt}, it appears challenging to achieve robust, high-fidelity gate operations with this higher level of charge noise. }
    \label{fig:vary_Ev_fidelity}
\end{figure}

Finally, in Fig.~\ref{fig:vary_Ev_fidelity}, we examine the pessimistic charge noise regime, $\sigma_\varepsilon = 15$~\SI{}{\micro\electronvolt}, following the same plotting and color-coding scheme as Fig.~\ref{fig:vary_Ev_fidelity_opt}, for the same two valley configurations.
For the favorable configuration $E_{vL} = E_{vR} = 100$~\SI{}{\micro\electronvolt}, in Fig.~\ref{fig:vary_Ev_fidelity}(a), a modest range of $\delta \phi \leq 0.5 \pi$ yields infidelities $<10^{-4}$, for the rectangular and cosine pulses, while poor fidelities are obtained for larger $\delta \phi$.
On the other hand, the charge-cosine pulse is not sensitive to valley fluctuations in this regime; however, due to its weaker driving at the anticrossing, it is more sensitive to charge noise.
Hence, this pulse shape is not preferable for stronger charge noise. 

In the unfavorable valley configuration $E_{vL} = E_{vR} = 20$~\SI{}{\micro\electronvolt}, the results are more challenging, overall.
For the rectangular and cosine pulse families, $\delta\Delta$ fluctuations dominate over $\delta\varepsilon$ fluctuations, yielding poor fidelities even for fairly small $\delta \phi$.
In fact, except for $\delta \phi = 0$, there are no configurations with reliably low total infidelities $<$$10^{-4}$.
In this regime, the charge-cosine pulse outperforms the other schemes, since it is relatively insensitive to valley fluctuations, resulting in total infidelities $\leq 3 \times 10^{-4} $ over the limited range of $\delta \phi\in[0,0.4\pi]$.
However, we cannot achieve robust, high-fidelity operations from any pulse shape, for this higher level of charge noise.

\subsection{Lessons for future pulse optimization schemes}

We have shown above that for favorable valley configurations and low levels of charge noise, detuning fluctuations are the dominant source of infidelity in strongly driven flopping mode qubits. 
In these circumstances, it suffices to model the noise as a detuning fluctuation and to optimize pulses as we have done in Sec.~\ref{sec:gate_op}.
However, in configurations with unfavorable valley splitting configurations or significant charge noise, infidelities caused by charge-noise-induced valley fluctuations can be larger than the infidelities caused by direct detuning fluctuations. 
This is especially true when low valley splittings are encountered in a high-Ge heterostructure, despite the larger average $E_v$ in these devices.
In these situations, pulse optimization schemes must account for such valley fluctuations.
This requires that we first characterize the valley splittings and valley phase differences in the double-dot system, as well as the valley response to nearby noise sources, which complicates the pulse optimization procedure, adding overhead to the tune-up process in a large quantum processor.
We view this as additional evidence that, to build scalable quantum processors, it is tremendously important to (1) reduce charge noise, and (2) reliably increase valley splittings.
Doing so not only boosts qubit fidelities, but also makes qubit optimization more straightforward.

\section{Scale-up considerations for flopping-mode qubits} \label{sec:scalability}

\begin{figure}
    \centering
    \includegraphics[width=8cm]{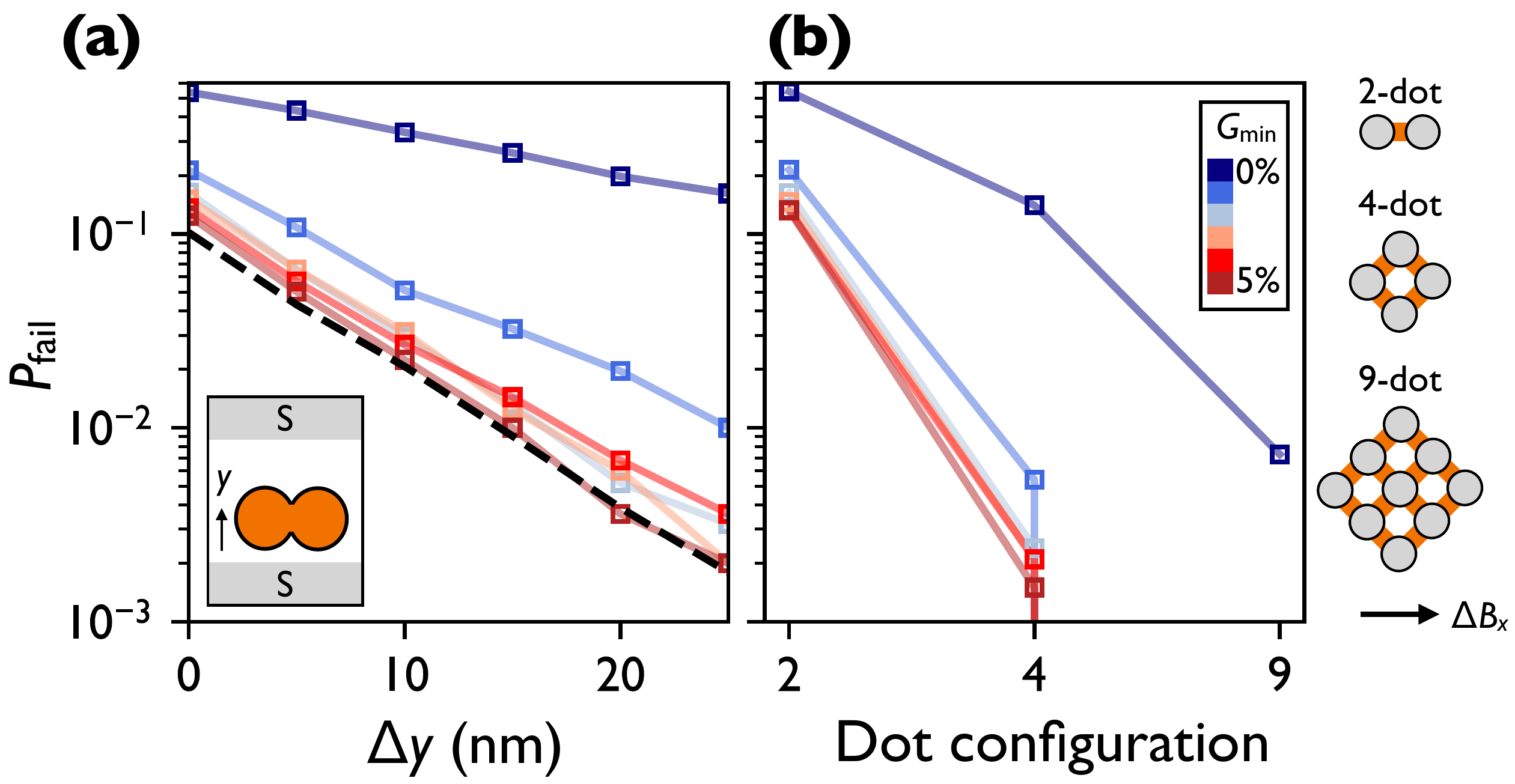}
    \caption{By engineering some tunability into the qubit architecture, we can significantly improve the probability of obtaining high-quality qubits. (a) We plot the probability of failure, $P_\text{fail}$, while allowing the dot position to shift along $\hat y$, over the range $y\in[0, \Delta y]$, as illustrated schematically in the inset. Each data point is computed from Eq.~(\ref{eq:pfail}) by considering probabilities determined from 2,500 instantiations of random-alloy disorder.
    We also consider heterostructures with minimum Ge concentrations of $G_\text{min}\in[0,5\%]$, as indicated by the color scale. For comparison, we include the lower bound $P_\text{phase}$, computed using Eq.~(\ref{eq:pphase}), which only takes into account valley phase disorder [black dashed line]. 
    (b) We plot $P_\text{fail}$ for quantum dot grids of various sizes, as illustrated on the right, for the same $G_\text{min}$ shown in (a). Each data point is computed from 10,000 random instantiations of the valley parameters in each dot. We find that, for grid sizes of four dots or higher, there are no instances of failure according to these definitions, when $G_\text{min} \geq 1$\%.}
    \label{fig:vertical_tunability}
\end{figure}

In this section, we discuss the implications of our results for the scalability of flopping-mode qubits.
First recall that, for heterostructures in the disordered valley splitting regime, the valley parameters $E_v$ and $\delta \phi$ are randomized from dot to dot~\cite{Losert:2023p125405} and, as noted above, flopping-mode qubit operations are often sensitive to these valley parameters.
So, a given double dot will not necessarily yield a successful flopping-mode qubit.
For example, in double dots with small $E_{vL(R)}$ or for valley phase differences $|\delta \phi| \approx \pi$, high-fidelity operations may not be possible.
However, given an initial unfavorable configuration, it may be possible to tune the system into a better-performing state by taking advantage of the random nature of the valley parameters.
Here, we explore two strategies to improve qubit fidelities, assuming an optimistic charge-noise regime of $\sigma_\varepsilon = 1$~\SI{}{\micro\electronvolt}.

First, we consider the possibility of laterally displacing the double dot.
For definiteness, we consider a displacement perpendicular to the detuning axis [i.e., along $\hat y$, as illustrated in the inset of Fig.~\ref{fig:vertical_tunability}(a)], which can be accomplished by modulating the screening gate potentials (S).
The shifted dots experience a rapidly varying random-alloy landscape.
Displacements of up to \SI{40}{\nano\meter} have been demonstrated experimentally [\citenum{Hollmann:2020p034068, Dodson:2022p146802, Volmer:2024p61, Volmer:2025arXiv}], so we view this as a realistic strategy for modifying the valley parameters, although there is no guarantee that the new valley configuration will be improved.
Alternatively, we consider sparse grids of quantum dots, as depicted in Fig.~\ref{fig:vertical_tunability}(b), where any two nearest neighbors may be used to form a flopping-mode qubit.
The scaling properties of this arrangement are favorable.
For example, in the 2-dot system, there is only one possible double dot configuration, while, for the 4-dot system, in a diamond arrangement, there are four possible double-dot configurations, as indicated by orange links.
Similarly, for the 9-dot grid, there are 12 potential double-dot configurations.
Furthermore, if a magnetic field gradient can be implemented along $\hat x$, each of the double dots should experience roughly the same magnetic parameters.
Thus, by increasing the grid size, we enhance the probability of realizing a high-fidelity flopping-mode qubit.
An obvious disadvantage of this strategy is that it achieves success at the expense of wasted dots. 

Finally, we consider the impact of high-Ge QWs on both the aforementioned strategies.
Adding Ge to the quantum well can boost the average $E_v$~\cite{Wuetz:2022p7730}, thereby increasing the likelihood of obtaining a high-fidelity valley configuration.
In Sec.~\ref{sec:valley_fluctuations}, we also showed that charge-noise-induced valley fluctuations can be mitigated by larger $E_v$ values.
Below, we explore these lateral-displacement and sparse-grid strategies in QWs with Ge concentrations in the range $G_\text{min}\in[0,5\%]$.

\subsection{Defining success}

Before analyzing these strategies, we define a method for estimating the likelihood of achieving high-fidelity gates.
Examining Fig.~\ref{fig:vary_Ev_fidelity_opt}(a), we observe a wide range of valley parameters that yield total infidelities $< 10^{-4}$ under optimistic charge-noise conditions, $\sigma_\varepsilon = 1$~\SI{}{\micro\electronvolt}.
Based on this, we conservatively estimate that, if $E_{vL}$ and $E_{vR}$ are both $\geq 100$~\SI{}{\micro\electronvolt}, and if $|\delta \phi| \leq 0.9\pi$, then a high-fidelity gate can be constructed for any of the three pulse families.
Additionally, from Fig.~\ref{fig:vary_Ev_fidelity_opt}(b), we estimate that, even for an unfavorable valley configuration, if $E_{vL}$ and $E_{vR}$ are both $\geq 20$~\SI{}{\micro\electronvolt} and $|\delta \phi| \leq 0.7 \pi$, then a relatively high-fidelity gate can still be constructed using the charge-cosine pulse family.
Thus, if either of these conditions is satisfied, our simulations suggest the qubit is likely to operate with high fidelity.
We therefore define the failure probability $P_\text{fail}$, as the probability that \textit{neither} of these conditions is met:
\begin{multline} \label{eq:pfail}
    P_\text{fail} = 1 - P \Big[ E_{vL}, E_{vR} \geq 100 \; \mathrm{\mu}\text{eV}, \delta \phi \leq 0.9\pi  \\
    \text{ or } E_{vL}, E_{vR} \geq 20 \; \mathrm{\mu}\text{eV}, \delta \phi \leq 0.7\pi \Big],
\end{multline}
%Equation~(\ref{eq:pfail}) defines our success criterion.
where $P[\cdot]$ represents the probability of satisfying these criteria for a large number of randomized instantiations of the valley parameters.

\subsection{Lateral displacement}

First, we consider lateral double-dot displacements.
%For a given heterostructure design (with $G_\text{min}$ ranging from 0 to 5\%), 
For each pair of dots, we simulate the valley splitting in each dot while displacing it along $\hat y$, as shown in the inset of Fig.~\ref{fig:vertical_tunability}(a).
Since the valley splitting correlation length is approximately equal to a dot diameter, we assume that $E_v$ is uncorrelated between the left and right dots; but for a given dot, $E_v$ is correlated with itself as the dot moves over small distances (see Appendix~\ref{app:vertical_tunability} for details).
We then evaluate the criteria given in Eq.~(\ref{eq:pfail}) while shifting the double-dot position $y$ over the range $y\in[ 0, \Delta y]$.
For each instance of random-alloy disorder, if the success condition defined in Eq.~(\ref{eq:pfail}) is met for \textit{any} $y$ within $[ 0, \Delta y]$, we label a given instance a success, since we can find a suitable configuration of the valley parameters.
Conversely, if the success condition is not met for \textit{all} $y$ within $[ 0, \Delta y]$, the given instance is labeled a failure.
We repeat this procedure for 2,500 instantiations of random-alloy disorder to determine the probability of failure given in Eq.~(\ref{eq:pfail}). 

The resulting estimates for $P_\text{fail}$ are plotted in Fig.~\ref{fig:vertical_tunability}(a) as a function of $\Delta y$ and for $G_\text{min}\in[0,5\%]$ (color scale).
We see that the ability to lateral shift the dots significantly improves our success rates, especially for heterostructures with a modest amount of Ge in the QW.
For example, for $G_\text{min} = 3$\%, we find that $P_\text{fail} \leq 1$~\% when $\Delta y \geq 20$~\SI{}{\nano\meter}.
Interestingly, we observe that improvements in the $P_\text{fail}$ curves saturate for $G_\text{min} \geq 3$\%.
This can be understood because, in this regime, the phase difference $\delta \phi$ becomes the limiting constraint, rather than the valley splitting values.
To emphasize this point, we repeat the previous analysis, while evaluating only the the valley phase criterion in Eq.~(\ref{eq:pfail}) and ignoring the valley splitting constraint: 
\begin{equation} \label{eq:pphase}
P_\text{phase} = P \left(|\delta \phi| > 0.9\pi \text{ for all } y \in [0, \Delta y] \right).
\end{equation}
Estimates of $P_\text{phase}$ from Eq.~(\ref{eq:pphase}) are plotted as a dashed black line in Fig.~\ref{fig:vertical_tunability}(a), where we see that $P_\text{phase}$ forms a lower-bound on $P_\text{fail}$.
This is expected since $P_\text{phase}$ is uncorrelated with $G_\text{min}$ in the disordered valley splitting regime, so that increasing the Ge content of the QW cannot reduce $P_\text{fail}$ beyond this bound.

\subsection{Sparse quantum dot grids}

Next, we consider sparse grids of quantum dots, within which we select two neighboring dots to house our qubit.
To estimate $P_\text{fail}$ in this case, we generate 10,000 values of the inter-valley coupling $\Delta$ for each dot in the grid.
Again, we assume that $\Delta$ is uncorrelated between neighboring dots.
We then evaluate whether any of the dot pairings highlighted in Fig.~\ref{fig:vertical_tunability}(b) meet the success criteria defined in Eq.~(\ref{eq:pfail}).
The results for $P_\text{fail}$ are plotted in Fig.~\ref{fig:vertical_tunability}(b) as a function of the grid size. 
Notably, we find that, if $G_\text{min} > 0$, $P_\text{fail}$ already drops below 1\% for the 4-dot configuration.
In the 9-dot configuration, we find that every one of the 10,000 disorder instantiations contains at least one suitable qubit, when $G_\text{min} > 0$.
Thus, by manufacturing sparse quantum dot grids, combined with a small amount of Ge in the quantum well to boost the average $E_v$, we can engineer systems with many high-fidelity qubits, with high success rates.
In Appendix~\ref{app:linear_array}, we also perform a similar analysis on linear qubit grids, finding that they yield slightly worse values of $P_\text{fail}$ than square grids of the same size; however, they still provide significant improvements over the base case, without a sparse grid.

\section{Conclusion}

We have analyzed the performance of flopping-mode quantum dot qubits in Si/SiGe by optimizing single-qubit rotations in the presence of charge noise and random valley disorder.
We have shown that, in the presence of weak detuning noise ($\sigma_\varepsilon= 1 $~\SI{}{\micro\electronvolt}), high-fidelity gates can be achieved for a wide range of valley parameters, provided that shapes of the gate pulses are fine-tuned for their specific valley parameters.
In the presence of larger detuning noise ($\sigma_\varepsilon= 15 $~\SI{}{\micro\electronvolt}), high-fidelity gates can still be performed, provided the valley splitting in each dot is large and the valley phase difference between dots is relatively small.
In addition to the typical detuning noise, we consider the impact of small charge-noise-induced fluctuations of other qubit parameters.
We find that, in some cases, small fluctuations of the valley parameters dominate over the direct detuning fluctuations, becoming the dominant source of infidelity.
Indeed, this situation is likely for qubits with unfavorable valley configurations (e.g., $E_v \sim 20$~\SI{}{\micro\electronvolt}), or when strong driving is employed, or when devices experience significant charge noise ($\sigma_\varepsilon= 15$~\SI{}{\micro\electronvolt}) or large valley splitting gradients (e.g., QWs containing 5\% Ge).
While we focus on single flopping-mode qubit operations in this work, we speculate that similar valley-induced infidelity dynamics may also exist for two-qubit gates in Si.
However, these valley-induced infidelities can be suppressed by introducing pulse shapes that drive more gradually through the valley anticrossing, such as the ``charge-cosine'' pulse.
We view this as evidence for the importance of engineering (1) large valley splittings, and (2) low charge noise in Si/SiGe spin qubits.
Finally, we have analyzed schemes to avoid unfavorable valley configurations, including allowing for lateral displacements of the double dot and sparse grids of quantum dots.
We find that both schemes can significantly enhance the probability of achieving high-fidelity qubit rotations.

\section{Data availability}

The data and code presented in this work is available in a Zenodo repository [\citenum{Losert:2025zenodoV2}].

\section{Acknowledgments}

This research was sponsored in part by the Army Research Office (ARO) under Awards No.\ W911NF-17-1-0274, No.\ W911NF-22-1-0090, and No.\ W911NF-23-1-0110.
The work was performed using the compute resources and assistance of the UW-Madison Center For High Throughput Computing (CHTC) in the Department of Computer Sciences [\citenum{chtc}]. The CHTC is supported by UW-Madison, the Advanced Computing Initiative, the Wisconsin Alumni Research Foundation, the Wisconsin Institutes for Discovery, and the National Science Foundation, and is an active member of the OSG Consortium, which is supported by the National Science Foundation and the U.S. Department of Energy's Office of Science.
The views, conclusions, and recommendations contained in this document are those of the authors and are not necessarily endorsed nor should they be interpreted as representing the official policies, either expressed or implied, of the Army Research Office (ARO) or the U.S. Government. The U.S. Government is authorized to reproduce and distribute reprints for Government purposes notwithstanding any copyright notation herein.

\appendix

\section{Additional analysis of the qubit Hamiltonian} \label{app:analysis}

In this Appendix, we include some further analysis of the qubit Hamiltonian.
Starting with the Hamiltonian of Eq.~(\ref{eq:ham_diag_valley}), we discard the spin terms, since $g \mu_B B \ll |\Delta|, t_c$ and they can be treated as perturbations.
The remaining valley-orbit Hamiltonian is given by
\begin{equation} \label{eq:ham_vo}
H_\text{vo} = 
\begin{pmatrix}
    |\Delta_L| + \frac{\varepsilon}{2} & 0 & t_{++} & t_{+-} \\
    0 & -|\Delta_L| + \frac{\varepsilon}{2} & t_{-+} & t_{--} \\
    t_{++}^* & t_{-+}^* & |\Delta_R| - \frac{\varepsilon}{2} & 0 \\
    t_{+-}^* & t_{--}^* & 0 & -|\Delta_R| - \frac{\varepsilon}{2}
\end{pmatrix}
\end{equation}
in the basis $\{ |L,+\rangle, |L,-\rangle, |R,+\rangle, |R,-\rangle \}$, where the $t_{ij}$ are defined in the main text.
The general solutions of $H_\text{vo}$ involve fourth-order polynomials, but we can simplify the problem in the limit where $\delta \phi = 0$.
In this case, $t_{++} = t_{--} = t_c$ and $t_{-+} = t_{+-} = 0$, and
\begin{equation}
H_\text{vo}^{\delta \phi = 0} = 
\begin{pmatrix}
    |\Delta_L| + \frac{\varepsilon}{2} & 0 & t_c & 0 \\
    0 & -|\Delta_L| + \frac{\varepsilon}{2} & 0 & t_c \\
    t_c & 0 & |\Delta_R| - \frac{\varepsilon}{2} & 0 \\
    0 & t_c & 0 & -|\Delta_R| - \frac{\varepsilon}{2}
\end{pmatrix}.
\end{equation}
In this regime, we can obtain simple expressions for the eigenvalues of the system:
\begin{equation} \label{eq:four_level_eigenvalues}
\begin{split}
    E_0 &= \frac{1}{2} \left( - \Delta_+ - \sqrt{ (\varepsilon - \Delta_-)^2 + 4 t_c^2} \right) \\
    E_1 &= \frac{1}{2} \left( \Delta_+ - \sqrt{ (\varepsilon + \Delta_-)^2 + 4 t_c^2} \right) \\
    E_2 &=  \frac{1}{2} \left( - \Delta_+ + \sqrt{ (\varepsilon - \Delta_-)^2 + 4 t_c^2} \right) \\
    E_3 &= \frac{1}{2} \left( \Delta_+ - \sqrt{ (\varepsilon + \Delta_-)^2 + 4 t_c^2} \right) ,
\end{split}
\end{equation}
where $\Delta_{\pm} = |\Delta_L| \pm |\Delta_R|$.
The corresponding (un-normalized) eigenstates are
\begin{equation} \label{eq:four_level_eigenstates}
\begin{split}
    |v_0\rangle & = 
    \begin{pmatrix}
        0 & \frac{1}{2 t_c}\left(-\Delta_- + \varepsilon - \sqrt{(\varepsilon-\Delta_-)^2 + 4 t_c)^2} \right) & 0 & 1
    \end{pmatrix}^T \\
    |v_1\rangle & = 
    \begin{pmatrix}
        \frac{1}{2 t_c}\left(\Delta_- + \varepsilon - \sqrt{(\varepsilon+\Delta_-)^2 + 4 t_c)^2} \right) & 0 & 1 & 0
    \end{pmatrix}^T \\
    |v_2\rangle & = 
    \begin{pmatrix}
        0 & \frac{1}{2 t_c}\left(-\Delta_- + \varepsilon + \sqrt{(\varepsilon-\Delta_-)^2 + 4 t_c)^2} \right) & 0 & 1
    \end{pmatrix}^T \\
    |v_3\rangle & = 
    \begin{pmatrix}
        \frac{1}{2 t_c}\left(\Delta_- + \varepsilon + \sqrt{(\varepsilon+\Delta_-)^2 + 4 t_c)^2} \right) & 0 & 1 & 0
    \end{pmatrix}^T .
\end{split}
\end{equation}
From these eigenstates, we can compute $\langle \tau_z \rangle_\epsilon$:
\begin{equation} \label{eq:tau_z_ep_theory_supp}
    \langle \tau_z \rangle_\epsilon = \frac{\langle v_0 | \tau_z | v_0 \rangle}{\langle v_0 | v_0 \rangle} = \frac{\Delta_- - \varepsilon}{\sqrt{ (\varepsilon-\Delta_-)^2 + 4 t_c^2) } }.
\end{equation}
The results for Eq.~(\ref{eq:tau_z_ep_theory_supp}) are included in Fig.~\ref{fig:valley_dependent_tuning}(b) in the main text.

We can also extract limits on charge excitations in the $\delta \phi = 0$ limit.
We define $R_0$ as the unitary transformation that diagonalizes $H_\text{vo}^{\delta \phi = 0}$, whose rows are given by (normalized) $\langle v_i|$ in Eq.~(\ref{eq:four_level_eigenstates}).
Since $\varepsilon$ is a function of time, the transformation $R_0$ is also time-dependent, and the transformed Hamiltonian can be computed as
\begin{equation}
H_{\delta \phi = 0}' = R_0 H_\text{vo}^{\delta \phi = 0} R_0^\dag - i \hbar R_0 \dot R_0^\dag,
\end{equation}
resulting in 
\begin{multline}
H_{\delta \phi = 0}' = \\
    \begin{pmatrix}
        E_0 & 0 & \frac{ i \hbar t_c \dot \epsilon}{(\varepsilon-\Delta_-)^2+ 4 t_c^2} & 0 \\
        0 & E_1 & 0 & \frac{ i \hbar t_c\dot \epsilon}{(\varepsilon+\Delta_-)^2+ 4 t_c^2} \\
        \frac{-i \hbar t_c\dot \epsilon}{(\varepsilon-\Delta_-)^2+ 4 t_c^2} & 0 & E_2 & 0 \\
        0 & \frac{-i \hbar t_c\dot \epsilon}{(\varepsilon+\Delta_-)^2+ 4 t_c^2} & 0 & E_3
    \end{pmatrix} ,
\end{multline}
where now the Hamiltonian is expressed in the basis of instantaneous eigenstates, and the $E_j$ are the eigenvalues given in Eq.~(\ref{eq:four_level_eigenvalues}).
Thus, the regime in which we can ignore charge excitations is given by $\dot \varepsilon \ll ((\varepsilon \pm \Delta_-)^2 + 4 t_c^2 )/ \hbar$.
Of course, in most scenarios when $t_{+-}$ and $t_{-+}$ are nonzero, valley excitations will create a much tighter limit on $\dot \varepsilon$.

Finally, we can also extract approximate expressions for the rate of valley excitation in this system in certain limits.
Namely, if $|\Delta_{L(R)}| \ll t_c$, then the valley terms in Eq.~(\ref{eq:ham}) can also be treated perturbatively.
In this limit, we can diagonalize the orbital (charge) Hamiltonian through the rotation $R_\text{orb} = \exp [\frac{i}{2} \arctan (\frac{2 t_c}{\varepsilon}) \tau_y] $.
Starting with Eq.~(\ref{eq:ham}), dropping the valley and spin terms as perturbations, and performing $H_{\text{orb}}' = R_\text{orb} H R_\text{orb}^\dag - i \hbar R_\text{orb} \dot R_\text{orb}^\dag$, we obtain
\begin{equation}
    H_{\text{orb}}' =  \sqrt{\varepsilon^2 + 4 t_c^2} \tilde \tau_z + \frac{\hbar t_c \dot \varepsilon}{4 t_c^2 + \varepsilon^2} \tilde \tau_y ,
\end{equation}
where $\tilde \tau_j$ are the Pauli matrices in the transformed basis.
Now, we assume evolution is adiabatic with respect to orbital levels, so we remain within the $\langle \tilde \tau_z \rangle = -1$ subspace, and we can replace $\tilde \tau_z \rightarrow -1$ and remove the term proportional to $\tilde \tau_y$.
We add the valleys as a perturbation, resulting in the Hamiltonian
\begin{equation}
\begin{split}
    H_{|\Delta| \ll t_c}' &= -\sqrt{\varepsilon^2 + 4 t_c^2} \\
    & + \frac{1 + \langle \tau_z \rangle_\varepsilon}{2}|\Delta_L| ( \gamma_x \cos \phi_L - \gamma_y \sin \phi_L ) \\
    & + \frac{1 - \langle \tau_z \rangle_\varepsilon}{2} |\Delta_R| ( \gamma_x \cos \phi_R - \gamma_y \sin \phi_R )  \\
    & \coloneqq V_x(t) \gamma_x + V_y(t) \gamma_y .
\end{split}
\end{equation} 
In this regime, we can use Eq.~(\ref{eq:tau_z_ep_theory_supp}) for $\langle \tau_z \rangle_\varepsilon$, setting $\Delta_- = 0$.
We diagonalize the valley space with the rotation
\begin{equation}
R_v = e^{-i \frac{\pi}{4} \gamma_y} e^{ \frac{i}{2} \arctan \left[ \frac{V_y(t)}{V_x(t)} \right] \gamma_z } ,
\end{equation}
resulting in another transformed Hamiltonian $H_{|\Delta| \ll t_c}'' = R_v H_{|\Delta| \ll t_c}' R_v^\dag - i \hbar R_v \dot R_v^\dag$, where 
\begin{equation}
    H_{|\Delta| \ll t_c}'' = -\sqrt{\varepsilon^2 + 4 t_c^2} + \sqrt{V_x(t)^2 + V_y(t)^2 } \tilde \gamma_z + S \tilde \gamma_x ,
\end{equation}
and where $\tilde \gamma_j$ are the Pauli matrices in the rotated valley space, and 
\begin{equation}
    S = \frac{V_x(t) V_y'(t) - V_y(t) V_x'(t)}{2 (V_x(t)^2 + V_y(t)^2)} .
\end{equation}
Solving for $S$, we obtain
\begin{equation} \label{eq:S_valley}
    S = \frac{  - \hbar |\Delta_L| |\Delta_R| \sin(\phi_L - \phi_R) \frac{d \langle\tau_z\rangle_\varepsilon}{dt} }{ |\Delta_+|^2  + 2 (|\Delta_L|^2 - |\Delta_R|^2) \langle \tau_z \rangle_\varepsilon + |\Delta_-|^2 \langle \tau_z \rangle_\varepsilon^2} .
\end{equation}
Equation~(\ref{eq:S_valley}) can be used to place approximate limits on the driving strength $d\langle \tau_z \rangle_\varepsilon/dt$ to avoid valley excitations, which can in turn place limits on $\dot \varepsilon$.

\section{Pulse optimization algorithm} \label{app:algorithm}

\begin{figure*}
    \centering
    \includegraphics[width=14cm]{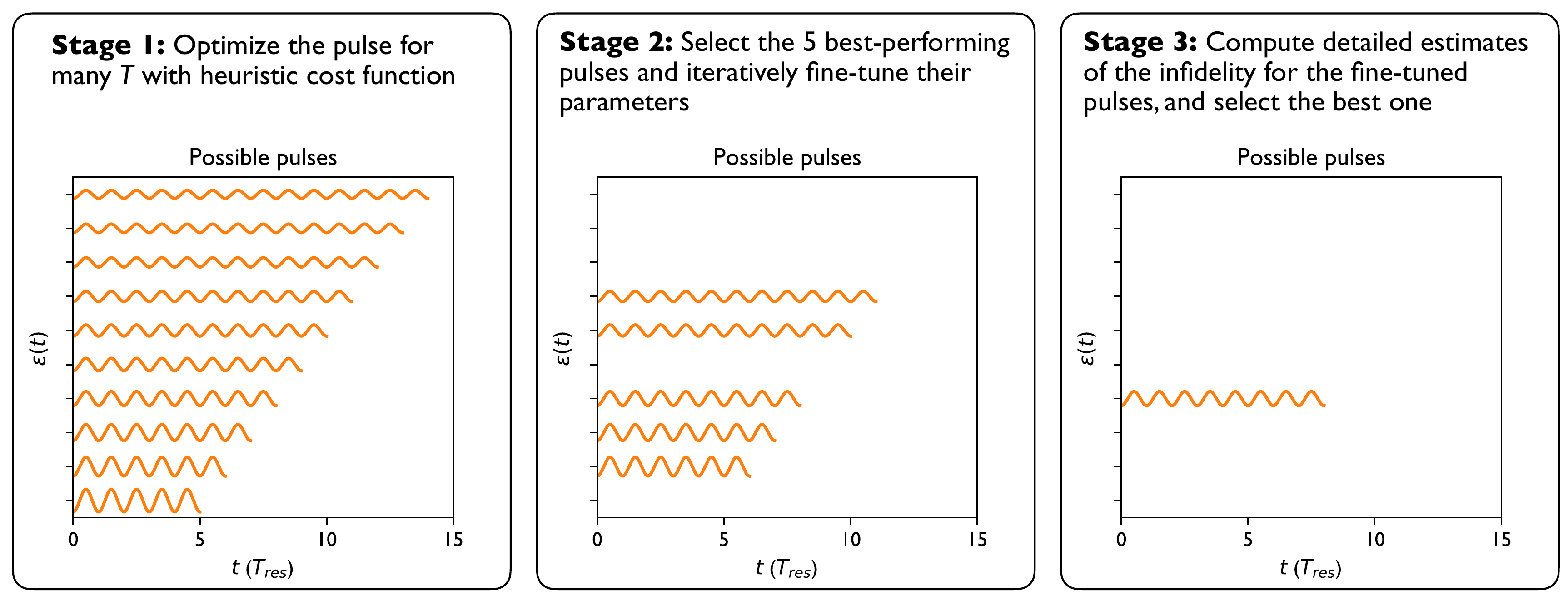}
    \caption{A graphical flow-chart of the optimization protocol described in Sec.~\ref{sec:gate_op} in the main text. In Stage 1, we optimize the pulse parameters for all reasonable pulse lengths $T$ using a heuristic cost function. In Stage 2, we select the five best-performing pulse lengths and iteratively fine-tune their parameters. Finally, in Stage 3, we compute a detailed estimate of the pulse infidelity for each of these five fine-tuned pulses, and we select the best one. The included graphics schematically indicate all the pulses under consideration at each stage, where we have vertically separated each possible pulse, $\varepsilon(t)$.}
    \label{fig:flowchart}
\end{figure*}

\begin{figure*}
    \centering
    \includegraphics[width=16cm]{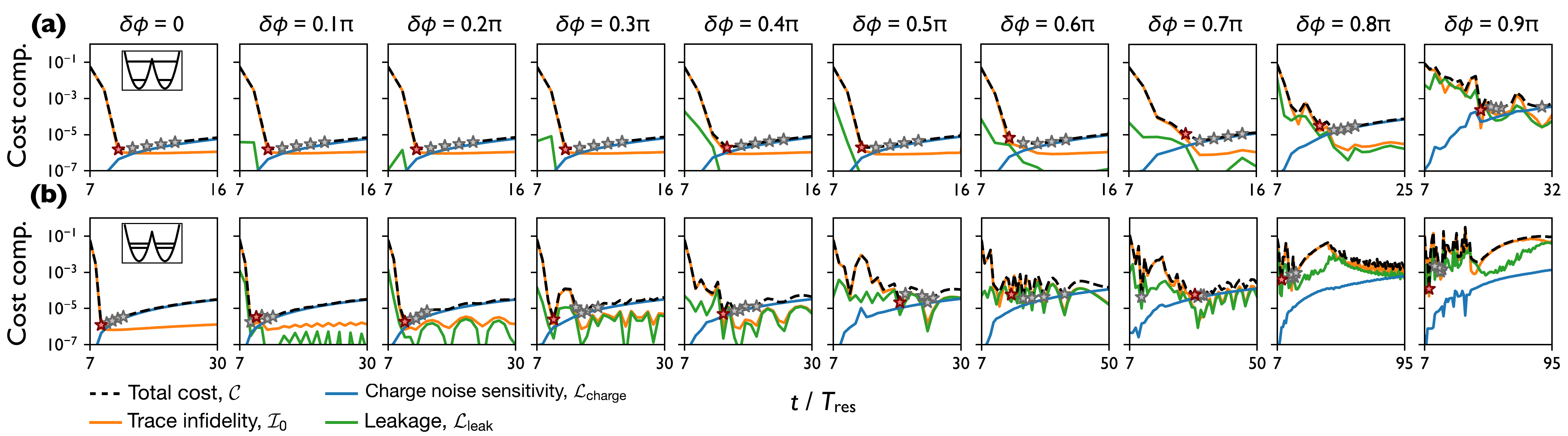}
    \caption{Here, we illustrate the behavior of the three components to our cost function, $\mathcal{C}$ in Eq.~(\ref{eq:cost_func}), across a variety of valley configurations. We show results for the cosine pulse family, assuming an optimistic charge noise regime, $\sigma_\varepsilon = 1$~\SI{}{\micro\electronvolt}. Each individual plot represents the cost components for gates optimized across a range of pulse lengths, for the given set of valley configurations. In the top row (a), we use $E_{vL} = E_{vR} = 100$~\SI{}{\micro\electronvolt}, and in the bottom row (b) we use $E_{vL} = E_{vR} = 20$~\SI{}{\micro\electronvolt}. Each column represents a value of $\delta \phi$, which we vary from $0$ to $0.9\pi$. The stars indicate the five pulses that minimized the total cost, and the red star indicates the pulse ultimately selected by the algorithm.}
    \label{fig:loss_components}
\end{figure*}

In this Appendix, we provide additional details on our pulse-optimization algorithm outlined in Sec.~\ref{sec:gate_op}, and illustrated schematically in Fig.~\ref{fig:flowchart}.
First, we provide details on the quantities $\mathcal{L}_\text{charge}$ and $\mathcal{L}_\text{leakage}$, used in our heuristic cost function in Eq.~(\ref{eq:cost_func}).

\subsection{Computing $\mathcal{L}_\text{charge}$}

As described in Sec.~\ref{sec:gate_op} in the main text, to develop high-fidelity pulses, we want to estimate the pulse sensitivity to charge noise. 
To do so, we utilize a formalism based on a general 2-level Hamiltonian~[\citenum{Gungordu:2022p023155, Buterakos:22021p010341}]
\begin{equation}
    H = H_c + \delta H ,
\end{equation}
where $H_c$ is the desired (time-dependent) control Hamiltonian, and the leakage term
\begin{equation} \label{eq:noise_general}
\delta H = \chi_x(t) \sigma_x + \chi_z(t) \sigma_z ,
\end{equation}
where $\chi_j(t)$ are time-dependent noise amplitudes.
For weak noise, the propagator associated with error can be solved from the first order Magnus expansion
\begin{equation} \label{eq:error_prop}
    U_\text{err} = \exp \left( -\frac{i}{\hbar} \sum_j \mathcal{E}_j(T) \right) ,
\end{equation}
where 
\begin{equation} \label{eq:error_prop_term}
    \mathcal{E}_j(T) = \int_0^T dt \; U_c^\dag(t) \left[ \chi_j(t) \sigma_j \right] U_c(t) ,
\end{equation}
and $U_c(t)$ is the propagator associated with $H_c$.
The noise sensitivity of the logical subspace to $\sigma_j$ errors can be estimated by 
\begin{equation} \label{eq:2_level_noise_sensitivity}
    N_j = \frac{1}{2}||\mathcal{E}_j(T) - \text{tr}[ \mathcal{E}_j(T)/2] ||^2
\end{equation}
where $\mathcal{E}_j$ represents the characteristic (average) strength of the time-dependent noise amplitude $\chi_j$ [\citenum{Gungordu:2022p023155}].

Now, we apply the above formalism to estimate the impact of charge noise in our flopping mode qubits.
For computational efficiency, we reduce the full 8-level Hamiltonian to an effective 2-level system, following Eq.~(\ref{eq:ham_2level}).
We note that, while this effective Hamiltonian does not exactly describe the spin dynamics of the system, it is good enough to estimate charge-noise sensitivity.
However, to compute the true gate fidelity $\mathcal{I}_0$, we retain the full 8-level system; see Sec.~\ref{sec:gate_op}.

Given a quasistatic fluctuation in the detuning, $\delta \varepsilon$, there is a resulting fluctuation in $\langle \tau_z \rangle$, which produces the effective noise Hamiltonian
\begin{align} \label{eq:H_eff_supp}
    \delta H_\text{eff} &= \frac{\delta \langle \tau_z \rangle}{2} \left( \delta E_x \sigma_x + \delta E_z \sigma_z \right),
\end{align}
where $\delta E_{x(z)} = g \mu_B \Delta B_{x(z)} $.
We can approximate 
\begin{equation} \label{eq:delta_tau_z}
    \delta \langle \tau_z \rangle \approx \delta \varepsilon \frac{d \langle \tau_z \rangle}{d\varepsilon},
\end{equation}
where the derivative $\frac{d\langle \tau_z \rangle}{d\varepsilon}$ is evaluated numerically from the 8-level Hamiltonian, and $\delta \varepsilon$ is the approximate size of a detuning fluctuation. 
In our optimization algorithm, we use $\delta \varepsilon = \sigma_\varepsilon$.
If we associate the noise amplitudes of Eq.~(\ref{eq:noise_general}) with our spin system, we have
\begin{align} \label{eq:noise_amps}
\chi_x^\text{spin}(t) &= \frac{\delta \varepsilon}{2} \frac{d\langle \tau_z \rangle}{d\varepsilon} \delta E_x \\
\chi_z^\text{spin}(t) &= \frac{\delta \varepsilon}{2} \frac{d\langle \tau_z \rangle}{d\varepsilon} \delta E_z.
\end{align}
We can obtain the resulting noise sensitivity by using the approximate noise amplitudes of Eq.~(\ref{eq:noise_amps}) in the above noise-sensitivity estimation.
So, in our cost function, we include
\begin{equation}
    \mathcal{L}_\text{charge} = \frac{1}{2} \left[ N_x(T) + N_z(T) \right].
\end{equation}
To estimate the error propagator $U_\text{err}$, we time-evolve the 2-level effective noise Hamiltonian of Eq.~(\ref{eq:H_eff_supp}).

\subsection{Computing $\mathcal{L}_\text{leak}$}

While we restrict ourselves to the 2-level spin subspace to approximate the gate sensitivity to charge noise, we need to consider leakage out of this subspace while performing gate operations. 
To do so, at each step, we simulate the evolution of the full 8-level system from $t = 0$ to $t = T \coloneqq n T_\text{res}$ under the Schrodinger equation, including a quasistatic detuning fluctuation, using Eq.~(\ref{eq:schrodinger_eq_propagator}).
For a given propagator, we can estimate the leakage outside of the ground state subspace by computing the quantity
\begin{equation}
    L_{\delta \varepsilon} = \frac{1}{2} \sum_{n = 2}^7 \left( |\langle n | U_{\delta \varepsilon} | 0 \rangle |^2 + | \langle n | U_{\delta \varepsilon} | 1 \rangle |^2 \right),
\end{equation}
where $\left\{ |n\rangle \right\}_{n=0}^7$ are the $8$ eigenstates of the system at $t = T$.
Finally, we obtain $\mathcal{L}_\text{leak}$ by averaging $L_{\delta \varepsilon}$ over 9 simulations using uniformly spaced $\delta \varepsilon$ between $-2 \sigma_\varepsilon$ and $2 \sigma_\varepsilon$, each weighted by the normal distribution $N(0, \sigma_\varepsilon)$.

\subsection{Comparison of cost components}

In Fig.~\ref{fig:loss_components}, we plot the components of the total cost $\mathcal{C}$ in Eq.~(\ref{eq:cost_func}) for a variety of valley configurations.
Here, we show results for the cosine pulse family, assuming an optimistic $\sigma_\varepsilon = 1$~\SI{}{\micro\electronvolt}.
In Fig.~\ref{fig:loss_components}(a), we assume $E_{vL} = E_{vR} = 100$~\SI{}{\micro\electronvolt}, and in (b) we assume $E_{vL} = E_{vR} = 20$~\SI{}{\micro\electronvolt}.
Each column represents a different valley phase difference $\delta \phi$, which we vary from $0$ to $0.9 \pi$.
We see that, for most configurations, the infidelity $\mathcal{I}_0$ and leakage $\mathcal{L}_\text{leak}$ fall as pulse time increases, while the charge noise sensitivity $\mathcal{I}_\text{charge}$ grows. 
For very short pulses, $\mathcal{I}_0$ and $\mathcal{L}_\text{leak}$ dominate $\mathcal{C}$, while for longer pulses, $\mathcal{L}_\text{charge}$ dominates $\mathcal{C}$.
This results in a minimum $\mathcal{C}$ for some optimial pulse length.
For each configuration, the five pulses that minimze $\mathcal{C}$ are marked with a star, and the red star indicates the pulse ultimately selected by our algorithm.
These are the five pulses that enter the next stage of randomization and re-optimization, described in the following section.
We also note that the loss landscape is significantly complicated for large $\delta \phi$, especially for small $E_v$. 
In this regime, the optimal $\mathcal{C}$ is sometimes much larger, and, in the very worst cases, realized at short pulse lengths.
We comment on this more in App.~\ref{app:pulse_behavior}.

\subsection{Randomization and re-optimization}

As described in the main text, in the second stage of our optimization procedure, we fine-tune the pulse parameters for each of the five best-performing pulse lengths through randomization and re-optimization.
We represent the pulse parameters as a vector $p$, which is optimized with the Nelder-Mead method~[\citenum{Nelder:1965p308}].
In order to restrict the range of the pulse parameters and normalize the optimization variables to a uniform scale, we define the relationship between the pulse parameters and the vector $p$ as follows.
For the rectangular pulse family, the amplitude is $A = \varepsilon_\text{max} |\sin(p_0)|$, the offset is $C = |\varepsilon_\text{max} - A| \sin(p_1)$, the steepness $R = R_\text{max} |\sin(p_2)|$, and the duty cycle coefficient $c_\text{dc} = |\sin(p_3)|$.
We use the same amplitude and offset definitions for the cosine pulse.
For the charge-cosine pulse, the amplitude $A = \langle \tau_z \rangle_\text{max} |\sin(p_0)|$ and the offset $C = |\langle \tau_z \rangle_\text{max} - A| \sin(p_1)$, where $\langle \tau_z\rangle_\text{max}$ is the maximum achievable $\langle \tau_z \rangle_\varepsilon$ for $|\varepsilon| \leq \varepsilon_\text{max}$.
To randomize and re-optimize a pulse, we take the $p$ obtained by the first optimization iteration, and we add a uniform random value between $\pm 0.1$ to each $p_n$.
This randomized $p$ serves as the initial condition for the next round of optimization, so that we may jump away from local minima in the cost landscape, while not moving too far in parameter space from a high-fidelity configuration.

\subsection{Additional details}

To perform our pulse optimization and simulation, we use the Julia programming language~[\citenum{Bezanson:2017julia}].
We select initial conditions such that the optimized pulse performs a single rotation from $\theta = 0$ to $\pi$, and not multiple rotations.
Data analysis, plotting, and other computations are performed in the Python programming language, including the libraries NumPy [\citenum{Harris:2020p357}], SciPy [\citenum{Virtanen:2020p261}], and Matplotlib [\citenum{Hunter:2007p90}].

\section{Optimized pulses in unfavorable valley configurations} \label{app:pulse_behavior}

\begin{figure*}
    \centering
    \includegraphics[width=15cm]{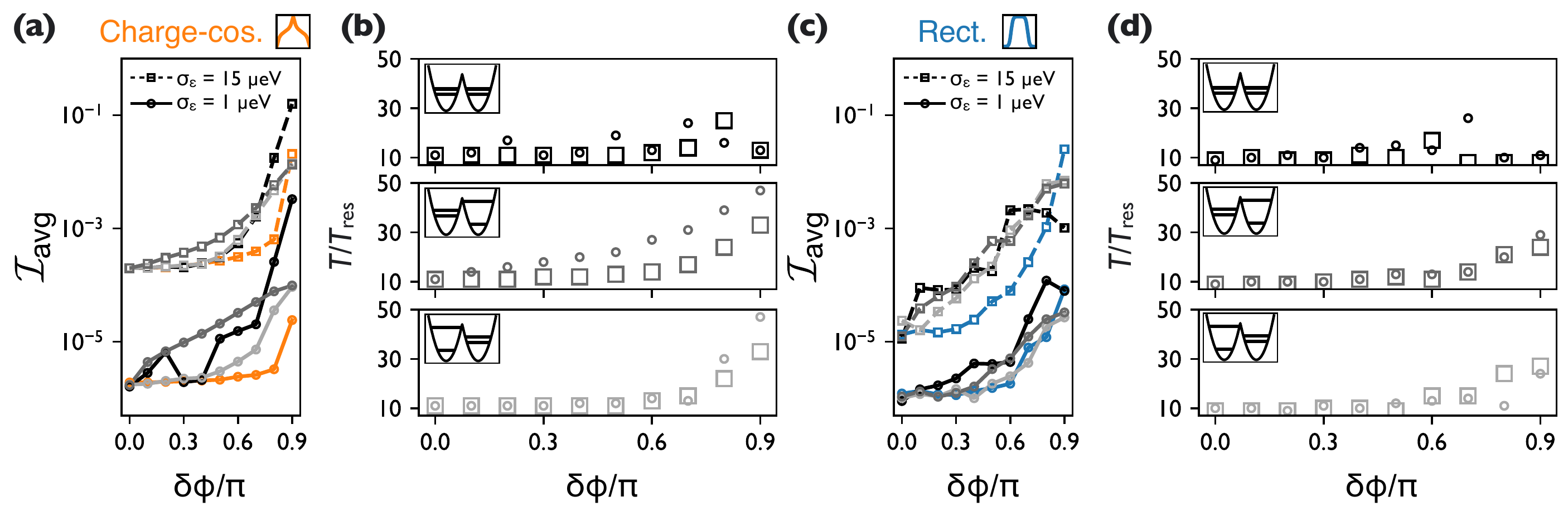}
    \caption{Comparison of infidelities and optimization results for different pulse shapes. For completeness, we include the same data as shown in Fig.~\ref{fig:valley_difference} of the main text, for the charge-cosine pulse family (a)-(b) and the rectangular pulse family (c)-(d). (a) Expected qubit infidelities due to charge noise computed with Eq.~(\ref{eq:tot_inf}), as we vary $\delta \phi$ from $0$ to $0.9\pi$, for valley splittings $E_{vL} = E_{vR} = 100$~\SI{}{\micro\electronvolt} [orange; same data as in Fig.~\ref{fig:phase_dependence}(a)], $E_{vL} = 100$~\SI{}{\micro\electronvolt} and $E_{vR} = 20$~\SI{}{\micro\electronvolt} [light gray lines], $E_{vL} = 20$~\SI{}{\micro\electronvolt} and $E_{vR} = 100$~\SI{}{\micro\electronvolt} [dark gray lines], and $E_{vL} = E_{vR} = 20$~\SI{}{\micro\electronvolt} [black lines]. We include data for both the optimistic (circles) and pessimistic (squares) charge noise regimes. (b) The total pulse times, for each of the pulses illustrated in (a). (c)-(d) The same data as in (a) and (b), for the rectangular pulse family. }
    \label{fig:valley_diff_supp}
\end{figure*}

\begin{figure}
    \centering
    \includegraphics[width=8cm]{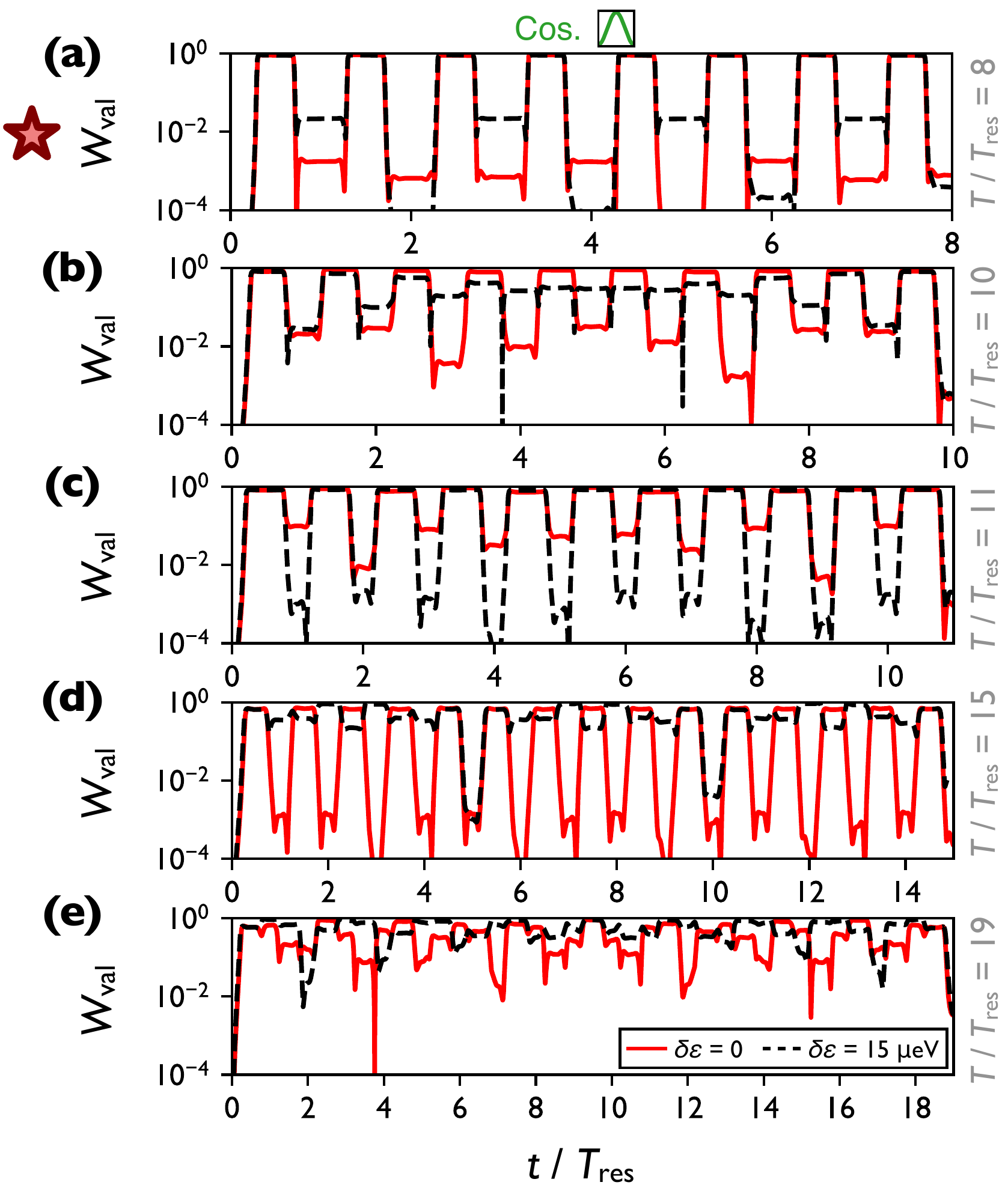}
    \caption{In very unfavorable valley conditions, complicated patterns of valley interference emerge. For each of the five pulses with minimal cost function $\mathcal{C}$ assuming $E_{vL} = E_{vR} = 20$~\SI{}{\micro\electronvolt} and $\delta \phi = 0.9\pi$, we plot the fraction of the wavefunction weight in the first excited valley eigenstate, $W_\text{val}$ as defined in Appendix~\ref{app:pulse_behavior}, for optimized pulses from the cosine family. The pulse plotted in (a), for $T = 8 T_\text{res}$, was the pulse ultimately selected by our algorithm and marked with a red star. For each of the five cases, we plot the excited valley weight assuming no detuning fluctuations, $\delta \varepsilon = 0$ (red solid lines), and a detuning fluctuation $\delta \varepsilon = 15$~\SI{}{\micro\electronvolt} (black dashed lines).}
    \label{fig:pulse_eigs_unfavorable}
\end{figure}

In this Appendix, we provide further analysis of the optimized pulses in favorable and unfavorable valley conditions.
In Fig.~\ref{fig:valley_diff_supp}, for both the charge-cosine (a-b) and rectangular (c-d) pulse families, we plot the average infidelity due to detuning fluctuations [$\mathcal{I}_\text{ave}$ defined in Eq.~(\ref{eq:tot_inf})] as a function of valley phase difference $\delta \phi$, for valley configurations $E_{vL} = 100$, $E_{vR} = 20$~\SI{}{\micro\electronvolt} (light gray), $E_{vL} = 20$, $E_{vR} = 100$~\SI{}{\micro\electronvolt} (dark gray), and $E_{vL} = E_{vR} = 20$~\SI{}{\micro\electronvolt} (black).
This is the same data as in the main text Fig.~\ref{fig:valley_difference}, for the charge-cosine and rectangular pulse families.
Qualitatively, all pulse families behave similarly. 
Worse valley configurations lead to worse fidelities, and larger $\delta \phi$ lead to worse fidelities and longer pulses.
We understand this as a leakage-noise tradeoff, as discussed in the main text.

We note that, in the very worst cases, such as $E_{vL} = E_{vR} = 20$~\SI{}{\micro\electronvolt} and $\delta \phi = 0.9\pi$, the optimized pulses are much shorter, as we see in Fig.~\ref{fig:valley_difference}(b) and Fig.~\ref{fig:valley_diff_supp}(b) and (d).
This may be due to a number of effects.
It is possible that the complicated patterns of valley interference make optimization difficult, and we are only able to find local minima in the cost function for long pulses. 
However, shorter pulses with fewer pulse oscillations may actually be less sensitive overall to detuning fluctuations in this regime.
In very unfavorable valley configurations, valley excitations are nearly unavoidable.
For example, in Fig.~\ref{fig:pulse_eigs_unfavorable}, we analyze the cosine pulse family, assuming the unfavorable valley configuration $E_{vL} = E_{vR} = 20$~\SI{}{\micro\electronvolt} and $\delta \phi = 0.9\pi$, under pessimistic charge noise $\sigma_\varepsilon = 15$~\SI{}{\micro\electronvolt}.
Assuming our qubit starts in the ground state, we plot the wavefunction weight in the first two excited valley eigenstates $W_\text{val} = W_2 + W_3$, where $W_n$ are defined in the main text.
We analyze all five of the total pulse lengths that yielded the lowest overall cost $\mathcal{C}$, ranging from 8 to 19 resonant periods.
(The shortest pulse, shown in (a) for $T = 8 T_\text{res}$, is the pulse selected by our algorithm and marked with a red star in Figs.~\ref{fig:loss_components}(b) and \ref{fig:pulse_eigs_unfavorable}(a).)
First, we plot the excited valley weight assuming zero detuning fluctuations, $\delta \varepsilon = 0$ (red solid lines).
We observe that, for all pulses regardless of length, the excited valleys are populated with weight near unity for a significant fraction of the pulse duration.
This is due to the particularly poor valley parameters considered here, and it is obviously very bad for pulse fidelity, as we see in Fig.~\ref{fig:valley_difference}.
Moreover, as the pulses get longer, more complicated patterns of valley interference emerge, which the pulse optimization algorithm must account for.
This picture is further complicated by the presence of detuning fluctuations.
In Fig.~\ref{fig:pulse_eigs_unfavorable}, we also plot the excited valley weight assuming a quasistatic detuning fluctuation $\delta \varepsilon = 15$~\SI{}{\micro\electronvolt} (black dashed lines).
We note that the valley interference pattern changes dramatically, compared with the $\delta \varepsilon = 0$ case.
This is true for all pulse lengths shown in Fig.~\ref{fig:pulse_eigs_unfavorable}, even the longer pulses that drive more slowly across the valley anticrossing.
Because the valley parameters are so unfavorable, there is no advantage to longer pulses in this case.
Of course, for very long and very weak pulses, we expect valley leakage to eventually die away. 
But in this case, charge noise will dominate the infidelity, resulting in worse pulses overall.

\section{Heterostructure parameters} \label{app:heterostructure_design}

Here, we elaborate on the parameters of the heterostructures we consider in the main text.
Following Ref.~[\citenum{Losert:2023p125405}], we define the expected Si concentration at position $z$ according to the following sigmoid function:
\begin{equation} \label{eq:avg_si_conc}
    \bar X_l = X_w + \frac{X_s - X_w}{1 - \exp[(z - z_t)/\tau]} + \frac{X_s - X_w}{1 - \exp[(z_b - z)/\tau]},
\end{equation}
where $z_{t(b)}$ mark the position of the top (bottom) QW interface, and $\lambda =  4\tau$ is the characteristic interface width.
We use interface width $\lambda = 1$~\SI{}{\nano\meter} and well width $|z_t - z_b| = 10$~\SI{}{\nano\meter} in this work.
In cases where the minimum Ge concentration $G_\text{min} = 1 - X_w > 0$, we set $X_s = X_w - 0.3$, maintaining a 30\% Ge concentration difference between the barrier and quantum well regions.

\section{Effective mass simulations} \label{app:effective_mass}

In this work, we use both 1D and 3D effective mass simulations. 
We use the 1D simulations to calculate $\sigma_\Delta$ (and therefore the average valley splitting) for the heterostructures considered in this work. 
In Appendices~\ref{app:orbital} and~\ref{app:tunnel_coupling}, we use 3D effective mass simulations to model the local ground-state energy of a single dot and the tunnel coupling in a double dot, respectively.
In this Appendix, we outline these effective mass simulations.

\subsection{1D simulations}

To obtain $\sigma_\Delta$ for a given heterostructure according to Eq.~(\ref{eq:sigma_delta}), we need to compute the envelope function $\psi_\text{env}$.
To do so, a 1D effective mass model suffices.
We solve the 1D effective mass Hamiltonian
\begin{equation} \label{eq:ham_em_1d}
    H_\text{EM}^\text{1D} = T_\text{1D} + U_\phi + U_\text{qw},
\end{equation}
where $T_\text{1D}$ is a discretized 1D kinetic energy operator, $U_\phi = e E_z z$ is the potential due to a vertical electric field $E_z$, and $U_\text{qw}$ is the quantum well potential.
We assume $E_z = 5$~\SI{}{\milli\electronvolt\per\nano\meter}.
The quantum well potential at position $\mathbf{r}$ is given by
\begin{equation} \label{eq:qw_pot}
    U_\text{qw} (\mathbf{r}) = \frac{X(\mathbf{r}) - X_s}{X_w - X_s} \Delta E_c,
\end{equation}
where $X(\mathbf{r})$ is the Si concentration at position $\mathbf{r}$.
In the 1D model, we can replace $\mathbf{r} \rightarrow z_l$, where $l$ is the layer index, and we use Eq.~(\ref{eq:avg_si_conc}) to define the Si concentration profile.
Following Refs.~[\citenum{Wuetz:2022p7730, Losert:2023p125405}], the conduction band offset is given by
\begin{multline} 
\Delta E_c = (X_w - X_s) \left[ \frac{X_w}{1-X_s}\Delta E_{\Delta_2}^\text{Si}(X_s) \right. \\
\left. - \frac{1 - X_w}{X_s} \Delta E_{\Delta_2}^\text{Ge}(X_s)  \right],
\end{multline}
where $\Delta E_{\Delta_2}^{\text{Si}(\text{Ge})}$ are the $\Delta_2$ conduction band offsets for strained Si (Ge) grown on unstrained $\text{Si}_X \text{Ge}_{1-X}$ substrate [\citenum{Schaffler:1997p1515}], approximated as
\begin{equation}
\begin{split}
\Delta E_{\Delta_2}^{\text{Si}}(X) &= -0.502 (1-X) \; (\text{eV}) \\
\Delta E_{\Delta_2}^{\text{Ge}}(X) &= 0.743 -0.625 (1-X) \; (\text{eV}).
\end{split}
\end{equation}
By diagonalizing Eq.~(\ref{eq:ham_em_1d}), we obtain the envelope function and can compute $\sigma_\Delta$ [defined in Eq.~(\ref{eq:sigma_delta})], which equals \SI{49}{\micro\electronvolt} for conventional QWs and \SI{360}{\micro\electronvolt} for 5\% Ge QWs, as discussed in the main text.
We note that, to compute $\psi_\text{env}$, we use the virtual crystal approximation and do not include random alloy disorder in the quantum well potential.
In this computation, unless otherwise specified, we assume the lateral confinement potential is parabolic and isotropic with characteristic level spacing $\hbar \omega_\text{orb} = 2$~\SI{}{\milli\electronvolt}.

\subsection{3D simulations}
We also employ 3D effective mass simulations to describe the local chemical potential and tunnel coupling fluctuations due to alloy disorder.
We note that these simulations contain no valley physics, since their purpose is only to describe the orbital energies of the system.
Following the methods of Ref.~[\citenum{Losert:2023p125405}], we use a coarse-grained model of the Si/SiGe heterostructure in three dimensions.
We discretize the heterostructure into cells of size $(\Delta x, \Delta y, \Delta z) = (a_0, a_0, a_0/4)$, where $a_0 = $~\SI{0.543}{\nano\meter} is the Si lattice constant.
In this model, each cell contains exactly two Si/Ge atoms.
On this discretized lattice, we solve the following effective-mass Hamiltonian:
\begin{equation} \label{eq:EM_ham}
    H_\text{EM}^\text{3D} = T_\text{3D} + U_\text{conf} + U_\phi + U_\text{qw},
\end{equation}
where $T_\text{3D}$ is the discretized 3D kinetic energy operator and $U_\text{qw}$ is the quantum well potential, given in Eq.~(\ref{eq:qw_pot}).
The lateral confinement potential $U_\text{conf}$ depends on whether we consider a single or a double-dot system.
For example, for a single quantum dot, we use the confinement potential
\begin{equation} \label{eq:single_dot_lateral_conf}
U_\text{conf}^\text{sd} = \frac{1}{2} m_t \omega_\text{orb}^2 \left((x-x_0)^2 + y^2 \right),
\end{equation}
where the dot is centered at $(x_0, 0)$. 
Again, unless otherwise specified, we consider single-dot unperturbed orbital splittings $\hbar \omega_\text{orb} = 2$~\SI{}{\milli\electronvolt}, and vertical fields $E_z = 5$~\SI{}{\milli\volt\per\nano\meter}.
We elaborate on the double-dot system in Appendix~\ref{app:tunnel_coupling}.

The quantum well potential is given by Eq.~(\ref{eq:qw_pot}), where we define $X(x_j, y_k, z_l) \coloneqq X_{j,k,l}$ as the Si concentration of a cell at $\mathbf{r} = (x_j, y_k, z_l)$, labeled with indices $(j,k,l)$.
In a system with alloy disorder, the atoms in any given cell will be randomly assigned as Si or Ge, where the probability of choosing a Si atom is given by the average (or expected) Si concentration at the position of the cell.
To separate the effects of alloy disorder from those intrinsic to the shape of the interface, we decompose $U_\text{qw}$ into two components, one due to the average Si concentrations in a given cell ($\bar U_\text{qw}$), and one due to the random fluctuations of the Si concentration in a given cell due to random alloy disorder ($U_\text{qw}^\text{dis}$).
Hence, $U_\text{qw} = \bar U_\text{qw} + U_\text{qw}^\text{dis}$, where
\begin{equation} \label{eq:qw_pot_split}
    \begin{split}
    \bar U_\text{qw}(x_j, y_k, z_l) &= \frac{\bar X_{j,k,l} - X_s}{X_w - X_s} \Delta E_c \\
    U_\text{qw}^\text{dis}(x_j, y_k, z_l) &= \frac{\delta X_{j,k,l} - X_s}{X_w - X_s} \Delta E_c,
    \end{split}
\end{equation}
where $\bar X_{j,k,l}$ is the expected Si concentration in a cell, and $\delta X_{j,k,l} = X_{j,k,l} - \bar X_{j,k,l}$ is the difference between the actual and expected Si concentrations in a cell.

We model the top and bottom quantum well interfaces as smooth sigmoid functions.
Without interface steps, the expected Si concentration at layer $l$, $\bar X_l$, is give by Eq.~(\ref{eq:avg_si_conc}).
With an interface step, we model the expected Si concentration for a cell at $(x_j, y_k, z_l)$ as 
\begin{equation} \label{eq:concs_step}
    \bar X_{j,k,l} = \bar X_l \Theta (x_j \geq x_\text{step}) + \bar X_{l+1} \Theta(x < x_\text{step}),
\end{equation}
where $x_\text{step}$ is the lateral position of the interface step, and $\Theta(\cdot)$ is the Heaviside step function.
Here, we consider a step oriented along the $\hat y$ direction.

Finally, in Appendices~\ref{app:orbital} and~\ref{app:tunnel_coupling}, we consider systems both with and without alloy disorder. 
To model a system without alloy disorder, we use the virtual crystal approximation, setting $U_\text{qw}^\text{dis} = 0$.
In systems with alloy disorder, we populate each cell with exactly two atoms.
The probability each atom in a cell is Si is given by $\bar X_{i,j,k}$, the expected Si concentration at that cell.
The resulting Si concentration in the cell defines $X_{i,j,k}$, as used in Eq.~(\ref{eq:qw_pot}).

\section{Local ground-state energy fluctuations} \label{app:orbital}

\begin{figure}
    \centering
    \includegraphics[width=8cm]{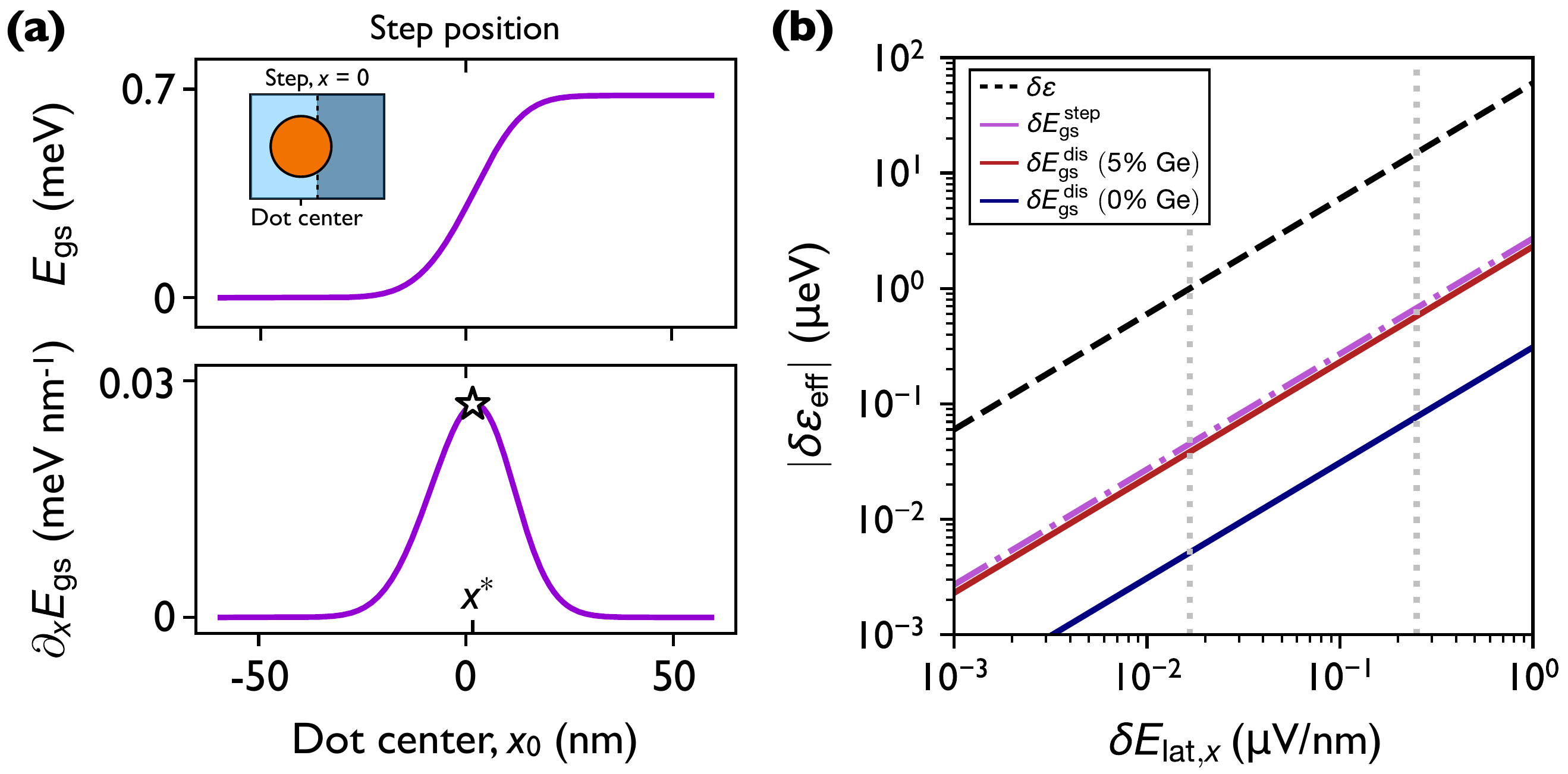}
    \caption{Comparison of effective detuning fluctuations due to the direct effects of fluctuating lateral fields and indirect effects due to alloy disorder and interface steps. (a) Upper panel: the ground-state energy of a single dot, centered at $x_0$, as it moves across a single interface step at $x_\text{step} = 0$, simulated with a 3D effective mass model using the virtual crystal approximation. Lower panel: the gradient of the potential energy with respect to the dot center, $\partial_{x_0} E_\text{gs}$, for the same system as in the upper panel, again computed with 3D effective mass simulations. (b) For a lateral field fluctuation along the $\hat x$ direction, we plot the resulting direct detuning fluctuations for the double dot (black dashed line), as well as the indirect effective detuning fluctuations due to alloy disorder, for both conventional (blue) and 5\% Ge QWs (red), and step disorder (pink dot-dashed line). We emphasize that the direct contribution to the fluctuations far exceeds the indirect contributions. }
    \label{fig:orb_variation}
\end{figure}

In this section, we elaborate on how fluctuations to the ground-state energy due to local disorder affect flopping mode qubit fidelities. 
In the presence of a quasistatic lateral field fluctuation $\delta \textbf{E}_\text{lat} = \delta E_{\text{lat},x} \hat x + \delta E_{\text{lat},y} \hat y$, there is a corresponding fluctuation in the inter-dot detuning, $\delta \varepsilon = e d \delta E_{\text{lat},x}$, where $d$ is the inter-dot spacing, and we have assumed the detuning axis of the double dot system is along $\hat x$.
Here, we assume an inter-dot spacing $d = 60$~\SI{}{\nano\meter} (see Appendix~\ref{app:tunnel_coupling}).
In Fig.~\ref{fig:orb_variation}(b), we plot this detuning fluctuation as a function of $\delta E_{\text{lat},x}$ (black dashed line).
However, this detuning fluctuation is not the only possible effect.
In the presence of fluctuating lateral fields, the center of a dot is shifted by a small amount $\delta x$, as described in Eq.~(\ref{eq:delta_x_E_field}) of the main text.
These lateral shifts cause the dot to sample slightly different disorder landscapes, both due to the presence of steps in the quantum well interface and alloy disorder.
This disorder can cause local fluctuations to the ground-state energy of the dot.
If these energy fluctuations are different for each dot in the double-dot qubit, the result is an effective detuning fluctuation, $\delta \varepsilon_\text{eff} = \delta E_\text{gs,L} - \delta E_\text{gs,R}$, where $\delta E_{\text{gs,L/R}}$ labels the ground-state energy shifts in the left/right dot.
We examine the effects of both sources below, where we find both steps and alloy disorder have much smaller impacts on the effective detuning than the direct detuning fluctuations described above.
Thus, we do not expect the local ground-state energy fluctuations due to alloy disorder or interface steps to materially impact flopping mode qubit fidelities.

\subsection{Alloy disorder}

As described above, alloy disorder can induce small shifts in the ground-state energy in the presence of fluctuating lateral electric fields.
From effective mass theory, the first-order correction to the ground-state energy due to alloy disorder is given by
\begin{equation} \label{eq:orb_dis}
     E_\text{gs}^{\text{dis}} = \int d\mathbf{r} |\psi_0(\mathbf{r})|^2 U_\text{qw}^\text{dis} (\mathbf{r}),
\end{equation}
where $\psi_0$ is the zeroth order envelope function, and $U_\text{qw}^\text{dis}$ is defined in Eq.~(\ref{eq:qw_pot_split}).
We assume $\psi_0(\mathbf{r}) = \phi_{0}(x) \phi_{0}(y) \psi_z(z)$, where $\phi_0$ is the ground state wavefunction of a 1D harmonic oscillator confinement potential, and $\psi_z$ is the ground state of the z-confinement potential, determined by the quantum well profile and the vertical field (see Appendix~\ref{app:effective_mass}).
Since we consider a coarse-grained model of heterostructure, we can replace the integral with a sum, according to the formula
\begin{equation} \label{eq:int_to_sum}
    \int d\mathbf{r} \rightarrow \frac{a_0^3}{4} \sum_{x_j,y_k,z_l},
\end{equation}
where the sum is over all cells in the heterostructure.
To understand the statistical properties of $E_\text{gs}^\text{dis}$, following Ref.~\citenum{Losert:2023p125405}] we take the variance in the resulting quantity, which we compute to be
\begin{equation} \label{eq:var_orb}
    \V [ E_\text{gs}^\text{dis} ] = \sigma_\Delta^2,
\end{equation}
where we have used
\begin{equation}
    a_0 \sum_{x_j} \phi_0^4(x_j) \approx \int dx \; \phi_0^4 (x) = \sqrt{\frac{m_t \omega }{2 \pi \hbar}}
\end{equation}
in evaluating the integral in Eq.~(\ref{eq:orb_dis}), and 
\begin{align}
    \V [ U_\text{qw}^\text{dis}(x_j, y_k, z_l)] & = \left(\frac{\Delta E_c}{X_w - X_s}\right)^2 \V [ \delta X_{j,k,l}] \\
    & = \frac{1}{2} \left( \frac{\Delta E_c}{X_w - X_s} \right)^2 \bar X_{j,k,l} (1 - \bar X_{j,k,l}).
\end{align}
Hence, energy fluctuation due to alloy disorder in any single dot is a normal random variable, with zero mean and variance given by Eq.~(\ref{eq:var_orb}).

Now, we introduce charge noise, in the form of fluctuating lateral electric fields.
As discussed above, these field fluctuations induce small shifts in the dot position $\delta x$, given by Eq.~(\ref{eq:delta_x_E_field}).
Following the methods of Ref.~[\citenum{Losert:2023p125405}] it can be shown that the gradient $\partial_x E_\text{gs}^\text{dis}$ is also a random variable, with variance given by $\V [ \partial_x E_\text{gs}^\text{dis} ] = \sigma_\Delta^2 / a_\text{dot}^2$.
So, in response to a small lateral perturbation $\delta x$, the resulting effective detuning fluctuation $\delta E_\text{gs}^\text{dis} =  E_\text{gs}^\text{dis} (\delta x) - E_\text{gs}^\text{dis} (0)$ will have the variance
\begin{equation}
    \V [ \delta E_\text{gs}^\text{dis}] = \V [ \delta x \; \partial_x  E_\text{gs}^\text{dis}] = \frac{\delta x^2 \sigma_\Delta^2}{a_\text{dot}^2}.
\end{equation}
Finally, since each dot in the double-dot qubit will have a nearly-uncorrelated alloy disorder profile, it will experience a different change in potential energy. 
The resulting effective detuning fluctuation $\delta \varepsilon_\text{eff} = \delta E_{\text{gs,L}}^\text{dis} - \delta  E_{\text{gs,R}}^\text{dis} $ will have variance
\begin{equation} \label{eq:var_dis_detun}
    \V [\delta \varepsilon_\text{eff}^\text{dis} ] = \frac{2 \delta x^2 \sigma_\Delta^2}{a_\text{dot}^2}.
\end{equation}
Once again, this is a normal random variable with zero mean and variance given by Eq.~(\ref{eq:var_dis_detun}).
Since these fluctuations can be positive or negative, we consider the magnitude $| \delta \varepsilon_\text{eff}^\text{dis} |$.
The distribution of $|\delta \varepsilon_\text{eff}^\text{dis} |$ is folded-normal, with mean 
\begin{equation} \label{eq:mean_abs_detun_dis}
    \E [ | \delta \varepsilon_\text{eff}^\text{dis} | ] = \sqrt{\frac{2}{\pi} \V [\delta \varepsilon_\text{eff}^\text{dis} ]} = 2\sqrt{\frac{2}{\pi}} \frac{\delta x \sigma_\Delta}{a_\text{dot}}.
\end{equation}
The predictions of Eq.~(\ref{eq:mean_abs_detun_dis}) for a given lateral field fluctuation $E_{\text{lat},x}$ are shown in Fig.~\ref{fig:orb_variation}(b), for both the conventional QW (blue) and the 5\% Ge QW (red). 
We see that the effects of alloy disorder are much smaller than the effects of direct detuning fluctuations for a double dot.

\subsection{Interface steps} 

Now, we consider the effect of an interface step on the effective detuning fluctuations in a flopping mode qubit. 
First, we determine the effects of an interface step on the ground-state energy of a single quantum dot.
We obtain the ground state energy $ E_\text{gs} $ by diagonalizing the effective mass Hamiltonian Eq.~(\ref{eq:EM_ham}), varying the dot center $x_0$ relative to the step at $x_\text{step} = 0$ [illustrated schematically in the inset of Fig.~\ref{fig:orb_variation}(a)].
We use the virtual crystal approximation described in Appendix~\ref{app:effective_mass} to avoid the effects of alloy disorder.
We plot the resulting values of $E_\text{gs} (x_0)$ in the top panel of Fig.~\ref{fig:orb_variation}(a) [solid line].
We see that, on either side of the step, the energy levels out, but near the step, $E_\text{gs}$ shifts by about \SI{0.7}{\milli\electronvolt}.
This offset on either side of the step is due to the constant vertical electric field: as the dot is pushed over the step, it's potential energy increases within this field gradient.
To confirm this, we note that a single-layer step will displace the dot $a_0 / 4 \approx 0.136$~\SI{}{\nano\meter}, which will increase the potential energy by \SI{0.679}{\milli\electronvolt} in a \SI{5}{\milli\volt\per\nano\meter} field gradient, agreeing well with our simulations.

Given $E_\text{gs}$ as a function of dot position $x$, we can also compute the gradient relative to the dot center, $\partial_{x_0} E_\text{gs}$, plotted in the lower panel of Fig.~\ref{fig:orb_variation}(a) [solid line].
The gradient peaks at the position labeled $x^*$, marked with a star in Fig.~\ref{fig:orb_variation}, which is nearly equal to $x_\text{step}$.
Given a lateral electric field fluctuation $\delta E_{\text{lat},x}$, the dot shifts by an amount $\delta x$, given in Eq.~(\ref{eq:delta_x_E_field}).
We can estimate the resulting detuning fluctuation at the position of maximal gradient, 
\begin{equation} \label{eq:delta_orb_step}
    \delta E_\text{gs} = \delta x \partial_{x_0} E_\text{gs} |_{x_0 = x^*}
\end{equation}
Furthermore, if we assume one dot in the double-dot qubit is centered along a step, Eq.~(\ref{eq:delta_orb_step}) describes an effective detuning shift $\delta \varepsilon_\text{eff}$ between the two dots.
This detuning shift is plotted in Fig.~\ref{fig:orb_variation}(b) [pink dot-dashed line] as a function of $\delta E_{\text{lat},x}$.
The resulting $\delta \varepsilon_\text{eff}$ is about as large as the expected detuning fluctuation due to alloy disorder in the 5\% Ge QW.
However, even at the position of the maximal $E_\text{gs}$ gradient, this shift is still about an order of magnitude smaller than the shift due to direct E-field induced detuning fluctuations.

\section{Tunnel coupling fluctuations} \label{app:tunnel_coupling}

\begin{figure*}
    \centering
    \includegraphics[width=14cm]{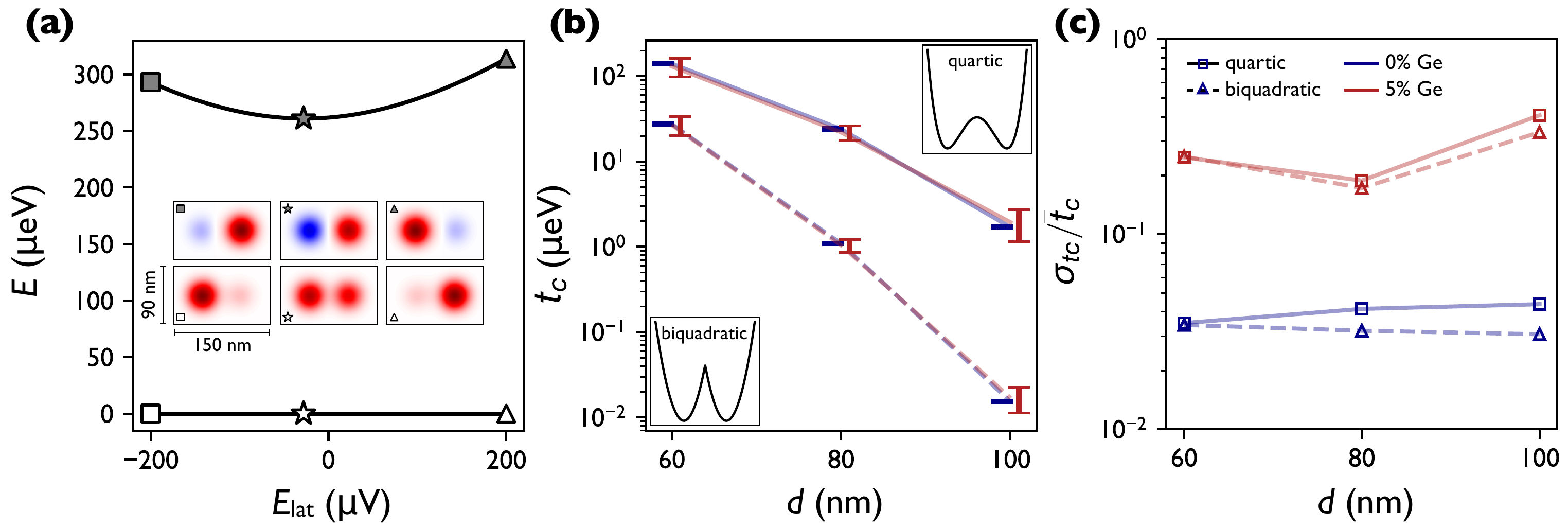}
    \caption{Computation of the tunnel coupling distribution using effective mass simulations. (a) An example diagram of the ground and first-excited orbital energies, for a single alloy disorder configuration, as we modulate the lateral field $E_\text{lat}$. Inset: the ground (open markers) and first-excited (closed markers) wavefunctions, plotted at three different $E_\text{lat}$ configurations, indicated by their shapes. (b) Error bars represent the 25-75 percentile range of 20 simulations of $t_c$, where each simulation has a different random-alloy disorder configuration, as we modulate the inter-dot distance $d$. We consider conventional QWs (blue) and QWs with 5\% Ge (red), and we consider quartic (upper series, solid lines) and biquadratic (lower series, dashed lines) double-dot potentials. Connecting lines are a guide to the eye. (c) For each configuration in (b), we plot the standard deviation of 20 tunnel coupling calculations ($\sigma_{tc}$), normalized by the average of the 20 calculations ($\bar t_c$). }
    \label{fig:tc_var_dis}
\end{figure*}

As well as fluctuations in the effective double-dot detuning, fluctuating lateral electric fields can also produce shifts in the double-dot tunnel coupling $t_c$.
In this section, we estimate the size of these shifts.
We stress that these are order-of-magnitude approximations to the fluctuations present in real devices. 
A true quantitative treatment of tunnel coupling in the presence of charge disorder is beyond the scope of this work.

If charge impurities are distant from the qubit, the effective lateral field fluctuations are constant across the double-dot.
As in Appendix~\ref{app:orbital}, these cause the double-dot system to displace slightly, sampling a slightly different disorder profile, leading to indirect tunnel coupling fluctuations, $\delta t_c$.
Since the dot potential minima are shifted by the same amount in the same direction, we ignore direct $t_c$ shifts in this case.
However, if charge impurities are nearby, lateral field fluctuations may displace two dots differently, effectively modulating the inter-dot spacing and causing direct shifts to the tunnel coupling.
In this Appendix, we estimate the resulting tunnel coupling fluctuations due to alloy disorder, interface steps, and inter-dot distance modulations.
In all cases, as expected, we find tunnel coupling fluctuations are not the dominant source of flopping mode qubit infidelity.

\subsection{Alloy disorder}

\begin{figure*}
    \centering
    \includegraphics[width=14cm]{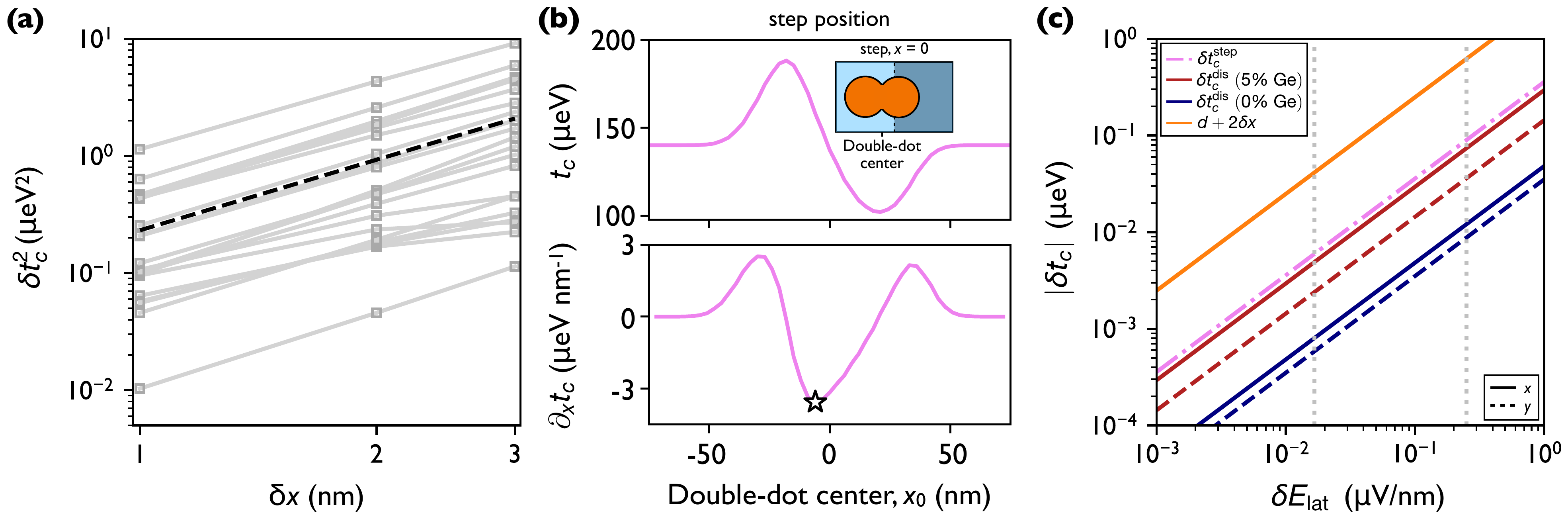}
    \caption{Estimating the size of tunnel coupling fluctuations due to charge noise. (a) Starting with a double-dot centered at $(x_0, y_0)$, we move the system slightly along the $\hat x$ direction, computing the resulting change in tunnel coupling, $\delta t_c$, as described in Appendix~\ref{app:tunnel_coupling}. We plot $\delta t_c^2$ against $\delta x$ for 20 instantiations of random-alloy disorder, as well as the best-fit line for $\E[\delta t_c^2]$ [see Eq.~(\ref{eq:tc_fluc_fit})]. For these simulations, we use a conventional QW, quartic double-dot potential, and inter-dot distnace $d = 60$~\SI{}{\nano\meter}. (b) For a system without alloy disorder, we compute $t_c$ as the double-dot potential moves across a single step in the QW interface at $x_\text{step} = 0$ (upper panel). We also indicate the gradient of $t_c$ relative to the double-dot center, $\partial_{x_0} t_c$ (lower panel). We indicate the position of maximal sensitivity to fluctuations in $x_0$ with a star. (c) For a given lateral field fluctuation $\delta E_\text{lat}$, we estimate the resulting $t_c$ fluctuation due to steps ($\delta t_c^\text{step}$) and alloy disorder ($\delta t_c^\text{dis}$). First, we consider a double-dot with a single interface step, at the position of maximal sensitivity to changes in $x_0$ (pink dot-dashed line), ignoring the effects of disorder. We also consider systems with random-alloy disorder and without interface steps, for fluctuations along $\hat x$ (solid lines) and $y$ (dashed lines), for conventional QWs (blue) and 5\% Ge QWs (red). Finally, we include the $\delta t_c$ due to fluctuations in the inter-dot position (orange line), where we assume $\delta E_\text{lat}$ is equal and opposite in each dot, arranged such that the inter-dot distance $d \rightarrow d + 2 \delta x$, for $\delta x$ defined in Eq.~(\ref{eq:delta_x_E_field}).}
    \label{fig:tc_comparison}
\end{figure*}

First, we evaluate the role of alloy disorder on the tunnel coupling in a double dot.
To do so, we consider two models of the lateral confinement potential: biquadratic and quartic, following Ref.~[\citenum{Li:2010p085313}], given by the following
\begin{equation} \label{eq:dd_conf}
    \begin{split}
        U_\text{bi} &= \frac{1}{2} m_t \omega^2 \Bigg\{ \text{Min} \left[ \left(x - x_0 - \frac{d}{2}\right)^2, \left( x - x_0 + \frac{d}{2} \right)^2 \right]   \\
        & \;\;\;\; \;\;\;\; \;\;\;\; \;\;\;\; + (y - y_0)^2 \Bigg\} \\
        U_\text{qu} &= \frac{1}{2} m_t \omega^2 \left[ \frac{1}{d^2} \left( (x - x_0)^2 - \frac{d^2}{4}\right)^2 + \left( y - y_0\right)^2 \right],
    \end{split}
\end{equation}
where the detuning axis is along $\hat x$, $d$ is the inter-dot spacing, the double-dot system is centered at $(x_0, y_0)$, and the orbital energy splitting in each dot is approximately given by $\hbar \omega = 2$~\SI{}{\milli\electronvolt}.
These potentials are illustrated schematically in Fig.~\ref{fig:tc_var_dis}(b).
We note that these models are fairly crude, and real devices allow for control over individual plunger gates and tunnel barriers.
Nonetheless, they allow us to make estimates of the impact of disorder on tunnel couplings.

The conventional way to measure the tunnel coupling in a double-dot system is to find the zero-detuning point and estimate the charge state energy gap at that point.
The models of Eq.~(\ref{eq:dd_conf}), however, have no explicit control of the inter-dot detuning.
In order to control the effective inter-dot detuning, we also include in our Hamiltonian the potential due to a lateral electric field, 
\begin{equation}\label{eq:lat_conf}
U_\text{lat} = e E_\text{lat} x.
\end{equation}
By modulating $E_\text{lat}$, we can precisely control the detuning in our system, allowing us to estimate $t_c$ as described below.

To obtain the tunnel coupling for a given disordered crystal lattice, we need to measure the energy gap between symmetric and antisymmetric charge states at the charge anticrossing.
To do so, we diagonalize the effective mass Hamiltonian Eq.~(\ref{eq:EM_ham}), with the double-dot confinement potential given by Eq.~(\ref{eq:dd_conf}) and lateral field potential given by Eq.~(\ref{eq:lat_conf}), obtaining energies for the ground and first-excited orbital states, $E_g$ and $E_e$.
In order to find the precise center of the anticrossing, we adjust the lateral field strength $E_\text{lat}$ to find the configuration that minimizes the ground state energy gap, $\Delta E = E_e - E_g$.
(This procedure is analogous to adjusting the relative detuning, until we reach the zero-detuning point.)
The tunnel coupling is approximately half this gap, $t_c \approx \Delta E / 2$.
This procedure is illustrated for one example in Fig.~\ref{fig:tc_var_dis}(a), where we have plotted both $E_g$ and $E_e$ as we modulate $E_\text{lat}$. 
We also plot the ground state wavefunction (open markers) and the first excited state wavefunctions (closed markers) at three different configurations.
We see that at the position of minimal $\Delta E$, indicated with stars, the wavefunctions are given by symmetric and antisymmetric combinations of left- and right-dot charge states with nearly equal weights.
Using this procedure, we compute $t_c$ for 20 instantiations of random-alloy disorder.

We perform the above procedure for inter-dot spacings $d = 60$, $80$, and $100$~\SI{}{\nano\meter}, for both the quartic and biquadratic confinement potentials defined in Eq.~(\ref{eq:dd_conf}). 
We consider both conventional QWs (blue) and QWs with 5\% Ge (red), and the resulting $t_c$ distributions are plotted in Fig.~\ref{fig:tc_var_dis}(b), where error bars represent the $25-75$ percentile range across the 20 instantiations of random-alloy disorder.
The upper series of data (connected with solid lines) indicate results for the quartic potential, and the lower series (dashed lines) the biquadratic potential.
We observe that the average $t_c$ varies over many orders of magnitude for the inter-dot spacings considered here.
This is the expected behavior, given the exponential dependence of $t_c$ on $d$. 
We also see that the variation in $t_c$ is much larger in the QWs with 5\% Ge than the conventional QWs.
To illustrate this further, we plot the size of $\sigma_{tc}$, the standard deviation of the $t_c$ distribution, relative to the average $\bar t_c$ in Fig.~\ref{fig:tc_var_dis}(c), showing that it is roughly independent of $d$.
We see that for both quartic and biquadratic models, $\sigma_{tc}$ is about 20\% of $\bar t_c$ in the 5\% Ge heterostructures, while only about 3\% of $\bar t_c$ for the conventional QWs.

Next, we consider the effects of charge noise.
We assume that a lateral field fluctuation $\delta E_\text{lat}$ will slightly modify the center coordinates $(x_0, y_0)$ of the double-dot qubit, causing the qubit to sample a slightly different disorder landscape.
Again, we assume $\delta E_\text{lat}$ is constant across the double-dot, which is true if the charge defects producing $\delta E_\text{lat}$ are far from the qubit.
We consider a quartic potential with dot spacing $d = 60$~\SI{}{\nano\meter}, which produces an average $\bar t_c$ of about \SI{100}{\micro\electronvolt} [see Fig.~\ref{fig:tc_var_dis}(b)].
For each of the 20 alloy disorder configurations considered above, for both the conventional and 5\% Ge QW, we slightly perturb the double-dot center coordinates.
We consider perturbations in both the $x$- and $y$-directions.
By computing the change in tunnel couplings due to these lateral shifts, labeled $\delta t_c$, we can extract the statistics of $t_c$ fluctuations.
Like other disorder-influenced quantities, we assume $\V[\delta t_c] = \E [\delta t_c^2]$ scales quadratically with displacement (for small displacements) and is proportional to the underlying tunnel coupling variance, $\sigma_{tc}^2$.
Hence, we fit the data to the relationship
\begin{equation} \label{eq:tc_fluc_fit}
    \E[\delta t_c^2] = \sigma_{tc}^2 \delta x^2 / a_{tc,x}^2,
\end{equation}
where $\sigma_{tc}$ is determined empirically from the initial 20 simulations, and $a_{tc,x}$ is the fitting parameter.
The parameter $a_{tc,x}$ has units of distance, so this parameter captures the length scales over which $t_c$ varies in a double dot, and the subscript $x$ indicates that the displacement $\delta x$ occurred along $\hat x$.
We plot one dataset of $\delta t_c^2$ vs $\delta x^2$ in Fig.~\ref{fig:tc_comparison}(a), for small displacements along the $\hat x$ axis in a conventional QW.
We also plot the fit given by Eq.~(\ref{eq:tc_fluc_fit}) [black dashed line].
Clearly, the tunnel coupling fluctuations are well-described by Eq.~(\ref{eq:tc_fluc_fit}).
We repeat this procedure for displacements along the $\hat y$ direction as well, for both conventional and 5\% Ge QWs.
The resulting fit parameters are summarized in Table~\ref{table:1}.
We note that the fluctuation length scales are all on the order of the dot size.

\begin{table}[]
\begin{center}
\def\arraystretch{1.3}
\begin{tabular}{c S[table-format=2.1] S[table-format=2.1] S[table-format=1.2] S[table-format=1.2]}% syntax for siunitx v2; for v1 use "tabformat"
 \hline \hline
  & {$\bar t_c$~(\SI{}{\micro\electronvolt})} & {$\sigma_{tc}$~(\SI{}{\micro\electronvolt})} & {$a_{tc,x}$~(\SI{}{\nano\meter})} & {$a_{tc,y}$~(\SI{}{\nano\meter})}   \\ [0.5ex] 
 \hline
 0\% Ge & 140 & 4.9 & 10 & 14  \\ [1ex] 
 5\% Ge & 131 & 32 & 11 & 22  \\
 \hline \hline
\end{tabular}
\caption{Numerical parameters describing $t_c$ sensitivity due to lateral perturbations of the double-dot center coordinates, as described in Appendix~\ref{app:tunnel_coupling}.}
\label{table:1}
\end{center}
\end{table}

Finally, we can relate these tunnel coupling fluctuations to a lateral field fluctuation.
As before, we assume that a lateral field fluctuation $\delta E_\text{lat}$ shifts the dot by a small distance $\delta x$, as given in Eq.~(\ref{eq:delta_x_E_field}), resulting in a small shift $\delta t_c$.
In Fig.~\ref{fig:tc_comparison}(c), we plot $\delta E_\text{lat}$ vs the expected rms tunnel coupling fluctuations, for both conventional (blue) and 5\% Ge (red) QWs and displacements along $\hat x$ (solid lines) and $\hat y$ (dashed lines).
Again, we note the expected $\delta E_\text{lat}$ corresponding to our optimistic and pessimistic charge noise configurations.
We see that, for the charge noise regimes considered in this work, we expect tunnel coupling fluctuations much less than \SI{1}{\micro\electronvolt} due to alloy disorder.

\subsection{Interface steps}

Next, we investigate the role of interface steps in tunnel coupling fluctuations. 
As above, we use 3D effective mass simulations to determine $t_c$ for different step configurations.
In this case, we exclude alloy disorder by using the virtual crytal approximation.
Again, we use the quartic potential model and an inter-dot separation $d = 60$~\SI{}{\nano\meter} in Eq.~(\ref{eq:dd_conf}).
We plot the resulting $t_c$ as we move the double-dot across a step at $x_\text{step} = 0$ in Fig.~\ref{fig:tc_comparison}(b) [top panel].
In the bottom panel, we plot the resulting gradient, $\partial_{x_0} t_c$.
At the position of maximal gradient $x^*$, we find $| \partial_{x_0} t_c |_{x_0 = x^*}  \approx 3.6$~\SI{}{\micro\electronvolt\per\nano\meter}.
Then, using Eq.~(\ref{eq:delta_x_E_field}), we relate the $t_c$ fluctuation to a lateral field fluctuation, using
\begin{equation}
    \delta t_c \approx \delta x \partial_{x_0} t_c |_{x_0 = x^*}.
\end{equation}
The resulting $\delta t_c$ are plotted in Fig.~\ref{fig:tc_comparison}(c) [pink dot-dashed line].
For the charge noise regime considered here, we expect $\delta t_c \ll 1$~\SI{}{\micro\electronvolt}.

\subsection{Fluctuations in the inter-dot distance}

Lastly, we investigate the role of fluctuations in the inter-dot distance due to charge noise.
In the case where charge defects live near the qubit, each dot could experience electric field fluctuations in opposite directions.
These could have the effect of moving the two dots slightly closer or slightly farther apart, creating a ``direct'' modulation of $t_c$, as opposed to the disorder-induced ``indirect'' effects.
To estimate the effect these fluctuations would have on the tunnel coupling, we use the data in Fig.~\ref{fig:tc_var_dis}(b) to estimate how $t_c$ scales with distance.
Using the data for the 0\% Ge QWs in quartic potentials at $d = 60$ and \SI{80}{\nano\meter}, we approximate $\log t_c \approx -3.57 - 0.088d$, for $t_c$ in eV and $d$ in nm.
Using this relationship, we can estimate $\delta t_c$ for small shifts in $d$.
To relate shifts in $d$ to lateral field fluctuations $\delta E_x$, we again use Eq.~(\ref{eq:delta_x_E_field}).
This time, we assume the lateral field fluctuations are equal and opposite for each dot, so the double-dot spacing $d$ is shifted by $\pm 2 \delta x$, for $\delta x$ given by Eq.~(\ref{eq:delta_x_E_field}).
We plot the resulting $|\delta t_c|$ as a function of $\delta E_\text{lat}$, assuming $d \rightarrow d + 2 \delta x$, in Fig.~\ref{fig:tc_comparison}(c) [orange line].

In our simulations above, we only find $\delta t_c$ approaching \SI{1}{\micro\electronvolt} when a fluctuating lateral field has opposite direction in each dot.
Alloy disorder and interface steps produce $\delta t_c$'s that are about an order of magnitude smaller, given the same lateral fields. 
Of course, precise quantification of the true $\delta t_c$ distribution will require atomistic simulations including charge traps and other impurities, combined with realistic modeling of the gate potentials.
However, we view this as an approximate upper bound on $\delta t_c$.
Using these estimates of the noise-induced tunnel coupling variability, we can estimate an upper bound on the contribution of these fluctuations to pulse infidelity, which we do in the following section.
\newline

\subsection{Impact of $\delta t_c$ on gate fidelity}

\begin{figure}
    \centering
    \includegraphics[width=7cm]{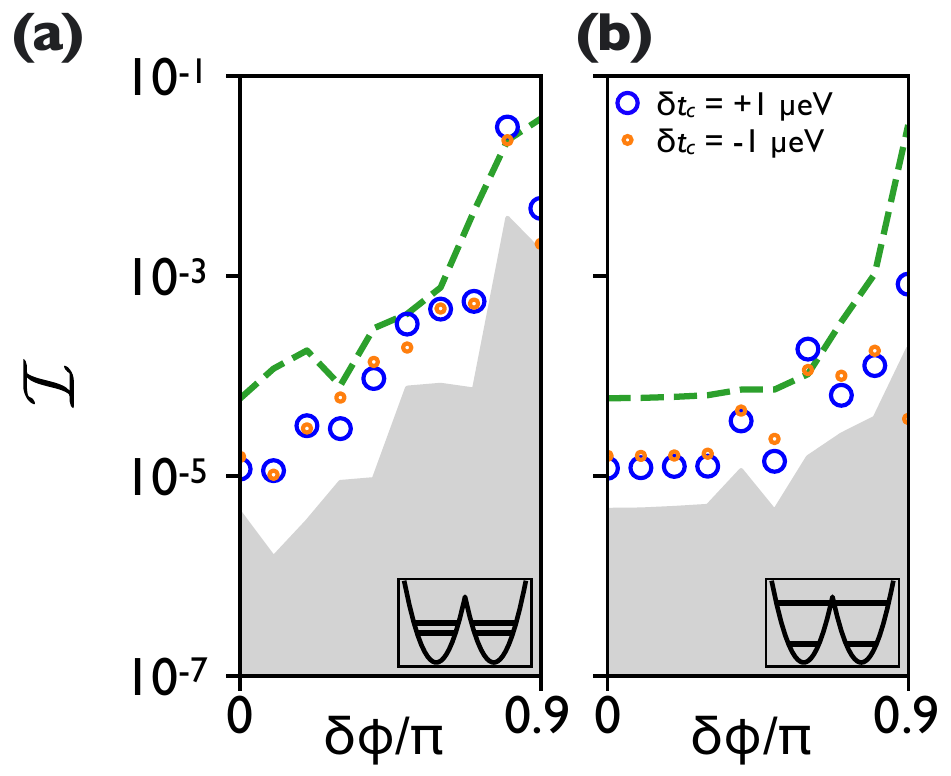}
    \caption{Estimating the pulse infidelity due to fluctuations in the tunnel coupling $t_c$. For $\delta \phi$ from $0$ to $0.9\pi$, we compute the infidelity resulting from tunnel coupling fluctuations $\delta t_c = \pm 1$~\SI{}{\micro\electronvolt}, for valley splittings (a) $E_{vL} = E_{vR} = 20$~\SI{}{\micro\electronvolt} and (b) $E_{vL} = E_{vR} = 100$~\SI{}{\micro\electronvolt}, for pulses optimized for $\sigma_\varepsilon = 15$~\SI{}{\micro\electronvolt}. We also plot the average infidelity due to detuning fluctuations for the cosine pulse, computed with Eq.~(\ref{eq:tot_inf}) [green dashed lines]. The gray regions indicate the infidelity without charge noise, $\mathcal{I}_0$, and serves as a lower bound on pulse fidelity. }
    \label{fig:tc_infidelity}
\end{figure}

\begin{figure}
    \centering
    \includegraphics[width=6cm]{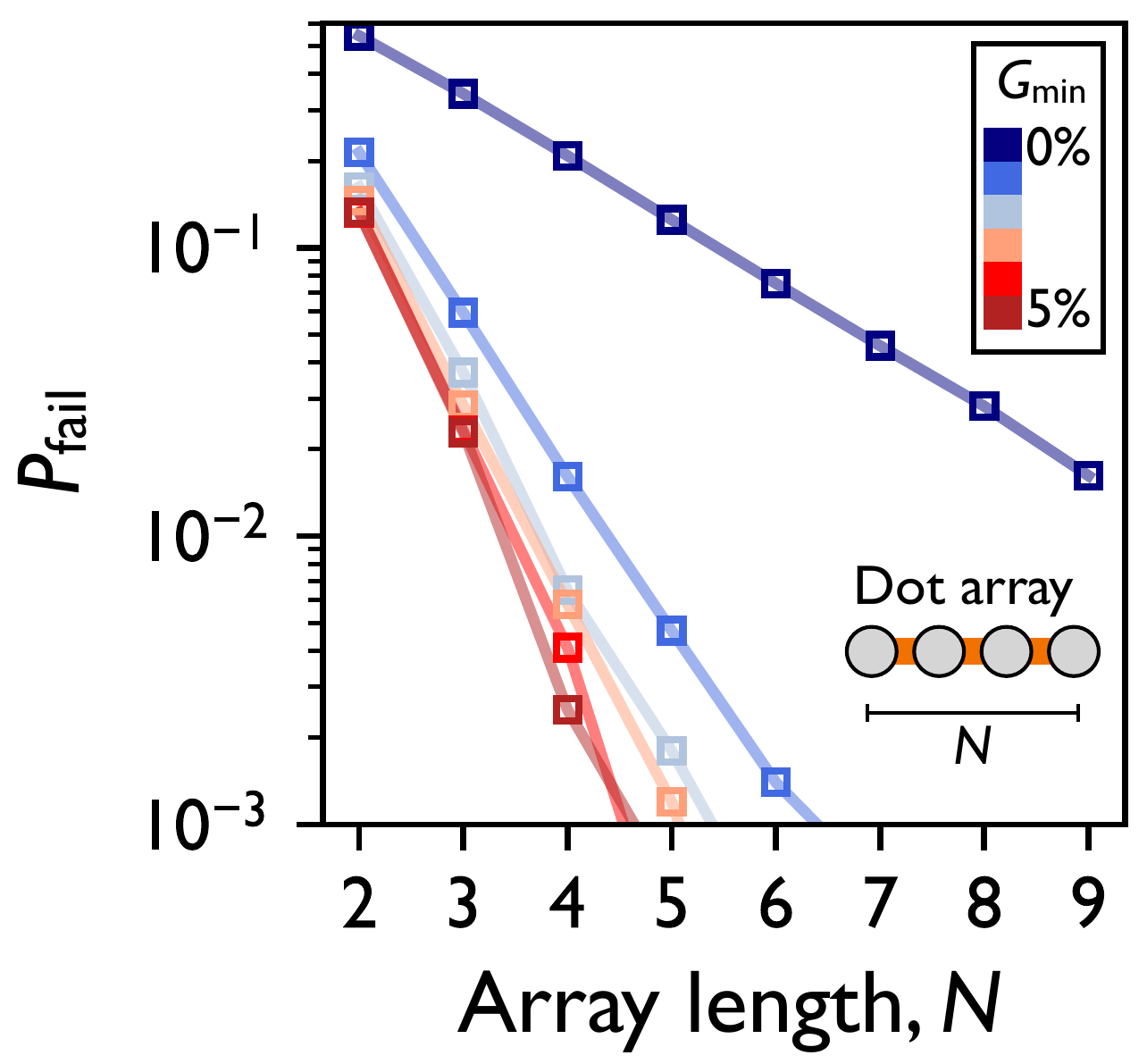}
    \caption{We plot $P_\text{fail}$, defined in Eq.~\ref{eq:pfail}, for linear arrays of quantum dots with size $N$. Each data point is computed from 10,000 instantiations of random alloy disorder, assuming $E_v$ is uncorrelated between neighboring dots. Colors indicate the minimum Ge concentration $G_\text{min}$, which we vary from $0$ to $5$~\%. }
    \label{fig:linear_array}
\end{figure}

To estimate the expected infidelity due to tunnel coupling fluctuations, we examine the cosine pulse family, in the pessimistic $\sigma_\varepsilon = 15$~\SI{}{\micro\electronvolt} charge noise regime.
We expect the effects of charge noise to be more apparent for large noise amplitudes.
Starting with a pulse optimized for either $E_{vL} = E_{vR} = 20$~\SI{}{\micro\electronvolt} or $E_{vL} = E_{vR} = 100$~\SI{}{\micro\electronvolt}, we then apply a $t_c$ fluctuation equal to $\pm 1$~\SI{}{\micro\electronvolt} to the Hamiltonian, and we compute the resulting pulse infidelity under the modified Hamiltonian.
Results are shown as circles in Fig.~\ref{fig:tc_infidelity} for $E_{vL} = E_{vR} = 20$~\SI{}{\micro\electronvolt} (a) and $E_{vL} = E_{vR} = 100$~\SI{}{\micro\electronvolt} (b), where large blue circles show infidelities for $\delta t_c = + 1$~\SI{}{\micro\electronvolt}, and small red circles show infidelities for $\delta t_c = - 1$~\SI{}{\micro\electronvolt}.
The green dashed lines in (a) and (b) indicate the infidelity due to detuning fluctuations, and the gray boundary indicates the baseline infidelity $\mathcal{I}_0$, as described in the main text.
We note that $\delta t_c = \pm 1$~\SI{}{\micro\electronvolt} is quite a large fluctuation, similar to the largest $\delta t_c$ observed above.
In Fig.~\ref{fig:tc_infidelity} (a) and (b), despite these very large fluctuations in $t_c$, we observe that the infidelity due to $t_c$ fluctuations is almost always smaller than the expected infidelity due to detuning fluctuations.
Hence, for the qubits considered in this work, we do not expect tunnel coupling fluctuations to be a dominant source of infidelity.

\section{Simulations of random $E_v$ landscapes} \label{app:vertical_tunability}

In both Sec.~\ref{sec:valley_fluctuations} and \ref{sec:scalability}, we have performed simulations of the spatial distribution of valley splittings along a 1D or 2D landscape. 
We use the GSTools python library [\citenum{Mueller:2023gstools}] to generate these landscapes, assuming the real and imaginary components of $\Delta$ are uncorrelated, and the spatial covariance of $\Delta$ is given by Eq.~(\ref{eq:delta_cov}).

\section{Linear quantum dot array} \label{app:linear_array}

In Section~\ref{sec:scalability} of the main text, we considered sparse grids of quantum dots, showing that they enable a much higher probability of finding a high-fidelity qubit.
Here, we perform the same analysis for linear arrays of quantum dots, illustrated in the inset to Fig.~\ref{fig:linear_array}. 
As in the main text, we estimate $P_\text{fail}$ for each configuration of dots, assuming the valley parameters are uncorrelated between neighboring dots, from 10,000 simulations of random alloy disorder, assuming a minimum QW Ge concentration $G_\text{min}$ from 0 to 5\%. 
Like we found for sparse grids of dots, we find that $P_\text{fail}$ falls dramatically as the number of dots in the grid is increased. 
For the same number of dots, we find the square grid configuration achieves slightly smaller $P_\text{fail}$ than the linear chain, since more qubits can be formed in the square grid.
For example, for 0\% Ge quantum wells, we find 9 dots arranged in a line achieves $P_\text{fail} \approx 1.6$~\%, while 9 dots arranged in a square grid achieves $P_\text{fail} \approx 0.7$~\%.

\section{The importance of large $t_c$} \label{app:weak_tc}

\begin{figure}
    \centering
    \includegraphics[width=6cm]{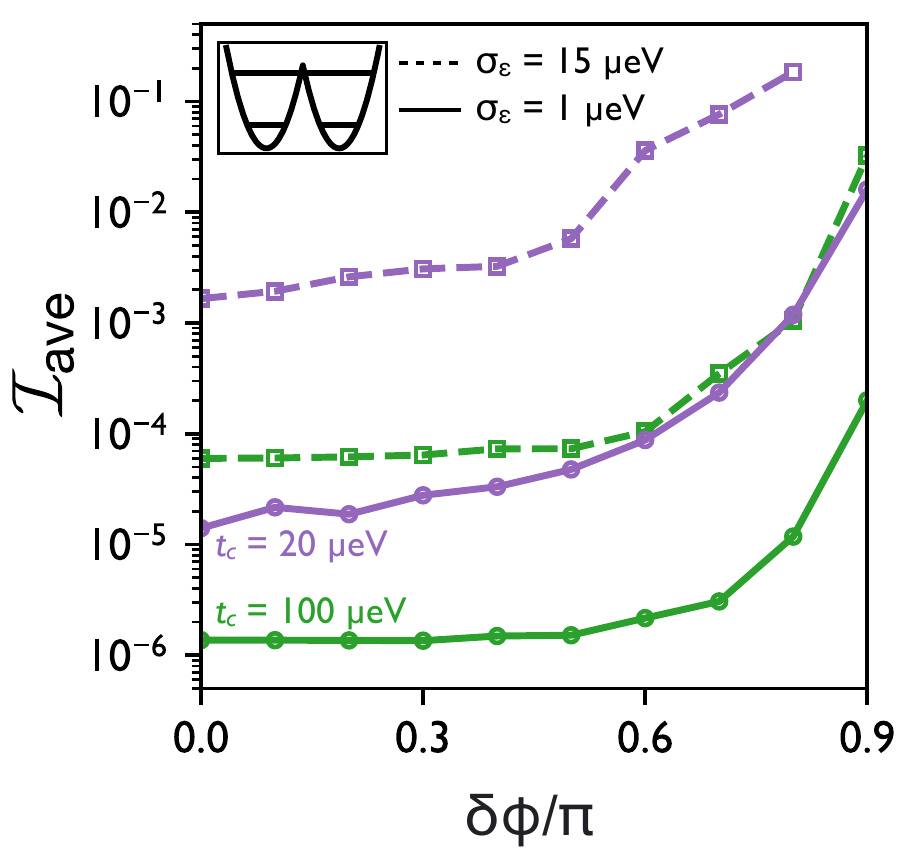}
    \caption{Large tunnel couplings significantly reduce pulse infidelities. We show average infidelities $\mathcal{I}_\mathrm{ave}$, defined in Eq.~(\ref{eq:tot_inf}), of optimized cosine pulses as a function of the valley phase difference $\delta \phi$, for $E_{vL} = E_{vR} = 100$~\SI{}{\micro\electronvolt}. Results are shown for a reduced $t_c = 20$~\SI{}{\micro\electronvolt} [purple lines], as well for the strongly coupled $t_c = 100$~\SI{}{\micro\electronvolt} [green lines, reproduced from Fig.~\ref{fig:phase_dependence}(a)]. In both cases, we optimize pulses either for the optimistic $\sigma_\varepsilon = 1$~\SI{}{\micro\electronvolt} (solid lines, circle markers) or the pessimistic $\sigma_\varepsilon = 15$~\SI{}{\micro\electronvolt} (dashed lines, square markers). We note that the data point for $t_c = 20$~\SI{}{\micro\electronvolt}, $\delta \phi = 0.9\pi$, and $\sigma_\varepsilon = 15$~\SI{}{\micro\electronvolt} is not included due to numerical difficulties optimizing the pulse under these conditions.}
    \label{fig:reduced_tc}
\end{figure}

In this section, we examine how the tunnel coupling $t_c$ impacts the fidelity of strongly-driven flopping-mode qubits. 
Using the same optimization procedure described above, we optimize cosine pulses assuming a smaller tunnel coupling $t_c = 20$~\SI{}{\micro\electronvolt}, for $E_{vL} = E_{vR} = 100$~\SI{}{\micro\electronvolt} and a range of valley phase differences $\delta \phi$. 
The resulting infidelities $\mathcal{I}_\mathrm{ave}$ for the reduced $t_c$ pulses are plotted for both the optimistic and pessimistic charge noise regimes in Fig.~\ref{fig:reduced_tc} (purple lines).
In addition, we include results from the strongly-tunnel-coupled $t_c = 100$~\SI{}{\micro\electronvolt} case, reproduced from Fig.~\ref{fig:phase_dependence}(a) [green lines].
With weaker tunnel coupling, the resulting infidelities are one to two orders of magnitude larger, across all valley phase configurations, for both the optimistic and pessimistic charge noise regimes.
Thus, we conclude that a large $t_c$ is critical for optimal charge noise resilience in the strongly-driven flopping mode qubit.

\bibliography{bibliography}

@article{Benito:2019p125430,
  title = {Electric-field control and noise protection of the flopping-mode spin qubit},
  author = {Benito, M. and Croot, X. and Adelsberger, C. and Putz, S. and Mi, X. and Petta, J. R. and Burkard, Guido},
  journal = {Phys. Rev. B},
  volume = {100},
  issue = {12},
  pages = {125430},
  numpages = {10},
  year = {2019},
  month = {Sep},
  publisher = {American Physical Society},
  doi = {10.1103/PhysRevB.100.125430},
  url = {https://link.aps.org/doi/10.1103/PhysRevB.100.125430}
}

@article{Bermeister:2014p192102,
    author = {Bermeister, Adam and Keith, Daniel and Culcer, Dimitrie},
    title = "{Charge noise, spin-orbit coupling, and dephasing of single-spin qubits}",
    journal = {Applied Physics Letters},
    volume = {105},
    number = {19},
    pages = {192102},
    year = {2014},
    month = {11},
    issn = {0003-6951},
    doi = {10.1063/1.4901162},
}

@article{Bezanson:2017julia,
  title={Julia: A fresh approach to numerical computing},
  author={Bezanson, Jeff and Edelman, Alan and Karpinski, Stefan and Shah, Viral B},
  journal={SIAM review},
  volume={59},
  number={1},
  pages={65--98},
  year={2017},
  publisher={SIAM},
  url={https://doi.org/10.1137/141000671}
}

@article{Borjans:2019p044063,
    title = {Single-Spin Relaxation in a Synthetic Spin-Orbit Field},
    author = {Borjans, F. and Zajac, D.M. and Hazard, T.M. and Petta, J.R.},
    journal = {Phys. Rev. Applied},
    volume = {11},
    issue = {4},
    pages = {044063},
    numpages = {9},
    year = {2019},
    month = {Apr},
    publisher = {American Physical Society},
    doi = {10.1103/PhysRevApplied.11.044063},
    url = {https://link.aps.org/doi/10.1103/PhysRevApplied.11.044063}
}

@article{Borselli:2011p123118,
    author = {Borselli,M. G.  and Ross,R. S.  and Kiselev,A. A.  and Croke,E. T.  and Holabird,K. S.  and Deelman,P. W.  and Warren,L. D.  and Alvarado-Rodriguez,I.  and Milosavljevic,I.  and Ku,F. C.  and Wong,W. S.  and Schmitz,A. E.  and Sokolich,M.  and Gyure,M. F.  and Hunter,A. T. },
    title = {Measurement of valley splitting in high-symmetry {S}i/{S}i{G}e quantum dots},
    journal = {Applied Physics Letters},
    volume = {98},
    number = {12},
    pages = {123118},
    year = {2011},
    doi = {10.1063/1.3569717}
}

@article{Burkard:2023p025003,
  title = {Semiconductor spin qubits},
  author = {Burkard, Guido and Ladd, Thaddeus D. and Pan, Andrew and Nichol, John M. and Petta, Jason R.},
  journal = {Rev. Mod. Phys.},
  volume = {95},
  issue = {2},
  pages = {025003},
  numpages = {58},
  year = {2023},
  month = {Jun},
  publisher = {American Physical Society},
  doi = {10.1103/RevModPhys.95.025003},
  url = {https://link.aps.org/doi/10.1103/RevModPhys.95.025003}
}

@article{Buterakos:22021p010341,
  title = {Geometrical Formalism for Dynamically Corrected Gates in Multiqubit Systems},
  author = {Buterakos, Donovan and Das Sarma, Sankar and Barnes, Edwin},
  journal = {PRX Quantum},
  volume = {2},
  issue = {1},
  pages = {010341},
  numpages = {12},
  year = {2021},
  month = {Mar},
  publisher = {American Physical Society},
  doi = {10.1103/PRXQuantum.2.010341},
  url = {https://link.aps.org/doi/10.1103/PRXQuantum.2.010341}
}

@article{Chen:2021p044033,
    title = {Detuning Axis Pulsed Spectroscopy of Valley-Orbital States in $\mathrm{Si}$/$\mathrm{Si}$-$\mathrm{Ge}$ Quantum Dots},
    author = {Chen, Edward H. and Raach, Kate and Pan, Andrew and Kiselev, Andrey A. and Acuna, Edwin and Blumoff, Jacob Z. and Brecht, Teresa and Choi, Maxwell D. and Ha, Wonill and Hulbert, Daniel R. and Jura, Michael P. and Keating, Tyler E. and Noah, Ramsey and Sun, Bo and Thomas, Bryan J. and Borselli, Matthew G. and Jackson, C.A.C. and Rakher, Matthew T. and Ross, Richard S.},
    journal = {Phys. Rev. Applied},
    volume = {15},
    issue = {4},
    pages = {044033},
    numpages = {13},
    year = {2021},
    month = {Apr},
    publisher = {American Physical Society},
    doi = {10.1103/PhysRevApplied.15.044033},
    url = {https://link.aps.org/doi/10.1103/PhysRevApplied.15.044033}
}

@misc{chtc,
  doi = {10.21231/GNT1-HW21},
  url = {https://chtc.cs.wisc.edu/},
  author = {{Center for High Throughput Computing}},
  title = {Center for High Throughput Computing},
  publisher = {Center for High Throughput Computing},
  year = {2006}
}

@article{Connors:2019p165305,
  title = {Low-frequency charge noise in {S}i/{S}i{G}e quantum dots},
  author = {Connors, Elliot J. and Nelson, JJ and Qiao, Haifeng and Edge, Lisa F. and Nichol, John M.},
  journal = {Phys. Rev. B},
  volume = {100},
  issue = {16},
  pages = {165305},
  numpages = {6},
  year = {2019},
  month = {Oct},
  publisher = {American Physical Society},
  doi = {10.1103/PhysRevB.100.165305},
  url = {https://link.aps.org/doi/10.1103/PhysRevB.100.165305}
}

@article{Connors:2022p940,
	Author = {Connors, Elliot J. and Nelson, J. and Edge, Lisa F. and Nichol, John M.},
	Da = {2022/02/17},
	Doi = {10.1038/s41467-022-28519-x},
	Id = {Connors2022},
	Isbn = {2041-1723},
	Journal = {Nature Communications},
	Number = {1},
	Pages = {940},
	Title = {Charge-noise spectroscopy of {S}i/{S}i{G}e quantum dots via dynamically-decoupled exchange oscillations},
	Ty = {JOUR},
	Url = {https://doi.org/10.1038/s41467-022-28519-x},
	Volume = {13},
	Year = {2022}
}

@article{Croot:2020p012006,
  title = {Flopping-mode electric dipole spin resonance},
  author = {Croot, X. and Mi, X. and Putz, S. and Benito, M. and Borjans, F. and Burkard, G. and Petta, J. R.},
  journal = {Phys. Rev. Res.},
  volume = {2},
  issue = {1},
  pages = {012006},
  numpages = {6},
  year = {2020},
  month = {Jan},
  publisher = {American Physical Society},
  doi = {10.1103/PhysRevResearch.2.012006}
}

@article{Culcer:2009p073102,
 author = {Culcer, Dimitrie and Hu, Xuedong and Das Sarma, S.},
 journal = {Appl. Phys. Lett.},
 pages = {073102},
 title = {Dephasing of {Si} spin qubits due to charge noise},
 volume = {95},
 year = {2009},
 doi = {10.1063/1.3194778},
 url = {https://doi.org/10.1063/1.3194778},
}

@article{Dodson:2022p146802,
  title = {How Valley-Orbit States in Silicon Quantum Dots Probe Quantum Well Interfaces},
  author = {Dodson, J. P. and Ercan, H. Ekmel and Corrigan, J. and Losert, Merritt P. and Holman, Nathan and McJunkin, Thomas and Edge, L. F. and Friesen, Mark and Coppersmith, S. N. and Eriksson, M. A.},
  journal = {Phys. Rev. Lett.},
  volume = {128},
  issue = {14},
  pages = {146802},
  numpages = {6},
  year = {2022},
  month = {Apr},
  publisher = {American Physical Society},
  doi = {10.1103/PhysRevLett.128.146802},
  url = {https://link.aps.org/doi/10.1103/PhysRevLett.128.146802}
}

@article{Dutta:1981p497,
 author = {Dutta, P. and Horn, P. M.},
 doi = {10.1103/RevModPhys.53.497},
 issue = {3},
 journal = {Rev. Mod. Phys.},
 link = {http://link.aps.org/doi/10.1103/RevModPhys.53.497},
 month = {Jul},
 numpages = {0},
 pages = {497--516},
 publisher = {American Physical Society},
 title = {Low-frequency fluctuations in solids: $\frac{1}{f}$ noise},
 volume = {53},
 year = {1981}
}

@article{Ercan:2022p247701,
  title = {Charge-Noise Resilience of Two-Electron Quantum Dots in $\mathrm{Si}/\mathrm{SiGe}$ Heterostructures},
  author = {Ercan, H. Ekmel and Friesen, Mark and Coppersmith, S. N.},
  journal = {Phys. Rev. Lett.},
  volume = {128},
  issue = {24},
  pages = {247701},
  numpages = {6},
  year = {2022},
  month = {Jun},
  publisher = {American Physical Society},
  doi = {10.1103/PhysRevLett.128.247701},
  url = {https://link.aps.org/doi/10.1103/PhysRevLett.128.247701}
}

@article{Esposti:2024p32,
	Author = {Degli Esposti, Davide and Stehouwer, Lucas E. A. and G{\"u}l, {\"O}nder and Samkharadze, Nodar and D{\'e}prez, Corentin and Meyer, Marcel and Meijer, Ilja N. and Tryputen, Larysa and Karwal, Saurabh and Botifoll, Marc and Arbiol, Jordi and Amitonov, Sergey V. and Vandersypen, Lieven M. K. and Sammak, Amir and Veldhorst, Menno and Scappucci, Giordano},
	Da = {2024/03/13},
	Doi = {10.1038/s41534-024-00826-9},
	Id = {Degli Esposti2024},
	Isbn = {2056-6387},
	Journal = {npj Quantum Information},
	Number = {1},
	Pages = {32},
	Title = {Low disorder and high valley splitting in silicon},
	Ty = {JOUR},
	Url = {https://doi.org/10.1038/s41534-024-00826-9},
	Volume = {10},
	Year = {2024}}

@article{Ferdous:2018p26,
    Author = {Ferdous, Rifat and Kawakami, Erika and Scarlino, Pasquale and Nowak, Micha{\l}P. and Ward, D. R. and Savage, D. E. and Lagally, M. G. and Coppersmith, S. N. and Friesen, Mark and Eriksson, Mark A. and Vandersypen, Lieven M. K. and Rahman, Rajib},
    Da = {2018/06/05},
    Doi = {10.1038/s41534-018-0075-1},
    Id = {Ferdous2018},
    Isbn = {2056-6387},
    Journal = {npj Quantum Information},
    Number = {1},
    Pages = {26},
    Title = {Valley dependent anisotropic spin splitting in silicon quantum dots},
    Ty = {JOUR},
    Url = {https://doi.org/10.1038/s41534-018-0075-1},
    Volume = {4},
    Year = {2018}
}

@article{Friesen:2007p115318,
 author = {M Friesen and S Chutia and C Tahan and S N Coppersmith},
 journal = {Phys. Rev. B},
 pages = {115318},
 title = {Valley splitting theory of {S}i{G}e/{S}i/{S}i{G}e quantum wells},
 volume = {75},
 year = {2007},
 doi = {10.1103/PhysRevB.75.115318},
 url = {https://link.aps.org/doi/10.1103/PhysRevB.75.115318}
}

@article{Gungordu:2022p023155,
  title = {Robust quantum gates using smooth pulses and physics-informed neural networks},
  author = {G\"ung\"ord\"u, Utkan and Kestner, J. P.},
  journal = {Phys. Rev. Res.},
  volume = {4},
  issue = {2},
  pages = {023155},
  numpages = {9},
  year = {2022},
  month = {May},
  publisher = {American Physical Society},
  doi = {10.1103/PhysRevResearch.4.023155},
  url = {https://link.aps.org/doi/10.1103/PhysRevResearch.4.023155}
}

@Article{Harris:2020p357,
 title         = {Array programming with {NumPy}},
 author        = {Charles R. Harris and K. Jarrod Millman and St{\'{e}}fan J.
                 van der Walt and Ralf Gommers and Pauli Virtanen and David
                 Cournapeau and Eric Wieser and Julian Taylor and Sebastian
                 Berg and Nathaniel J. Smith and Robert Kern and Matti Picus
                 and Stephan Hoyer and Marten H. van Kerkwijk and Matthew
                 Brett and Allan Haldane and Jaime Fern{\'{a}}ndez del
                 R{\'{i}}o and Mark Wiebe and Pearu Peterson and Pierre
                 G{\'{e}}rard-Marchant and Kevin Sheppard and Tyler Reddy and
                 Warren Weckesser and Hameer Abbasi and Christoph Gohlke and
                 Travis E. Oliphant},
 year          = {2020},
 month         = sep,
 journal       = {Nature},
 volume        = {585},
 number        = {7825},
 pages         = {357--362},
 doi           = {10.1038/s41586-020-2649-2},
 publisher     = {Springer Science and Business Media {LLC}},
 url           = {https://doi.org/10.1038/s41586-020-2649-2}
}

@article{Hollmann:2020p034068,
  title = {Large, Tunable Valley Splitting and Single-Spin Relaxation Mechanisms in a $\mathrm{Si}$/$\text{Si}_{x}$$\text{Ge}_{1\ensuremath{-}x}$ Quantum Dot},
  author = {Hollmann, Arne and Struck, Tom and Langrock, Veit and Schmidbauer, Andreas and Schauer, Floyd and Leonhardt, Tim and Sawano, Kentarou and Riemann, Helge and Abrosimov, Nikolay V. and Bougeard, Dominique and Schreiber, Lars R.},
  journal = {Phys. Rev. Applied},
  volume = {13},
  issue = {3},
  pages = {034068},
  numpages = {8},
  year = {2020},
  month = {Mar},
  publisher = {American Physical Society},
  doi = {10.1103/PhysRevApplied.13.034068},
  url = {https://link.aps.org/doi/10.1103/PhysRevApplied.13.034068}
}

@article{Hu:2023p134002,
    author = {Hu, Rui-Zi and Ma, Rong-Long and Ni, Ming and Zhou, Yuan and Chu, Ning and Liao, Wei-Zhu and Kong, Zhen-Zhen and Cao, Gang and Wang, Gui-Lei and Li, Hai-Ou and Guo, Guo-Ping},
    title = "{Flopping-mode spin qubit in a Si-MOS quantum dot}",
    journal = {Applied Physics Letters},
    volume = {122},
    number = {13},
    pages = {134002},
    year = {2023},
    month = {03},
    issn = {0003-6951},
    doi = {10.1063/5.0137259}
}

@Article{Hunter:2007p90,
  Author    = {Hunter, J. D.},
  Title     = {Matplotlib: A 2D graphics environment},
  Journal   = {Computing in Science \& Engineering},
  Volume    = {9},
  Number    = {3},
  Pages     = {90--95},
  publisher = {IEEE COMPUTER SOC},
  doi       = {10.1109/MCSE.2007.55},
  year      = 2007
}

@article{Kranz:2020p2003361,
author = {Kranz, Ludwik and Gorman, Samuel Keith and Thorgrimsson, Brandur and He, Yu and Keith, Daniel and Keizer, Joris Gerhard and Simmons, Michelle Yvonne},
title = {Exploiting a Single-Crystal Environment to Minimize the Charge Noise on Qubits in Silicon},
journal = {Advanced Materials},
volume = {32},
number = {40},
pages = {2003361},
doi = {https://doi.org/10.1002/adma.202003361},
year = {2020}
}

@article{Krzywda:2021p075439,
  title = {Interplay of charge noise and coupling to phonons in adiabatic electron transfer between quantum dots},
  author = {Krzywda, Jan A. and Cywi\ifmmode \acute{n}\else \'{n}\fi{}ski, \L{}ukasz},
  journal = {Phys. Rev. B},
  volume = {104},
  issue = {7},
  pages = {075439},
  numpages = {21},
  year = {2021},
  month = {Aug},
  publisher = {American Physical Society},
  doi = {10.1103/PhysRevB.104.075439},
  url = {https://link.aps.org/doi/10.1103/PhysRevB.104.075439}
}

@article{Li:2010p085313,
 author = {Li, Qiuzi and Cywi\ifmmode \acute{n}\else \'{n}\fi{}ski, \L{}ukasz and Culcer, Dimitrie and Hu, Xuedong and Das Sarma, S.},
 doi = {10.1103/PhysRevB.81.085313},
 issue = {8},
 journal = {Phys. Rev. B},
 link = {http://link.aps.org/doi/10.1103/PhysRevB.81.085313},
 month = {Feb},
 numpages = {17},
 pages = {085313},
 publisher = {American Physical Society},
 title = {Exchange coupling in silicon quantum dots: Theoretical considerations for quantum computation},
 volume = {81},
 year = {2010}
}

@article{Lima:2024p036202,
  title = {Valley splitting depending on the size and location of a silicon quantum dot},
  author = {Lima, Jonas R. F. and Burkard, Guido},
  journal = {Phys. Rev. Mater.},
  volume = {8},
  issue = {3},
  pages = {036202},
  numpages = {9},
  year = {2024},
  month = {Mar},
  publisher = {American Physical Society},
  doi = {10.1103/PhysRevMaterials.8.036202},
  url = {https://link.aps.org/doi/10.1103/PhysRevMaterials.8.036202}
}

@article{Lima:2023p025004,
    author = {Jonas R. F. Lima and Guido Burkard},
    title = {Interface and electromagnetic effects in the valley splitting of {S}i quantum dots},
    journal = {Mater. Quantum. Technol.},
    volume = {3},
    pages = {025004},
    year = {2023},
    doi = {10.1088/2633-4356/acd743},
    url = {https://doi.org/10.1088/2633-4356/acd743},
}

@article{Losert:2023p125405,
  title = {Practical strategies for enhancing the valley splitting in {S}i/{S}i{G}e quantum wells},
  author = {Losert, Merritt P. and Eriksson, M. A. and Joynt, Robert and Rahman, Rajib and Scappucci, Giordano and Coppersmith, Susan N. and Friesen, Mark},
  journal = {Phys. Rev. B},
  volume = {108},
  issue = {12},
  pages = {125405},
  numpages = {31},
  year = {2023},
  month = {Sep},
  publisher = {American Physical Society},
  doi = {10.1103/PhysRevB.108.125405},
  url = {https://link.aps.org/doi/10.1103/PhysRevB.108.125405}
}

@article{Losert:2024p040322,
  title = {Strategies for Enhancing Spin-Shuttling Fidelities in $\mathrm{Si}$/$\mathrm{Si}$$\mathrm{Ge}$ Quantum Wells with Random-Alloy Disorder},
  author = {Losert, Merritt P. and Oberl\"ander, Max and Teske, Julian D. and Volmer, Mats and Schreiber, Lars R. and Bluhm, Hendrik and Coppersmith, S.N. and Friesen, Mark},
  journal = {PRX Quantum},
  volume = {5},
  issue = {4},
  pages = {040322},
  numpages = {26},
  year = {2024},
  month = {Nov},
  publisher = {American Physical Society},
  doi = {10.1103/PRXQuantum.5.040322},
  url = {https://link.aps.org/doi/10.1103/PRXQuantum.5.040322}
}

@misc{Losert:2025zenodoV2,
      author = {Losert, Merritt P. and G{\"u}ng{\"o}rd{\"u}, Utkan and Coppersmith, S. N. and Friesen, Mark and Tahan, Charles},
      title = {Source data and code for the publication ``{T}he effects of alloy disorder on strongly-driven flopping mode qubits in {S}i/{S}i{G}e''},
      doi = {10.5281/zenodo.19390791},
      url = {http://dx.doi.org/10.5281/zenodo.19390791},
      note = {10.5281/zenodo.19390791},
    }

@article{Loss:1998p120,
 affiliation = {Univ Calif Santa Barbara, Inst Theoret Phys, Santa Barbara, CA 93106 USA},
 author = {D. Loss and D. P. DiVincenzo},
 journal = {Phys. Rev. A},
 number = {1},
 pages = {120--126},
 title = {Quantum computation with quantum dots},
 volume = {57},
 year = {1998},
 doi = {10.1103/PhysRevA.57.120},
 url = {https://link.aps.org/doi/10.1103/PhysRevA.57.120}
}

@article{McJunkin:2022p7777,
      title={Si{G}e quantum wells with oscillating {Ge} concentrations for quantum dot qubits}, 
      author={Thomas McJunkin and Benjamin Harpt and Yi Feng and Merritt Losert and Rajib Rahman and J. P. Dodson and M. A. Wolfe and D. E. Savage and M. G. Lagally and S. N. Coppersmith and Mark Friesen and Robert Joynt and M. A. Eriksson},
  journal = {Nature Commun.},
  volume = {13},
  pages = {7777},
  numpages = {7},
  year = {2022},
  month = {Dec},
  doi = {10.1038/s41467-022-35510-z},
}

@article{Mi:2017p176803,
    title = {High-Resolution Valley Spectroscopy of {S}i Quantum Dots},
    author = {Mi, X. and P\'eterfalvi, Csaba G. and Burkard, Guido and Petta, J. R.},
    journal = {Phys. Rev. Lett.},
    volume = {119},
    issue = {17},
    pages = {176803},
    numpages = {5},
    year = {2017},
    month = {Oct},
    publisher = {American Physical Society},
    doi = {10.1103/PhysRevLett.119.176803},
    url = {https://link.aps.org/doi/10.1103/PhysRevLett.119.176803}
}

@article{Mi:2018p161404,
 author = {X. Mi and S. Kohler and J. R. Petta},
 journal = {Phys. Rev. B},
 month = {Oct},
 pages = {161404(R)},
 title = {Landau-{Z}ener interferometry of valley-orbit states in {S}i/{S}i{G}e double quantum dots},
 volume = {98},
 year = {2018},
 doi = {10.1103/PhysRevB.98.161404},
 url = {https://link.aps.org/doi/10.1103/PhysRevB.98.161404}
}

@article{Mills:2022p5130,
author = {Adam R. Mills  and Charles R. Guinn  and Michael J. Gullans  and Anthony J. Sigillito  and Mayer M. Feldman  and Erik Nielsen  and Jason R. Petta },
title = {Two-qubit silicon quantum processor with operation fidelity exceeding 99\%},
journal = {Science Advances},
volume = {8},
number = {14},
pages = {eabn5130},
year = {2022},
doi = {10.1126/sciadv.abn5130},
}

@software{Mueller:2023gstools,
  author       = {Sebastian Müller and
                  Lennart Schüler},
  title        = {{G}eo{S}tat-{F}ramework/{G}{S}{T}ools: v1.5.0 `{N}ifty {N}eon'},
  month        = jun,
  year         = 2023,
  publisher    = {Zenodo},
  version      = {v1.5.0},
  doi          = {10.5281/zenodo.8044720},
  url          = {https://doi.org/10.5281/zenodo.8044720}
}

@article{Nelder:1965p308,
 author = {J. A. Nelder and R. Mead},
 journal = {Computer Journal},
 pages = {308-313},
 title = {A simplex method for function minimization},
 volume = {7}
}

@article{Neyens:2018p243107,
 author = {Neyens, Samuel F. and Foote, Ryan H. and Thorgrimsson, Brandur and Knapp, T. J. and McJunkin, Thomas and Vandersypen, L. M. K. and Amin, Payam and Thomas, Nicole K. and Clarke, James S. and Savage, D. E. and Lagally, M. G. and Friesen, Mark and Coppersmith, S. N. and Eriksson, M. A.},
 title = {The critical role of substrate disorder in valley splitting in {S}i quantum wells},
 journal = {Appl. Phys. Lett.},
 volume = {112},
 pages = {243107},
 year = {2018},
 doi = {10.1063/1.5033447},
 url = {https://doi.org/10.1063/1.5033447},
}

@article{Noiri:2022p338,
	Author = {Noiri, Akito and Takeda, Kenta and Nakajima, Takashi and Kobayashi, Takashi and Sammak, Amir and Scappucci, Giordano and Tarucha, Seigo},
	Da = {2022/01/01},
	Doi = {10.1038/s41586-021-04182-y},
	Id = {Noiri2022},
	Isbn = {1476-4687},
	Journal = {Nature},
	Number = {7893},
	Pages = {338--342},
	Title = {Fast universal quantum gate above the fault-tolerance threshold in silicon},
	Ty = {JOUR},
	Volume = {601},
	Year = {2022},
}

@article{Oh:2021p125122,
	Author = {Oh,Seong Woo and Denisov,Artem O. and Chen,Pengcheng and Petta,Jason R.},
	Booktitle = {AIP Advances},
	Da = {2021/12/01},
	Date = {2021/12/01},
	Doi = {10.1063/5.0056648},
	Journal = {AIP Advances},
	M3 = {doi: 10.1063/5.0056648},
	Month = {2022/03/28},
	Number = {12},
	Pages = {125122},
	Publisher = {American Institute of Physics},
	Title = {Cryogen-free scanning gate microscope for the characterization of {S}i/$\text{Si}_{0.7}\text{Ge}_{0.3}$ quantum devices at milli-{K}elvin temperatures},
	Ty = {JOUR},
	Url = {https://doi.org/10.1063/5.0056648},
	Volume = {11},
	Year = {2021},
	Year1 = {2021}}

@article{Saraiva:2022p2105488,
author = {Saraiva, Andre and Lim, Wee Han and Yang, Chih Hwan and Escott, Christopher C. and Laucht, Arne and Dzurak, Andrew S.},
title = {Materials for Silicon Quantum Dots and their Impact on Electron Spin Qubits},
journal = {Advanced Functional Materials},
volume = {32},
number = {3},
pages = {2105488},
doi = {https://doi.org/10.1002/adfm.202105488},
year = {2022}
}

@article{Scarlino:2017p165429,
  title={Dressed photon-orbital states in a quantum dot: Intervalley spin resonance},
  author={Scarlino, P. and Kawakami, E. and Jullien, T. and Ward, D. R. and Savage, D. E. and Lagally, M. G. and Friesen, Mark and Coppersmith, S. N. and Eriksson, M. A. and Vandersypen, L. M. K.},
  journal={Physical Review B},
  volume={95},
  number={16},
  pages={165429},
  year={2017},
  doi = {10.1103/PhysRevB.95.165429},
  url = {https://link.aps.org/doi/10.1103/PhysRevB.95.165429}
}

@article{Schaffler:1997p1515,
 author = {F Sch\"{a}ffler},
 journal = {Semicond. Sci. Tech.},
 number = {12},
 pages = {1515-1549},
 title = {High-mobility {S}i and {G}e structures},
 volume = {12},
 year = {1997},
 doi = {10.1088/0268-1242/12/12/001},
 url = {https://doi.org/10.1088/0268-1242/12/12/001},
}

@article{Shehata:2023p045305,
  title = {Modeling semiconductor spin qubits and their charge noise environment for quantum gate fidelity estimation},
  author = {Shehata, M. Mohamed El Kordy and Simion, George and Li, Ruoyu and Mohiyaddin, Fahd A. and Wan, Danny and Mongillo, Massimo and Govoreanu, Bogdan and Radu, Iuliana and De Greve, Kristiaan and Van Dorpe, Pol},
  journal = {Phys. Rev. B},
  volume = {108},
  issue = {4},
  pages = {045305},
  numpages = {28},
  year = {2023},
  month = {Jul},
  publisher = {American Physical Society},
  doi = {10.1103/PhysRevB.108.045305},
  url = {https://link.aps.org/doi/10.1103/PhysRevB.108.045305}
}

@article{Shi:2011p233108,
 author = {Shi, Z. and Simmons, C. B. and Prance, J.R. and Gamble, John King and Friesen, Mark and Savage, D. E. and Lagally, M. G. and Coppersmith, S. N. and Eriksson, M. A.},
 journal = {Appl. Phys. Lett.},
 pages = {233108},
 title = {Tunable singlet-triplet splitting in a few-electron {Si/SiGe} quantum dot},
 volume = {99},
 year = {2011},
 doi = {10.1063/1.3666232},
 url = {https://doi.org/10.1063/1.3666232},
}

@article{Shiau:2007p195345,
  title = {Valley {K}ondo effect in silicon quantum dots},
  author = {Shiau, Shiue-yuan and Chutia, Sucismita and Joynt, Robert},
  journal = {Phys. Rev. B},
  volume = {75},
  issue = {19},
  pages = {195345},
  numpages = {14},
  year = {2007},
  month = {May},
  publisher = {American Physical Society},
  doi = {10.1103/PhysRevB.75.195345},
  url = {https://link.aps.org/doi/10.1103/PhysRevB.75.195345}
}

@article{Struck:2020p40,
    Author = {Struck, Tom and Hollmann, Arne and Schauer, Floyd and Fedorets, Olexiy and Schmidbauer, Andreas and Sawano, Kentarou and Riemann, Helge and Abrosimov, Nikolay V. and Cywi{\'n}ski, {\L}ukasz and Bougeard, Dominique and Schreiber, Lars R.},
    Da = {2020/05/15},
    Doi = {10.1038/s41534-020-0276-2},
    Id = {Struck2020},
    Isbn = {2056-6387},
    Journal = {npj Quantum Information},
    Number = {1},
    Pages = {40},
    Title = {Low-frequency spin qubit energy splitting noise in highly purified $^{28}${S}i/{S}i{G}e},
    Ty = {JOUR},
    Volume = {6},
    Year = {2020},
}

@article{Teske:2023p035302,
  title = {Flopping-mode electron dipole spin resonance in the strong-driving regime},
  author = {Teske, Julian D. and Butt, Friederike and Cerfontaine, Pascal and Burkard, Guido and Bluhm, Hendrik},
  journal = {Phys. Rev. B},
  volume = {107},
  issue = {3},
  pages = {035302},
  numpages = {17},
  year = {2023},
  month = {Jan},
  publisher = {American Physical Society},
  doi = {10.1103/PhysRevB.107.035302},
  url = {https://link.aps.org/doi/10.1103/PhysRevB.107.035302}
}

@ARTICLE{Virtanen:2020p261,
  author  = {Virtanen, Pauli and Gommers, Ralf and Oliphant, Travis E. and
            Haberland, Matt and Reddy, Tyler and Cournapeau, David and
            Burovski, Evgeni and Peterson, Pearu and Weckesser, Warren and
            Bright, Jonathan and {van der Walt}, St{\'e}fan J. and
            Brett, Matthew and Wilson, Joshua and Millman, K. Jarrod and
            Mayorov, Nikolay and Nelson, Andrew R. J. and Jones, Eric and
            Kern, Robert and Larson, Eric and Carey, C J and
            Polat, {\.I}lhan and Feng, Yu and Moore, Eric W. and
            {VanderPlas}, Jake and Laxalde, Denis and Perktold, Josef and
            Cimrman, Robert and Henriksen, Ian and Quintero, E. A. and
            Harris, Charles R. and Archibald, Anne M. and
            Ribeiro, Ant{\^o}nio H. and Pedregosa, Fabian and
            {van Mulbregt}, Paul and {SciPy 1.0 Contributors}},
  title   = {{{SciPy} 1.0: Fundamental Algorithms for Scientific
            Computing in Python}},
  journal = {Nature Methods},
  year    = {2020},
  volume  = {17},
  pages   = {261--272},
  adsurl  = {https://rdcu.be/b08Wh},
  doi     = {10.1038/s41592-019-0686-2},
}

@article{Volmer:2024p61,
	Author = {Volmer, Mats and Struck, Tom and Sala, Arnau and Chen, Bingjie and Oberl{\"a}nder, Max and Offermann, Tobias and Xue, Ran and Visser, Lino and Tu, Jhih-Sian and Trellenkamp, Stefan and Cywi{\'n}ski, {\L}ukasz and Bluhm, Hendrik and Schreiber, Lars R.},
	Da = {2024/06/19},
	Doi = {10.1038/s41534-024-00852-7},
	Id = {Volmer2024},
	Isbn = {2056-6387},
	Journal = {npj Quantum Information},
	Number = {1},
	Pages = {61},
	Title = {Mapping of valley splitting by conveyor-mode spin-coherent electron shuttling},
	Ty = {JOUR},
	Url = {https://doi.org/10.1038/s41534-024-00852-7},
	Volume = {10},
	Year = {2024}
}

@misc{Volmer:2025arXiv,
      title={Reduction of the impact of the local valley splitting on the coherence of conveyor-belt spin shuttling in $^{28}${S}i/{S}i{G}e}, 
      author={Mats Volmer and Tom Struck and Jhih-Sian Tu and Stefan Trellenkamp and Davide Degli Esposti and Giordano Scappucci and Łukasz Cywiński and Hendrik Bluhm and Lars R. Schreiber},
      year={2025},
      eprint={2510.03773},
      archivePrefix={arXiv},
}

@article{Woods:2024p54,
	Author = {Woods, Benjamin D. and Soomro, Hudaiba and Joseph, E. S. and Frink, Collin C. D. and Joynt, Robert and Eriksson, M. A. and Friesen, Mark},
	Da = {2024/05/31},
	Doi = {10.1038/s41534-024-00853-6},
	Id = {Woods2024},
	Isbn = {2056-6387},
	Journal = {npj Quantum Information},
	Number = {1},
	Pages = {54},
	Title = {Coupling conduction-band valleys in {S}i{G}e heterostructures via shear strain and {G}e concentration oscillations},
	Ty = {JOUR},
	Url = {https://doi.org/10.1038/s41534-024-00853-6},
	Volume = {10},
	Year = {2024}
}

@article{Wuetz:2022p7730,
	Author = {Paquelet Wuetz, Brian and Losert, Merritt P. and Koelling, Sebastian and Stehouwer, Lucas E. A. and Zwerver, Anne-Marije J. and Philips, Stephan G. J. and M{\k a}dzik, Mateusz T. and Xue, Xiao and Zheng, Guoji and Lodari, Mario and Amitonov, Sergey V. and Samkharadze, Nodar and Sammak, Amir and Vandersypen, Lieven M. K. and Rahman, Rajib and Coppersmith, Susan N. and Moutanabbir, Oussama and Friesen, Mark and Scappucci, Giordano},
	Da = {2022/12/13},
	Doi = {10.1038/s41467-022-35458-0},
	Id = {Paquelet Wuetz2022},
	Isbn = {2041-1723},
	Journal = {Nature Communications},
	Number = {1},
	Pages = {7730},
	Title = {Atomic fluctuations lifting the energy degeneracy in {S}i/{S}i{G}e quantum dots},
	Ty = {JOUR},
	Volume = {13},
	Year = {2022}
}

@article{Xue:2022p343,
    Author = {Xue, Xiao and Russ, Maximilian and Samkharadze, Nodar and Undseth, Brennan and Sammak, Amir and Scappucci, Giordano and Vandersypen, Lieven M. K.},
    Da = {2022/01/01},
    Doi = {10.1038/s41586-021-04273-w},
    Id = {Xue2022},
    Isbn = {1476-4687},
    Journal = {Nature},
    Number = {7893},
    Pages = {343--347},
    Title = {Quantum logic with spin qubits crossing the surface code threshold},
    Ty = {JOUR},
    Url = {https://doi.org/10.1038/s41586-021-04273-w},
    Volume = {601},
    Year = {2022}
}

@article{Yoneda:2018p102,
author = {Yoneda, Jun and Takeda, Kenta and Otsuka, Tomohiro and Nakajima, Takashi and Delbecq, Matthieu R. and Allison, Giles and Honda, Takumu and Kodera, Tetsuo and Oda, Shunri and Hoshi, Yusuke and Usami, Noritaka and Itoh, Kohei M. and Tarucha, Seigo},
title = {{A quantum-dot spin qubit with coherence limited by charge noise and fidelity higher than 99.9\%}},
journal = {Nature Nanotechnol.},
year = {2018},
pages = {102},
volume = {13},
Url = {https://doi.org/10.1038/s41565-017-0014-x},
Doi = {10.1038/s41565-017-0014-x},
}

@article{Young:2025p064042,
  title = {Benchmarking low-power flopping-mode spin-qubit fidelities in {S}i/{S}i$_{0.7}${G}e$_{0.3}$ devices with alloy disorder},
  author = {Young, Steve and Brickson, Mitchell and Petta, Jason R. and Jacobson, N. Tobias},
  journal = {Phys. Rev. Appl.},
  volume = {24},
  issue = {6},
  pages = {064042},
  numpages = {9},
  year = {2025},
  month = {Dec},
  publisher = {American Physical Society},
  doi = {10.1103/j4ww-lt1l},
  url = {https://link.aps.org/doi/10.1103/j4ww-lt1l}
}

@article{Zajac:2015p223507,
 adsurl = {http://adsabs.harvard.edu/abs/2015ApPhL.106v3507Z},
 author = {{Zajac}, D.~M. and {Hazard}, T.~M. and {Mi}, X. and {Wang}, K. and
{Petta}, J.~R.},
 doi = {10.1063/1.4922249},
 eid = {223507},
 journal = {Appl. Phys. Lett.},
 pages = {223507},
 title = {A reconfigurable gate architecture for {S}i/{S}i{G}e quantum dots},
 volume = {106},
 year = {2015}
}

@article{Zhao:2022p064043,
  title = {Measurement of Tunnel Coupling in a Si Double Quantum dot Based on Charge Sensing},
  author = {Zhao, Xinyu and Hu, Xuedong},
  journal = {Phys. Rev. Appl.},
  volume = {17},
  issue = {6},
  pages = {064043},
  numpages = {14},
  year = {2022},
  month = {Jun},
  publisher = {American Physical Society},
  doi = {10.1103/PhysRevApplied.17.064043},
  url = {https://link.aps.org/doi/10.1103/PhysRevApplied.17.064043}
}

@article{Zwanenburg:2013p961,
 author = {Zwanenburg, F. A. and Dzurak, A. S. and Morello, A. and Simmons, M. Y. and Hollenberg, L. C. L. and Klimeck, G. and Rogge, S. and Coppersmith, S. N. and Eriksson, M. A.},
 journal = {Rev. Mod. Phys.},
 pages = {961},
 title = {Silicon Quantum Electronics},
 volume = {85},
 year = {2013},
 doi = {10.1103/RevModPhys.85.961},
 url = {https://link.aps.org/doi/10.1103/RevModPhys.85.961}
}

\end{document}